\documentclass[10pt,journal,compsoc]{IEEEtran}
\usepackage{graphicx}
\usepackage{textcomp}
\usepackage{xcolor}
\usepackage{amsmath}
\usepackage{booktabs}
\usepackage{algorithm}

\usepackage{amsfonts} 
\usepackage{float}
\usepackage{lipsum}
\usepackage{ragged2e}

\usepackage{algpseudocode}
\usepackage{array}
\usepackage{multirow}

\usepackage{longtable}
\usepackage{rotating}
\algnewcommand\Assert[1]{\State \algorithmicassert(#1)}

\makeatletter
\newenvironment{breakablealgorithm}
{% \begin{breakablealgorithm}
    \begin{center}
        \refstepcounter{algorithm}% New algorithm
        \hrule height.8pt depth0pt \kern2pt% \@fs@pre for \@fs@ruled
        \renewcommand{\caption}[2][\relax]{% Make a new \caption
            {\raggedright\textbf{\ALG@name~\thealgorithm} ##2\par}%
            \ifx\relax##1\relax % #1 is \relax
            \addcontentsline{loa}{algorithm}{\protect\numberline{\thealgorithm}##2}%
            \else % #1 is not \relax
            \addcontentsline{loa}{algorithm}{\protect\numberline{\thealgorithm}##1}%
            \fi
            \kern2pt\hrule\kern2pt
        }
    }{% \end{breakablealgorithm}
        \kern2pt\hrule\relax% \@fs@post for \@fs@ruled
    \end{center}
}
\makeatother

\usepackage[colorlinks,linkcolor=black,anchorcolor=black,citecolor=black]{hyperref}
\usepackage{subfigure}

\ifCLASSOPTIONcompsoc
  \usepackage[nocompress]{cite}
\else
  \usepackage{cite}
\fi

\ifCLASSINFOpdf
\else
\fi

\begin{document}

\title{Adaptive Resource Allocation for Workflow Containerization on Kubernetes}

\author{SHAN Chenggang,
        WU Chuge,
        XIA Yuanqing,%~\IEEEmembership{Senior Member,~IEEE,}
        GUO Zehua,%~\IEEEmembership{Senior Member,~IEEE,}
        LIU Danyang,
        and ZHANG Jinhui % <-this % stops a space
\IEEEcompsocitemizethanks{
\IEEEcompsocthanksitem C.~Shan and D.~Liu are with the School of Automation, 
Beijing Institute of Technology, Beijing 100081, China. C.~Shan is also with 
the School of Artificial Intelligence, Zaozhuang University, Zaozhuang 277100, China. 
E-mail:~\{uzz\_scg,~liudanyang093\}@163.com

\IEEEcompsocthanksitem 

C.~Wu, Y.~Xia, Z.~Guo and J.~Zhang are with the School of Automation, 
Beijing Institute of Technology, Beijing 100081, China.
E-mail:~\{wucg, xia\_yuanqing, guo, zhangjinh\}@bit.edu.cn
\protect\\
}
\thanks{
%%Manuscript received April 19, 2005; revised August 26, 2015.
(Corresponding author: Jinhui Zhang)
}
}

\markboth{}%
{Shan \MakeLowercase{\textit{et al.}}: Adaptive Resource Allocation for Workflow Containerization on Kubernetes}

\IEEEtitleabstractindextext{%

\justifying
\begin{abstract}
In a cloud-native era, the Kubernetes-based workflow engine enables workflow containerized execution through the inherent abilities of Kubernetes. 
However, when encountering continuous workflow requests and unexpected resource request spikes, the engine is limited to the current workflow load 
information for resource allocation, which lacks the agility and predictability of resource allocation, resulting in over and under-provisioning resources. 
This mechanism seriously hinders workflow execution efficiency and leads to high resource waste. 
To overcome these drawbacks, we propose an adaptive resource allocation scheme named ARAS for the Kubernetes-based workflow engines. 
Considering potential future workflow task requests within the current task pod's lifecycle, the ARAS uses a resource scaling strategy to allocate resources 
in response to high-concurrency workflow scenarios. 
The ARAS offers resource discovery, resource evaluation, and allocation functionalities and serves as a key component for our tailored workflow engine (KubeAdaptor). 
By integrating the ARAS into KubeAdaptor for workflow containerized execution, we demonstrate the practical abilities of KubeAdaptor and 
the advantages of our ARAS. 
Compared with the baseline algorithm, experimental evaluation under three distinct workflow arrival patterns shows that ARAS gains time-saving 
of $9.8\%$ to $40.92\%$ in the average total duration of all workflows, time-saving of $26.4\%$ to $79.86\%$ in the average duration of 
individual workflow, and an increase of $1\%$ to $16\%$ in CPU and memory resource usage rate.
\end{abstract}

\begin{IEEEkeywords}
Resource Allocation, Workflow Containerization, Kubernetes, Workflow Management Engine.
\end{IEEEkeywords}}

\maketitle

\IEEEdisplaynontitleabstractindextext

\IEEEpeerreviewmaketitle

\IEEEraisesectionheading{\section{Introduction}\label{sec:introduction}}

\IEEEPARstart{W}{ith} the advent of a cloud-native era, the most popular virtualization solution is using Docker~\footnote[1]{http://docs.docker.com} 
for container encapsulation with Kubernetes~(K8s)~\cite{burns2016borg} for multi-host container orchestrating. Docker and K8s~\footnote[2]{https://kubernetes.io/} have become mainstream tools for 
cloud resource management and dominated the whole cloud-native technology ecosystem~\cite{bernstein2014containers}.
Workflows have been widely applied in scientific computing communities such as astronomy, bioinformatics, material science, 
and earth science~\cite{juve2013characterizing}. 
A scientific workflow is commonly formulated as a directed acyclic graph~(DAG), which consists of dozens of workflow tasks~(represented by nodes) and 
dependencies among tasks~(indicated by directed edges). 
A DAG abstracts a particular scientific computing process through shared data files between tasks and predefined task 
dependencies~\cite{lee2013stretch,zheng2017deploying}. 
Powered by Docker and K8s, cloud infrastructure features the scalability and high availability of computational 
resources\cite{silver2017software} and is especially suitable as a running platform for scientific workflows.

Scientific workflows usually serve large-scale applications and require a considerable amount of resources to execute. 
Efficient resource allocation is a key issue in workflow execution. Existing workflow management engines like 
Nextflow~\cite{di2017nextflow}, Pagasus~\cite{deelman2015pegasus,deelman2019evolution}, Galaxy~\cite{jalili2020galaxy}, 
and Argo workflow engine~\footnote[3]{http://github.com/argoproj/argo}
can execute hundreds of workflows on cloud infrastructure and be responsible for assigning computational resources to 
workflow tasks~\cite{bader2021tarema}. 
When encountering continuous workflow requests and unexpected resource request spikes, the computational resource requirements of 
workflows can be highly dynamic.
The ever-changing resource requirements of workflows bring a great administrative burden to the workflow engines for resource allocation 
and seriously decrease the execution efficiency of workflows. 
On the one hand, the permanent provision of fixed computational resources will cope with peak loads in a resource-intensive scenario 
but incur high costs and resource over-provisioning, as resources are not fully utilized during off-peak times. On the other hand, some workflows 
may not be executed at all and suffer from a poor Quality of Service~(QoS) due to insufficient resource provisions.

In order to avoid over and under-provisioning of resources, some existing works propose reasoning~\cite{hoenisch2013self,hoenisch2013workflow}, 
feedback~\cite{witt2019feedback}, heuristics~\cite{khatua2014heuristic}, 
learning and prediction models~\cite{abdullah2020diminishing,chen2021adaptive,chen2016statistical,schuler2021ai} to cope with resource allocation in cloud environment. 
Although these solutions can partially address the cloud resource allocation problem, they commonly use prior knowledge of cloud systems to cope with 
resource allocation. As a result, these solutions might play to their strengths in a specific application scenario, but they are not fully adaptable to the 
K8s-based cloud environment with dynamic resource requirements. 
Besides, numerous iteration training may result in high computational complexity and resource overheads in learning and prediction models.
Therefore, with the application platform and technology stack in mind, they do not fit with the K8s-based workflow management engines. 
The bottleneck here is the absence of a high-efficiency adaptive resource allocation scheme that can help the K8s-based workflow management 
engines to make appropriate resource provisions in response to continuous workflow requests and unexpected request resource spikes. 

In our former work~\cite{shan2021workflow,shan2022kubeadaptor}, we presented the customized K8s-based workflow management engine~(KubeAdaptor), 
able to integrate workflow systems with K8s and implement workflow containerization on K8s cluster. 
In this paper, we present an adaptive resource allocation scheme~(ARAS) that follows the 
Monitor-Analyse-Plan-Execute~Knowledge~(MAPE-K) model~\cite{iglesia2015mape,arcaini2015modeling}. 
The ARAS periodically responds to the task pod's resource request and uses the resource discovery algorithm, resource evaluation algorithm, 
and resource allocation algorithm to complete the resource allocation of this round task by the resource scaling strategy.
We reconstruct and extend KubeAdaptor, and implement ARAS as the \textit{Resource Manager} component of KubeAdaptor, which consists of a \textit{Resource Discovery} module, 
\textit{Resource Evaluator} module, and \textit{Allocator} module. 
Three modules complement each other to achieve the adaptive resource allocation~(refers to Fig.~\ref{fig_arch}).
First, the \text{Resource Discovery} module invokes the resource discovery algorithm to obtain the remaining resources~(such as CPU and memory) of K8s cluster nodes 
and the resource usage of running task pods. 
Then the \textit{Resource Evaluator} module integrates the remaining resources of the K8s cluster and workflow workloads from the Redis database and 
evaluates resource adequacy for the K8s cluster nodes. 
Finally, the \textit{Allocator} module uses a resource scaling strategy~(i.e., vertical autoscaling)~\cite{rzadca2020autopilot} to make resource provisions 
for current active task pods in response to continuous workflow requests and sudden request spikes. 
We have open-sourced the proposed ARAS. 
The source code is publicly available on GitHub \footnote[4]{https://github.com/CloudControlSystems/ResourceAllocation}.

This paper focuses on adaptation, that is, the adaptive adjustment of resource allocation in the context of changing workflow resource requirements.
Compared with the baseline algorithm, experimental evaluation of running four scientific workflows under three different workflow arrival modes 
shows that ARAS gains time-saving of $9.8\%$ to $40.92\%$ in the average total duration of all workflows, 
time-saving of $26.4\%$ to $79.86\%$ in the average duration of individual workflow, and an increase of $1\%$ to $16\%$ in CPU and memory resource usage rate.
The main contributions of this paper are summarized as follows.

\begin{itemize}
  \item \textit{MAPE-K architecture.} 
  With the MAPE-K mechanism as a core, we decouple and reconstruct the KubeAdaptor and integrate our ARAS 
  into four phases of the MAPE-K model to equip the KubeAdaptor with self-healing and self-configuration abilities. 

  \item \textit{A novel monitoring mechanism.} 
 We devise and develop a resource discovery algorithm through the K8s resource characteristics and \textit{Informer} component. 
 The \textit{Resource Discovery} module uses this algorithm to build a novel monitoring mechanism to 
 collect all related data in K8s clusters. 
 
  \item \textit{Automation deployment.} 
  We modularize and implement the four steps of the proposed ARAS with loose coupling in mind 
  so that the users can easily mount a newly designed algorithm module to replace an existing one with minimal intrusion 
  into the workflow management engine. 

  \item \textit{Better performance.} 
   With the help of K8s and the MAPE-K mechanism, we use ARAS to conduct a wealth of experiments on 
   four scientific workflows on K8s clusters. 
   Our ARAS shows better performance compared to the baseline algorithm.
  \end{itemize}

The rest of the paper is organized as follows.
Section 2 introduces related work. Section 3 elaborates on the system model and problem formulation, 
while Section 4 further describes our system architecture and components.
Section 5 illustrates the implementation of our adaptive resource allocation scheme. 
Section 6 describes the experimental setup and discusses the evaluation results. Finally, 
Section 7 concludes this paper.

\section{Related Work}
The resource allocation scheme in the workflow management engine is influenced by the virtualization technology of cloud infrastructure, 
which is directly related to whether workflow tasks are hosted by VM instances or containers.
In this section, we review the development of resource allocation strategy and discuss three categories of resource allocation strategy 
from the perspective of the evolution of virtualization technology, namely VM-based, Container-based, and Cloud-native-based. 
Note that the analysis of each aspect is not completely limited to the scope of the workflow management engine.  

\subsection{VM-based resource allocation}
In the VM-based era, Lee et al.~\cite{lee2009adaptive} propose an adaptive scheduling approach to adjust resource allocation 
and scheduling in the Pegasus workflow management system. 
This approach utilizes batch queues to assign jobs to cluster's VMs and optimizes job scheduling across cluster's VMs through 
the average queue time of each available VM. 
Islam et al.~\cite{islam2012empirical} develop prediction-based resource measurement and provisioning strategies 
using neural networks and linear regression to satisfy upcoming resource demands. 
The sliding window approach for predicting resource usage in the cloud fits with dynamic and proactive resource management 
for interactive e-commerce applications. 
As for Business Process Management System~(BPMS) field, Hoenisch et al.~\cite{hoenisch2013self} present 
a self-adaptive resource allocation approach to automatically lease and release cloud resources for workflow executions 
based on knowledge~(resource usage in VMs) about current and future process landscape. 
This approach has been implemented as part of ViePEP, a BPMS able to manage and schedule workflows in the cloud. 
By monitoring resource usage of VMs and the QoS of individual service invocations in VMs, ViePEP uses a prediction 
model to provide resource provisioning for elastic process execution of workflows. 
Subsequently, Hoenisch et al.~\cite{hoenisch2013workflow} extend ViePEP by dynamic workflow scheduling and resource 
allocation algorithms. The proposed algorithm not only provides a complete schedule plan based on their former 
predicting model but also moves the service invocations~(workflow task) from one timeslot to another to fully utilize the acquired resources.

Although these solutions present appropriate cloud resource allocation schemes to some extent, 
the predictive models commonly require the collection and modeling according to former data. 
These upfront preparations consume unnecessary resources and block the automatic operation flow 
of the workflow management engines. 
Besides, VMs-based resource allocation schemes are commonly limited to VM's features such as slow startup, 
clumsy deployment, and high resource consumption. 
Therefore, these schemes are not suitable for performing workflows with dynamic and ever-changing resource requirements 
in cloud infrastructures.

\subsection{Container-based resource allocation}
In the Container era, container-based resource allocation schemes gradually become the mainstream of cloud resource management. 
Considering the absolute resource isolation and security features of VMs, 
most container-based resource allocation scenarios adopt the deployed model of VM hosting containers. 
For instance, Mao et al.~\cite{mao2022differentiate} propose a differentiated quality of experience scheduler to adjust resource 
provisioning for deep learning applications. 
This scheduler is implemented into Docker Swarm~\footnote[5]{https://docs.docker.com/engine/swarm/} and can accept the targeted quality of 
experience specifications from clients and dynamically adjust resource limits of containers to approach performance targets. 
Abdullah et al.~\cite{abdullah2020diminishing} introduce a new deep learning-based approach to estimate the execution time 
of the jobs through the collected performance traces. 
This approach also predicts the execution time for different CPU pins and uses the laws of diminishing marginal returns to provide 
optimal CPU allocation to Docker containers. 
In the fog computing community, Yin et al.~\cite{yin2018tasks} propose a container-based task-scheduling algorithm with task delay 
constraint in mind. 
Herein, a resource reallocation mechanism works to achieve resource-utilization maximization on fog nodes by modifying the resource 
quota of task containers. 
Hu et al.~\cite{hu2022cec} propose \textit{CEC}, a containerized edge computing framework for dynamic resource 
provisioning in response to multiple intelligent applications. 
The \textit{CEC} first makes resource provisioning for containers in advance based on the workload prediction for the edge cluster 
formed by Docker Swarm and then uses the idea of control theory to achieve dynamic resource adjustments~
(meaning a sufficient number of containers) for hosted service applications. 

Containers, an efficient and lightweight virtualization technology, bring significant technology change to VMs-based resource 
allocation strategies. 
But in fact, it needs a container orchestration tool~(e.g., Docker Swarm) to manage a wealth of containers 
across cluster nodes in several scenarios. 
In practice, resource limits' adjustment and reallocation of resource quotas in running containers have brought significant 
administrative burdens to Docker Swarm. 
Also, the adjustments to the number of containers will cause a delay in the startup of new containers. 
In addition, the preparation for predictive models is also not conducive to the automation of workflow management 
engines. Due to the shortcomings of the above solutions and the task dependencies and high concurrency of workflow, 
these resource allocation strategies can not provide efficient ideas for container-based workflow management engines.

\subsection{Cloud-native-based resource allocation}
As the first hosted project by Cloud Native Computing Foundation (CNCF)~\footnote[6]{https://www.cncf.io/}, 
K8s has become the de-facto standard container orchestration system. 
Docker and K8s are reshaping resource management strategies for cloud infrastructures in the cloud-native era. 
For example, Chang et al.~\cite{chang2017kubernetes} propose a generic platform to facilitate dynamic resource 
provisioning based on K8s. 
The platform employs open-source tools to retrieve all the resource utilization metrics (such as CPU and memory) while 
integrating the application QoS metrics into monitoring. 
The resource scheduler module in the platform makes dynamic resource provisioning by horizontal scaling of task pods 
according to the K8s cluster's workload. 
Mao et al.~\cite{mao2020resource} investigate the performance of using cloud-native frameworks~(Docker and K8s) 
for big data and deep learning applications from the perspective of resource management fields. 
Together with Prometheus~\footnote[7]{https://github.com/prometheus/prometheus} and 
Grafana~\footnote[8]{https://github.com/grafana/grafana}, the authors build a container monitoring system to keep tracking 
the resource usage of each job on worker nodes. 
To address massive aggregate resource wastage, Google uses \textit{Autopilot} to configure resources automatically, 
adjusting both the number of concurrent tasks in a job~(horizontal scaling) and the CPU/memory limits for 
individual tasks~(vertical scaling)~\cite{rzadca2020autopilot}. 
Subsequently, Bader et al.~\cite{bader2021tarema} propose \textit{Tarema}, a system for allocating task instances to 
heterogeneous K8s cluster resources during the execution of scalable scientific workflows. Using a scoring algorithm 
to determine the best match between a task and the available resources, \textit{Tarema} provides the near-optimal task-resource allocation.

However, most of these resource allocation solutions in the cloud-native era use open source tools (from the CNCF community) to 
build resource monitoring systems, obtain the required resource utilization of the cluster, 
and provide corresponding resource provisioning strategies. 
It brings high deployment costs to the workflow management engine, which is inconsistent with the simple deployment 
and automatic system characteristics.
In addition, these tools put too much pressure on the K8s cluster because of frequent access to 
kube-apiserver~\footnote[9]{The API server serves as the front end and is a component of the K8s control plane 
that exposes the K8s API. Its performance is a weathervane for the overall K8s cluster performance.} 
for acquiring cluster resources~\cite{chakraborty2020enabling}.
 
To summarize the related work, we can conclude that resource allocation policies change with the evolution of virtualization 
technologies to adapt to different application scenarios and technology platforms. 
The most typical example is ViePEP-C~\cite{waibel2019viepep}, which evolved from 
former work~\cite{schulte2012introducing,hoenisch2013workflow,hoenisch2015optimization,hoenisch2013self} in the VMs era 
to a container-based resilient BPMS platform in the container era, using containers instead of VMs for the execution of business process activities.
Considering automation and flexible deployment of the integrated platform, resource allocation technology in the cloud-native era 
is more focused on Docker and K8s platforms. 
The design of our workflow management engine follows this idea. 
K8s, with its unique technical advantages and ecology in scheduling, automatic recovery, horizontal scalability, 
resource monitoring, and other aspects, makes its integration with workflow management engines 
far beyond the capabilities of container-based workflow management engines.  
Inspired by the work in~\cite{mao2022differentiate,mao2020resource,rzadca2020autopilot}, 
our ARAS takes into account workflow loads in K8s clusters and uses vertical scaling technology of 
containers to cope with continuous workflow requests and sudden resource spikes.

\section{System Model}
This section describes how to use our ARAS to cope with continuous workflow requests 
and unexpected resource request spikes and maximize resource utilization while meeting workflow Service Level Objectives~(SLOs).

\subsection{System description}
For the clarity of presentation, we consider the scenario of a single K8s cluster with a set of nodes~(VMs), 
donated by $V=\{v_{1},v_{2},...,v_{m}\}$, where $m$ represents the number of K8s cluster nodes. 
As for $m$ nodes, we have a set of available CPU cores $C=\{c_{1},c_{2},...,c_{m}\}$ and a set of availble memory 
capacity $M=\{mem_{1},mem_{2},...mem_{m}\}$ correspondingly. 
The workflow set injected into the KubeAdaptor is represented as $W=\{w_{1},w_{2},...,w_{k}\}$, 
where $k$ indicates the amount of workflows. 
Herein, a workflow is abstractly defined as $w_{i}=\{sla_{w_i},s_{i,1},s_{i,2},...,s_{i,n}\}$, wherein $i$ indicates 
the $ID$ of a workflow, $sla_{w_i}$ represents a Service Level Agreement~(SLA) of a workflow and 
$s_{i,1},s_{i,2},...,s_{i,n}$ indicates steps~(i.e., tasks) of workflow $w_{i}$. 
Each workflow task is defined as
\begin{equation}
  \label{eq_s_ij}
  \begin{split}
s_{i,j}=\{sla_{s_{i,j}},id,image,cpu,mem,duration,\\min_{cpu},min_{mem}\},~1\leq i \leq k~and~1\leq j \leq n.   
  \end{split}
\end{equation}
Herein, $id$ is the unique identifier of this workflow task in workflow $w_i$, and $image$ represents the Docker Image address 
of this workflow task. The $cpu$ is the amount of CPU Milli cores required by the users, and $mem$ is the amount of memory capacity 
required by the users. $duration$ indicates the duration of the task pod running, and $min_{cpu}$ and $min_{mem}$ represent a minimum 
of CPU and memory resources required to run the task container of $s_{i,j}$ in workflow $w_i$, respectively. 
Generally, a workflow can have an optional SLA~($sla_{i}$) composed of several SLOs expressed 
by $slo_{1},slo_{2},...,slo_{n}$ on workflow~($sla_{w_i}$) or workflow task~($sla_{s_{i,j}}$) as following: 
\begin{eqnarray}
  \label{eq_sla_i}
sla_{i}=\{slo_{1},slo_{2},...,slo_{n}\},i\in\{w_{i},s_{i,j}\}.
\end{eqnarray}

Herein, we only consider the deadline as the single SLO, 
meaning that each task in the workflow must be completed before its respective deadlines. Likewise, this workflow is no exception.
\begin{eqnarray}
  \label{eq_sla}
  \begin{split}
  &sla_{w_i}=deadline_{w_i},\\
  &sla_{s_{i,j}}=deadline_{s_{i,j}}.
  \end{split}
\end{eqnarray}
Note that the deadline for the last task $s_{i,last}$ in a workflow is identical to this workflow execution deadline:
\begin{equation}
  \label{eq_deadline}
  deadline_{s_{i,last}}=deadline_{w_i}.
\end{equation}

\subsection{Problem formulation}
We assume that SLAs and deadlines defined by users are valid and achievable, i.e., a properly completed workflow means that 
all of its tasks must be completed by the deadline.  
With maximizing the resource utilization in the K8s cluster as a goal, as long as a task request arrives, our ARAS
uses the resource scaling method to provide computational resources to the task container. 
Herein, the resource provision of the task container must not be less than a minimum of running resources to ensure the smooth operation 
of the task container. 

\begin{figure}[!t]
\centering
\includegraphics[width=3.5in]{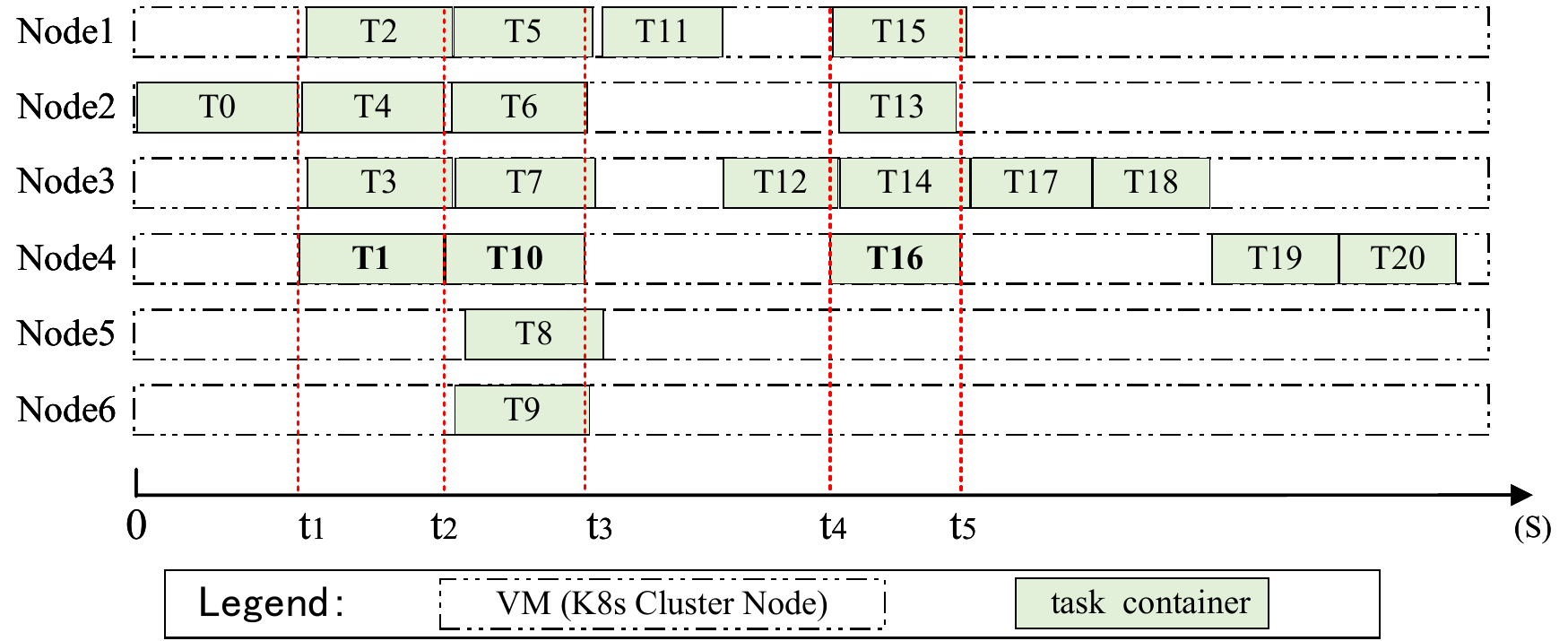}

\caption{Resource allocation example. A small-scale Montage workflow with $21$ tasks is used to illustrate the resource scaling method in 
our ARAS. The test environment uses our experimental setup in Section~\ref{sec:senario}.}
\label{fig_formulaiton}
\end{figure}

Fig.~\ref{fig_formulaiton} depicts an execution process of a small-scale Montage workflow. At $t_{1}$ seconds, task request of $T_{1}$ arrives, 
and our ARAS has abilities to acquire concurrent tasks within its lifecycle~(from $t_1$ to $t_2$) considering
predefined deadlines. 
As can seen from Fig.~\ref{fig_formulaiton}, $T_2$, $T_3$ and $T_4$ will be launched within $T_{1}$'s lifecycle and four workflow tasks
will compete for computing resources each other. 
To ensure that four concurrent tasks have enough resources to run smoothly, 
our ARAS employs the resource scaling method to 
reasonably allocate resources, i.e., scaling down resource requirements of the current task $T_1$ according to the ratio of 
the total resource requirements of four tasks to the remaining resources in the K8s cluster~(refers to Eq.~(\ref{eq:rate})).
Similarly, $T_{10}$ executes between $t_2$ and $t_3$, $T_{16}$ executes between $t_4$ and $t_5$. 
Their respective lifecycle all contains several concurrent tasks. 
The arrival of each task request also requires the resource scaling method to allocate resources in line with Eq.~(\ref{eq:rate}).

In the following, we elaborate on the optimal problem in our ARAS. 
The allocated CPU and memory resources for each requested task in workflow $w_{i}$ are respectively defined as follows:
\begin{eqnarray}
  \label{eq_ur}
  \begin{split}
  &U=\{u_{i,1},u_{i,2},...,u_{i,n}\},\\
  &R=\{r_{i,1},r_{i,2},...,r_{i,n}\}.    
  \end{split}
\end{eqnarray}

$x_{y,z}^{i} \in \{0,1\}$ with workflow identifier $i$ is adopted as a decision variable for task placement, 
where $1\leq y\leq n$ and $1\leq z\leq m$ and defined as 
  $$ x_{y,z}^{i} = \begin{cases}
    1& \text{if $y^{th}$ task in $w_{i}$ is scheduled on Node $v_{z}$},\\
    0& \text{if $y^{th}$ task in $w_{i}$ is not scheduled on Node $v_{z}$}.
  \end{cases} $$

We assume that each node in the K8s cluster is always active and that workflows are continuously injected into 
our workflow management engine. 
$Mem_{total}$ indicates the total number of remaining memory resources of the K8s cluster.  
Since CPU is a compressible resource and memory is an incompressible resource, we only consider memory 
to maximize resource allocation in the optimization model. So our objective function is as follows:

\begin{eqnarray}
  \label{eq_mini}
  \begin{split}
&Maximize:~~~~\sum_{i=1}^{k}\sum_{j=1}^{n} r_{i,j}/Mem_{total}~~~~~~~~~~~~~~
\end{split}
\end{eqnarray}

Subject to: 
\begin{eqnarray}
  \label{eq_subject}
  \begin{split}
  &\sum_{z=1}^{m}x_{y,z}^{i}=1\\
  &\sum_{i=1}^{k}\sum_{y=1}^{n}x_{y,j}^{i} \cdot u_{i,y} \leq c_{j}\\
&\sum_{i=1}^{k}\sum_{y=1}^{n}x_{y,j}^{i} \cdot r_{i,y} \leq mem_{j}.    
  \end{split}
\end{eqnarray}

Eq.~(\ref{eq_mini}) shows the objective function in our model, 
which maximizes the resource utilization for the remaining memory of the K8s cluster at each moment of the task request. 
Eq.~(\ref{eq_subject}) shows three constraints of our model. 
The first constraint indicates that a task can be scheduled on only one cluster node. 
The last two constraints imply that the total CPU and memory resources consumed by all task pods on the hosted node $v_{j}$ must 
be less than or equal to the amount of the respective available resources on that node.

\begin{figure*}[htbp]
\centering
\includegraphics[scale=0.55]{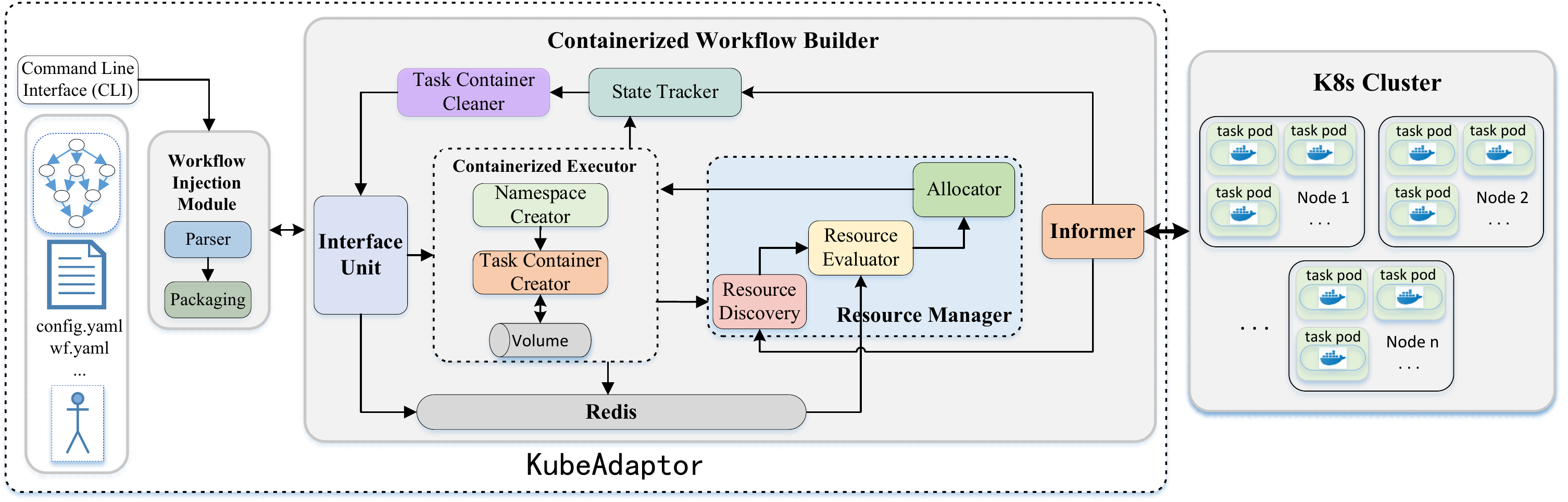}
\caption{KubeAdaptor architecture.}
\vspace{-0.5cm}
\label{fig_arch}
\end{figure*}

\section{Architecture}
This section presents the system architecture of KubeAdaptor in detail, including its framework, design logic, and key modules. 
Subsequently, the MAPE-K model is elaborated around the resource allocation mechanism of KubeAdaptor. 

\subsection{KubeAdaptor framework}
The KubeAdaptor for the ARAS is illustrated in Fig.~\ref{fig_arch}. 
As a workflow management engine, it works to administrate, schedule, and execute containerized workflow tasks. 
Its core functionalities are as follows:
\begin{itemize}
\item Provide an interface to the public or private cloud, allowing to customize workflows on demand
\item Implement the containerized execution of workflows following the precedence and dependency relationships
\item Adaptively allocate resource quotas for requested workflow tasks and maximize resource utilization when ensuring the SLAs of workflows
\item Provide the ability of flexible deployment and the automatic operation flow and integrate into the K8s platform  
\end{itemize}

With the assistance of ARAS in this paper, KubeAdaptor equips with the functionalities 
of a cloud resource management system to elegantly manage a potentially highly volatile cloud workflow application scenario.

\subsection{KubeAdaptor modules}
\label{kube-modules}
As depicted in Fig.~\ref{fig_arch}, KubeAdaptor consists out of three main top level entities, including a 
\textit{Command Line Interface}~(CLI), a \textit{Workflow Injection Module} and a \textit{Containerized Workflow Builder}. 
The \textit{Containerized Workflow Builder} comprises seven sub-components responsible for workflow reception, containerization, 
resource allocation, resource monitoring, and task container cleanup. 
We focus on the \textit{Resource manager} module related to ARAS.

\textbf{CLI:} It aims to define SLAs-based workflows and offer configuration files of 
one-key deployment. 
In addition, the users may request many workflows consecutively or even simultaneously through the
\textit{CLI} module.

\textbf{Workflow Injection Module:} 
Its \textit{Parser} and \textit{Packaging} modules serve as an independent function pod and work to read variable configuration information of workflow definition 
from the mounted directory, parse and encapsulate workflows in response to generating request of the subsequent workflow from the \textit{Interface Uint}. 

\textbf{Interface Uint:} 
This module works on receiving the workflow generating request, decomposing the workflow tasks, 
watching the state changes of task pods or workflows from the \textit{Task Container Cleaner}, 
invoking the \textit{Containerized Executor} to generate workflow namespaces and task pods, 
and writing workflow status into the Redis database. 
Once the creation of the task pod fails, this module turns to fault tolerance management~\cite{shan2022kubeadaptor}, 
also known as \textit{self-healing}, the ability of a system to detect and recover from potential problems and continue to operate smoothly. 

\textbf{Containerized Executor:} 
Its two subcomponents work on generating workflow namespaces and task pods. 
This module creates task pods through allocated resources from the \textit{Resource Manager}. 
In addition, the states of workflows and task pods are timely written into the Redis database. 

\textbf{Resource Manager:} 
It contains three subcomponents, such as \textit{Resource Discovery}, \textit{Resource Evaluator} and \textit{Allocator}. 
The \textit{Resource Discovery} is responsible for acquiring the remaining resource of the overall K8s cluster from the \textit{Informer}. 
The \textit{Resource Evaluator} obtains workflow resource requirements and workflow execution states from the Redis database, 
assesses the adequacy of the current remaining resources of the K8s cluster, and launches corresponding countermeasures, if necessary. 
The \textit{Allocator} module uses resource scaling strategy to make resource allocation for current active
task pods in response to continuous workflow requests and sudden request spikes.
It is also known as \textit{self-configuration}, the ability of a system to reconfigure itself under changing and unpredictable circumstances. 

\textbf{Informer:} 
As a core toolkit in Client go~\footnote[10]{https://github.com/kubernetes/client-go}, 
the \textit{Informer} is in charge of synchronizing resource objects and events between K8s core components and \textit{Informer} 
local cache. 
It provides the \textit{Resource Discovery} with the remaining resources of the K8s cluster and responds to the \textit{State Tracker} 
for the state changes of the resource objects.

\textbf{State Tracker:}
It hosts the monitoring program based on the List-Watch mechanism and responds state queries of various resource objects to 
each module anytime. 

\textbf{Task Container Cleaner:}
It works on deleting the pods in a state of \textit{Succeeded} or \textit{Failed} or \textit{OOMKilled} and 
workflow namespaces without uncompleted task pods. 
Once receiving successful feedback on the just-deleted workflow or task pods, this module proceeds to 
\textit{Interface Unit} and triggers the following workflow or subsequent task. 

\textbf{Redis:} 
The Redis database is to be deployed within or outside the cluster in advance and is responsible for storing 
workflow execution status and predefined resource requirements of workflow tasks.

KubeAdaptor is implemented by the Go language and provides an \textit{CLI} interface to K8s clusters. 
With just a few tweaks to the configuration file, the users can go out of the box and smoothly deploy the KubeAdaptor on K8s clusters. 
The deployment and uninstalling of KubeAdaptor are non-intrusive and clean to the cluster, 
and its workflow containerization execution works in an automated way. 
For further details about KubeAdaptor, we refer to~\cite{shan2022kubeadaptor}.
\begin{figure}[!t]
\centering
\includegraphics[width=3.5in]{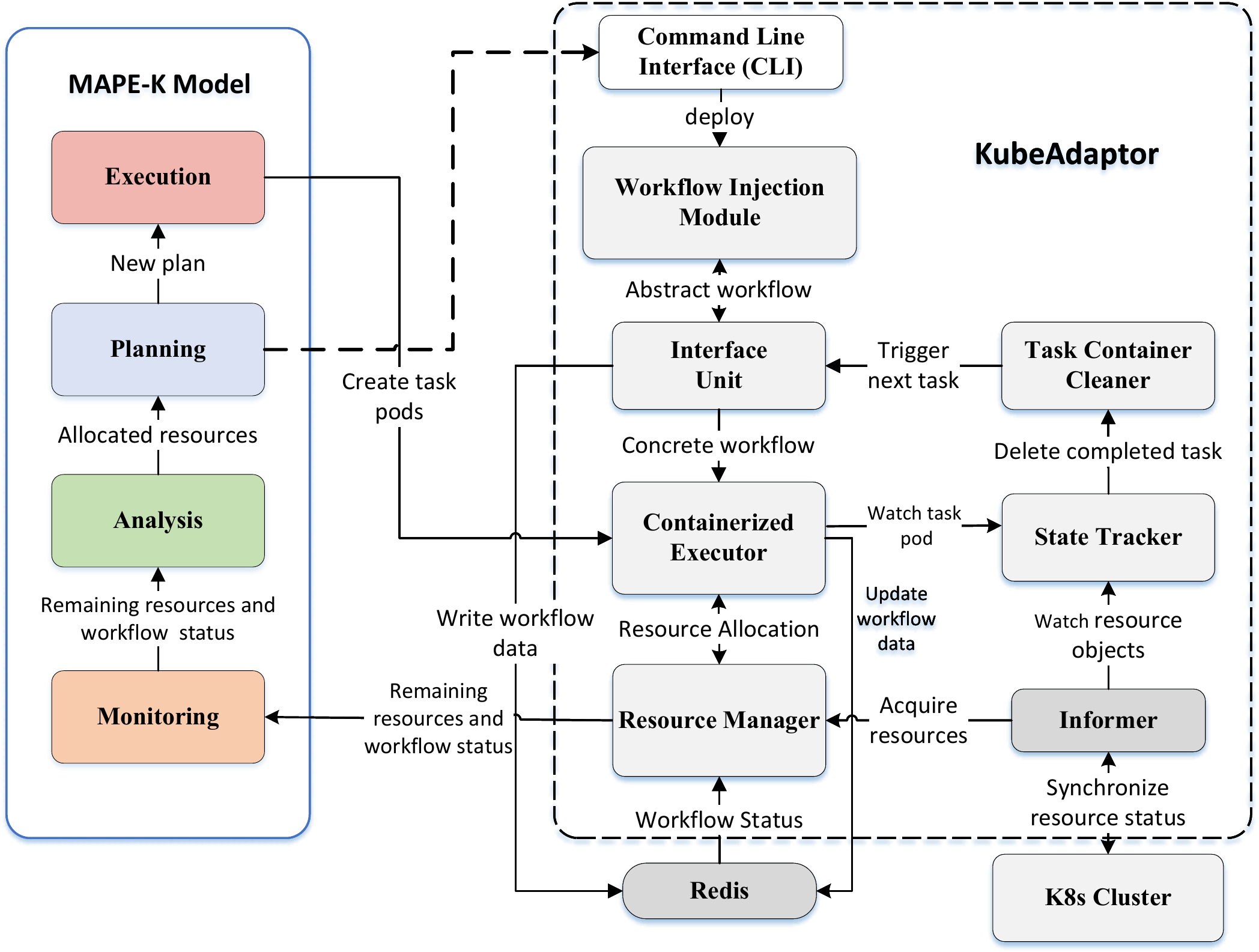}

\caption{Resource allocation scheme based on MAPE-K model.}
\vspace{-0.5cm}
\label{fig_mape}
\end{figure}

\subsection{MAPE-K model}
\label{sec:mape}
The MAPE-K model~\cite{kephart2003vision}, originated from the field of Automatic Computing, is an instrumental framework for systematic 
development of adaptive systems, including resource allocation and workflow adaptation.  
Adaptive strategy within the KubeAdaptor works to realize the self-optimization of resource utilization maximization. 
Herein, the self-optimization ability, along with self-healing and self-configuration~(elaborated in \ref{kube-modules}), 
enables our KubeAdaptor to become a self-management system. 

To deal with persistent workflow requests and ever-changing resource requirements, 
we use the MAPE-K model to retrofit with minimal intervention to the KubeAdaptor, 
which forms an adaptive execution cycle as depicted in Fig.~\ref{fig_mape}. In the following, 
we will briefly discuss the four steps of this cycle and how they influence the self-management capabilities of KubeAdaptor:

 \textbf{Monitoring:} 
 The Monitoring functionalities stem from the \textit{Informer} and Redis database and work to provide workflow status and 
 the remaining resources in K8s Clusters to the next step. 
 Workflow status data includes which workflows have been executed, which tasks have been completed, 
 resource requirements of workflow tasks, as well as the SLOs of the workflow and task. 
 The remaining resources refer to the residual CPU and memory resources of the K8s cluster and each node in the K8s cluster. 

 \textbf{Analysis:}
The functionality of this step comes into play in the \textit{Resource Evaluator} and \textit{Allocator} within \textit{Resource Manager}. 
For adaptive resource allocation, it is necessary to analyze the monitored data and reason on the general knowledge about the system. 
It is done so that we can adapt according to countermeasures to deal with dynamic workflow requests and SLA violations.

 \textbf{Planning:}
The planning step takes full account of the resource requirements for future workflow tasks to be launched 
within the current task lifecycle and SLAs of workflows to carry out reasoning and generate a resource allocation plan.
The planning results also provide sufficient prior knowledge for subsequent workflow input to the CLI module.

  \textbf{Execute:}
 The execution steps are put into practice through \textit{Containerized Executor}, 
 aiming to finish the creation of a new round of task pods combined with the analysis results of the MAPE-K model.
 
 \textbf{Knowledge Base:}
While not really a part of the cycle, the Knowledge Base stores the configuration information about the system and 
provides workflow execution status and the remaining resources in K8s clusters to the MAPE-K model running. 

In short, the KubeAdaptor needs to provide further analysis of the resource allocation scheme to 
self-optimize application scenarios in response to persistent workflow requests and sudden resource spikes.

% \begin{table}[!t]
\begin{table*}
\centering
\caption{Major notations used in adaptive resource allocation scheme}
\label{table:notation}
% \begin{tabular}{|p{2.5cm}|p{5.5cm}|}
  \centering
  \setlength{\tabcolsep}{6 mm}
  \begin{tabular}{p{2.cm}p{13.cm}}
\toprule[1pt]
Notation     & Meaning                                                                                                                                        \\ \hline
$v_i$ & a K8s node~(VM) $v_{i} \in \mathcal{V}$ \\ 
$s_{i,j}$				& the $j$th task in workflow $w_{i}$, $1 \leq i \leq k$~and~$1 \leq j \leq n$		\\ 
$allocated_{cpu}$      & the allocated CPU resource number through the Adaptive Resource Allocation Algorithm \\ 
$allocated_{mem}$     & the allocated memory resource number through the Adaptive Resource Allocation Algorithm \\ 
$request.cpu$      & the accumulated CPU resource number for many task requests \\ 
$request.mem$      & the accumulated memory resource number for many task requests \\ 
$Re_{max}^{cpu}$ & the maximum remaining resource amount of CPU among K8s cluster nodes \\ 
$Re_{max}^{mem}$ & the maximum remaining resource amount of memory among K8s cluster nodes \\ 
$totalResidual.cpu$   & the total of residual CPU resource across K8s cluster nodes \\ 
$totalResidual.mem$ & the total of residual memory resource across K8s cluster nodes \\ 
$task^{req}$ & the current task request with respects to $s_{i,j}$ \\ 
$task_{i,j}^{redis}$ & a record of task-state data from Redis database \\  
$PodLister$ & an interface of acquiring pod list in \textit{Informer} component \\ 
$NodeLister$ & an interface of acquiring node list in \textit{Informer} component \\  
$ResidualMap$ & a data dictionary for storing remaining resources~(CPU and memory) of each node \\ 
$nodeReq.cpu$ & the accumulated CPU resource requirements for all pods on a node \\ 
$nodeReq.mem$ & the accumulated memory resource requirements for all pods on a node \\ 
$allocatable.cpu$ & the accumulated allocatable CPU resource across K8s cluster nodes \\ 
$allocatable.mem$ & the accumulated allocatable memory resource across K8s cluster nodes\\ 
$residual.cpu$ & the residual CPU resource in K8s cluster \\
$residual.mem$ & the residual memory resource in K8s cluster \\ 
$pod_i$ & a data struct from Podlist, which contains many key fields about container's features \\ 
$cpu_{cut}$ & the allocated CPU resource amount for task request based on Eq.~(\ref{eq:rate}) \\ 
$mem_{cut}$ & the allocated memory resource amount for task request based on Eq.~(\ref{eq:rate}) \\  
$\alpha$ & a proportional value derived from experience, $\alpha \in (0,1)$ \\ 
$\beta$ & a constant value derived from experience, $\beta \geq 20$ \\ 
\bottomrule[1pt]
\end{tabular}
\end{table*}

\section{Adaptive Resource Allocation Scheme}
Once \textit{Resource Manager} receives a resource request of workflow task from \textit{Containerized Executor}, 
its three subcomponents immediately launch resource discovery, resource evaluation, and resource allocation in turn. 
The entire execution process responds to the workflow task's resource request iteratively.
Next, we introduce the adaptive resource allocation algorithm, resource discovery algorithm, 
and resource evaluation algorithm. All notations used in our algorithm are illustrated in Table~\ref{table:notation}. 
In addition, workflow execution states are represented as a set of state data for all tasks of 
workflow $w_{i}~(1\leq i\leq k)$ and a record of task-state data is defined as 

\begin{eqnarray}
    \label{eq_taski}
  \begin{split}
 task_{i,j}^{redis}=\{t_{start},duration,t_{end},cpu,mem,flag\},\\
  1\leq i\leq k,~1\leq j\leq n~and~flag \in \{false,true\}. 
  \end{split}
\end{eqnarray}

Note that as long as KubeAdaptor starts, $task_{i,j}^{redis}$ is stored in Redis database through \textit{Interface Unit}, 
and then is continuously updated by the \textit{Containerized Executor}. 
Herein, $t_{start}$ is the start time of the current task pod in K8s cluster, 
$duration$ is similar to the definition of $s_{i,j}.duration$ and represents the duration of the current task pod's running, 
$t_{end}$ is the completed time of the current task in the K8s cluster, $cpu$ and $mem$ are equivalent to the definition of $s_{i,j}.cpu$ 
and $s_{i,j}.mem$, and $flag$ is a boolean variable that identifies the execution status of the current task pod. 
Herein, the boolean variable \textit{false} indicates that the current task is not complete. 
We use \textit{Dictionary} data structure $Map<task_{i,j}.id,~task_{i,j}^{redis}>$ to indicate state data of the 
current task, and $task_{i,j}.id$ is the unique identifier passed by task $s_{i,j}$ of workflow $w_{i}$~(refers to Eq.(\ref{eq_s_ij})). 

Because the pod is the minimum execution unit of the K8s container orchestrator, coupled with KubeAdaptor's non-invasive automated 
execution process, \textit{Resource Manager} allocates resources only once throughout the requested task pod's lifecycle in response to 
task pod's resource request. 
The users can initially set $min_{cpu}$ and $min_{mem}$ of the task pod in \textit{CLI} module, by which this task pod ensures its 
hosted container runs smoothly. 

% \begin{algorithm}[!t]
\begin{breakablealgorithm}
\caption{AdaptiveResourceAllocationAlgorithm}
\label{alg}
\begin{flushleft}
	{\bf{Input:}} $s_{i,j}$;\\
	{\bf{Output:}} $allocated_{cpu}$, $allocated_{mem}$;
\end{flushleft}
\begin{algorithmic}[1]
  \State Initialization: $request.cpu$, $request.mem$, $Re_{max}^{cpu}$, $Re_{max}^{mem}$, $totalResidual.cpu$, $totalResidual.mem$ $\leftarrow$ 0;
  \For{each task pod's resource request}
    \State /*\textit{ Access the Redis and get the total of requested resources of all pods to be launched within $s_{i,j}$'s lifecycle }*/
    \State Get $task^{req}$ with respects to $s_{i,j}$ of $w_{i}$ from Redis;
    \State $request.cpu$ $\leftarrow$ $task^{req}.cpu$;
    \State $request.mem$ $\leftarrow$ $task^{req}.mem$;
    \State Get all $task_{i,j}^{redis}$ for all workflows from Redis;
    \For{each $task$ $\in$ \{$task_{i,j}^{redis}$\}}  
      \If{$task.t_{start}$$\in$$\left[ task^{req}.t_{start},~task^{req}.t_{end} \right)$}
        \State $request.cpu$ += $task.cpu$;
        \State $request.mem$ += $task.mem$;
      \EndIf
    \EndFor
    \State /*\textit{ Call the ResourceDiscoveryAlgorithm }*/
    \State $ResidualMap$ $\leftarrow$ \textbf{ResourceDiscoveryAlgorithm};
    \For{each $item$ $\in$ $ResidualMap$}
      \State $totalResidual.cpu$ += $item.residual.cpu$;
      \State $totalResidual.mem$ += $item.residual.mem$; 
      \If{$item.residual.cpu$ \textgreater $Re_{max}^{cpu}$}
        \State $Re_{max}^{cpu}$ = $item.residual.cpu$;
        \State $Re_{max}^{mem}$ = $item.residual.mem$;
      \EndIf    
    \EndFor
    \State /*\textit{ Call the ResourceEvaluationAlgorithm }*/
    \State $allocated_{cpu}$, $allocated_{mem}$ $\leftarrow$\\
   ~~~~~~~~~~~~~~~~~~~~~~~~~~~~~~~~~\textbf{ResourceEvaluationAlgorithm};
    \If{$(allocated_{cpu} \geq s_{i,j}.min_{cpu})~and~$\\~~~~~~~~ $(allocated_{mem} \geq s_{i,j}.min_{mem} + \beta)$
    \State break;
    \EndIf}
  \EndFor
  \State return $allocated_{cpu}$, $allocated_{mem}$;
\end{algorithmic}
% \end{algorithm}
\end{breakablealgorithm}

\subsection{Adaptive resource allocation algorithm}
\label{sec:adaptive}
Algorithm 1 presents an adaptive resource allocation algorithm. 
It initializes these parameters to be 0 in line 1 and takes workflow task $s_{i,j}$ as input. 
Once the \textit{Containerized Executor} sends a task pod's resource request, the \textit{AdaptiveResourceAllocationAlgorithm} performs 
the following process. 
Lines 4-13 work to access the Redis database and get the total of requested resources of all task pods to be launched 
within $s_{i,j}$'s lifecycle. 
These task pods have resource competition with the currently requested task pod. 
Subsequently, Algorithm 1 uses \textit{ResourceDiscoveryAlgorithm}~(\ref{sec:discovery}) to obtain the remaining resources about 
the K8s cluster and each node in the K8s cluster. 
Lines 16-23 traverse the remaining resource structure $ResidualMap$ and accumulate to get the total remaining resources of all nodes 
across the K8s cluster. 
Meanwhile, the proposed algorithm obtains the maximal remaining CPU and memory resources. 
Herein, we assume that one node with the maximal remaining CPU resources also has the maximal remaining memory to facilitate 
the conditional comparison in \textit{ResourceEvaluationAlgorithm}~(prioritize CPU resource for allocation). 
Then, the proposed algorithm calls \textit{ResourceEvaluationAlgorithm} to present the allocated resources~(line 25). 

To ensure the task pod run properly in our experimental testbed, with the running program via \textit{Stress} tool within the task pod in mind, 
we add a constant $\beta$ to the minimum running resource for the task pod. 
It is because that \textit{Stress} tool in the running program of task pod uses $min_{mem}$ to allocate and release memory resources 
for resource loads. 
Resource amount $min_{mem}+\beta$ as a minimum of memory resource is just enough to run the task pods. 
Finally, the allocated resources returned by the proposed algorithm meet the conditions of minimum resource demands~(line 27).

% \begin{algorithm}[!t]
\begin{breakablealgorithm}
\caption{ResourceDiscoveryAlgorithm}
\label{discovery}
\begin{flushleft}
	{\bf{Input:}} $PodLister$, $NodeLister$, $ResidualMap$,;\\
	{\bf{Output:}} $ResidualMap$;
\end{flushleft}
\begin{algorithmic}[1]
  \State Initialization: $nodeReq.cpu$, $nodeReq.mem$,  $allocatable.cpu$, $allocatable.mem$, $residual.cpu$, $residual.mem$ $\leftarrow$ 0;  
	\State Get the $PodList$ from $PodLister$ through \textit{Informer};
	\State Get the $NodeList$ from $NodeLister$ through \textit{Informer};
	\For {$node~v_{i} \in V$}
    \State/*\textit{obtain the total resource requests of all pods on $v_{i}$}*/
    \For {each pod $p_{i}$ in $PodList$}
      \If{$p_{i}$ is hosted in $v_{i}$}
        \If{$pod_{i}.phase$ $\in$ $\{Running,Pending\}$}
          \State $nodeReq.cpu$ += $pod_{i}.request.cpu$;
          \State $nodeReq.mem$ += $pod_{i}.request.mem$;
        \EndIf
      \EndIf
    \EndFor
    \State/*\textit{obtain the allocatable resources on each node $v_{i}$}*/
    \State Obtain $node_{i}$ from $NodeList$ corresponding to $v_{i}$. 
    \State $allocatable.cpu$ = $node_{i}.allocatable.cpu$;
    \State $allocatable.mem$ = $node_{i}.allocatable.mem$;
    \State/*\textit{acquire the remaining resources on node $v_{i}$}*/
    \State $residual.cpu$ = $allocatable.cpu$ - $nodeReq.cpu$;
    \State $residual.mem$ = $allocatable.mem$ - $nodeReq.mem$;
    \State /*\textit{encapsulate Dictionary $ResidualMap$}*/
    \State ResidualMap[$v_{i}.ip$]=$\{residual.cpu,residual.mem\}$;
	\EndFor
\State return $ResidualMap$;
\end{algorithmic}
% \end{algorithm}
\end{breakablealgorithm}

\subsection{Resource discovery algorithm}
\label{sec:discovery}
Algorithm 2 shows how our resource discovery algorithm acquires the remaining resources of the K8s cluster and 
returns the remaining resource Map$ResidualMap$.
At first step, this algorithm initializes related parameters to be $0$ in line 1 and respectively get the $PodList$ 
and $NodeList$ of K8s cluster from $PodLister$ and $NodeLister$ through \textit{Informer} component.
Then the algorithm traverses all nodes in the K8s cluster and uses for-loop~(lines 6-13) to acquire the total number of occupied resources 
for all pods with \verb|Running| and \verb|Pending| states on the current node $v_i$. 
Lines 15-17 obtain the residual CPU and memory resources on the current $v_i$. 
Next, the algorithm encapsulates the $ResidualMap$ for each $v_i$. 
Once the iteration is complete for all the K8s cluster nodes, the algorithm returns the residual resource $ResidualMap$. 

\subsection{Resource evaluation algorithm}
\label{sec:evaluation}
Algorithm 3 elaborates the resource evaluation process in detail. 
The algorithm takes $task_{req}$, $Re_{max}^{cpu}$, $Re_{max}^{mem}$, $totalResidual.cpu$, $totalResidual.mem$, $request.cpu$ 
and $request.mem$ as input and finally return the allocated CPU and memory resources. 
As mentioned in (\ref{sec:adaptive}), some task pods to be launched during the requested task pod's lifecycle will compete for 
computational resources with the current task request $task^{req}$, 
and we use a resource scaling method to provide resource allocation based on a proportional value of the total remaining resources 
over the total amount of resource requests, defined as follows.
\begin{eqnarray}
  \label{eq:rate}
  \begin{split}
&cpu_{cut} = \left(task^{req}.cpu\right) \cdot \frac{totalResidual.cpu}{request.cpu},\\
&mem_{cut} = \left(task^{req}.mem\right) \cdot \frac{totalResidual.mem}{request.mem}. 
  \end{split} 
\end{eqnarray}
In addition, we define a resource allocation factor $\alpha$ for each node with a maximum of residual 
resources~(CPU or memory). 
Through lots of experimental evaluations, we use $\alpha = 0.8$, which means that the algorithm only 
allocates $80\%$ of the remaining resources in response to insufficient residual resource scenario on the current 
node while ensuring $20\%$ residual resources for its other loads.

\begin{breakablealgorithm}
\caption{ResourceEvaluationAlgorithm}
\label{evaluation}
\begin{flushleft}
	{\bf{Input:}}  $task^{req}$, $Re_{max}^{cpu}$, $Re_{max}^{mem}$, $totalResidual.cpu$,\\
  ~~~~~~~~~~~$totalResidual.mem$; $request.cpu$, $request.mem$;\\
	{\bf{Output:}} $allocated.cpu$, $allocated.mem$;
\end{flushleft}
\begin{algorithmic}[1]
  \State Get $cpu_{cut}$ and $mem_{cut}$ through Eq.~(\ref{eq:rate});
  \State define conditions $request.cpu$\textless$totalResidual.cpu$ as $A_1$, $request.mem$\textless$totalResidual.mem$ as $A_2$, 
  $task^{req}.cpu$\textless$Re_{max}^{cpu}$ as $B_1$, $task^{req}.mem$\textless$Re_{max}^{mem}$ as $B_2$, 
  $cpu_{cut}$\textless$Re_{max}^{cpu}$ as $C_1$, and $mem_{cut}$\textless$Re_{max}^{mem}$ as $C_2$;
  \State define the symbol $\neg$ as the negation of a condition and the symbol $\wedge$ as the logical and.
  \State /*\textit{~(1) The remaining resources are sufficient} */
  \If{$A_1$ $\wedge$ $A_2$}
    \If{$B_1$ $\wedge$ $B_2$}
      \State $allocated.cpu$ = $task^{req}.cpu$
      \State $allocated.mem$ = $task^{req}.mem$
    \Else
     \If{$\neg$$B_1$ $\wedge$ $B_2$}
      \State $allocated.cpu$ = $Re_{max}^{cpu} \cdot \alpha$
      \State $allocated.mem$ = $task^{req}.mem$
     \Else
       \If{$B_1$ $\wedge$ $\neg$$B_2$}
          \State $allocated.cpu$ = $task^{req}.cpu$ 
          \State $allocated.mem$ = $Re_{max}^{mem} \cdot \alpha$
       \Else
          \State $allocated.cpu$ = $Re_{max}^{cpu} \cdot \alpha$ 
          \State $allocated.mem$ = $Re_{max}^{mem} \cdot \alpha$
       \EndIf
      \EndIf
    \EndIf
  \EndIf

  \State /*\textit{~(2) The remaining cpu resource is unsufficient} */
  \If{$\neg$$A_1$ $\wedge$ $A_2$}
    \If{$C_1$ $\wedge$ $B_2$}
      \State $allocated.cpu$ = $cpu_{cut}$
      \State $allocated.mem$ = $task^{req}.mem$
    \Else
      \If{$\neg$$C_1$ $\wedge$ $B_2$}
        \State $allocated.cpu$ = $Re_{max}^{cpu} \cdot \alpha$
        \State $allocated.mem$ = $task^{req}.mem$   
      \Else
        \If{$C_1$ $\wedge$ $\neg$$B_2$}
          \State $allocated.cpu$ = $cpu_{cut}$
          \State $allocated.mem$ = $Re_{max}^{mem} \cdot \alpha$       
        \Else
          \State $allocated.cpu$ = $Re_{max}^{cpu} \cdot \alpha$
          \State $allocated.mem$ = $Re_{max}^{mem} \cdot \alpha$ 
        \EndIf
      \EndIf
    \EndIf
  \EndIf

  \State /*\textit{~(3) The remaining memory resource is unsufficient} */
  \If{$A_1$ $\wedge$ $\neg$$A_2$}
    \If{$B_1$ $\wedge$ $C_2$}
      \State $allocated.cpu$ = $task^{req}.cpu$
      \State $allocated.mem$ = $mem_{cut}$
    \Else
      \If{$\neg$$B_1$ $\wedge$ $C_2$}
        \State $allocated.cpu$ = $Re_{max}^{cpu} \cdot \alpha$
        \State $allocated.mem$ = $mem_{cut}$  
      \Else
        \If{$B_1$ $\wedge$ $\neg$$C_2$}
          \State $allocated.cpu$ = $task^{req}.cpu$
          \State $allocated.mem$ = $Re_{max}^{mem} \cdot \alpha$       
        \Else
          \State $allocated.cpu$ = $Re_{max}^{cpu} \cdot \alpha$
          \State $allocated.mem$ = $Re_{max}^{mem} \cdot \alpha$ 
        \EndIf
      \EndIf
    \EndIf
  \EndIf

  \State /*\textit{~(4) Both residual resources are unsufficient} */
  \If{$\neg$$A_1$ $\wedge$ $\neg$$A_2$}
      \State $allocated.cpu$ = $cpu_{cut}$
      \State $allocated.mem$ = $mem_{cut}$
    \EndIf
\State return $allocated.cpu$, $allocated.mem$;
\end{algorithmic}
\end{breakablealgorithm}

Algorithm 3 first uses resource scaling to obtain allocated CPU and memory resources by Eq.~(\ref{eq:rate}) 
and defines six comparative conditions $A_{1}$, $A_{2}$, $B_{1}$, $B_{2}$, $C_{1}$ and $C_{2}$~(lines 1-2). 
The symbol $\neg$ denotes the negation of the condition and the symbol $\wedge$ denotes the logical {\itshape{and}}~(line 3).
Then the algorithm implements four resource allocation scenarios according to comparative conditions between accumulated resource 
requests~(concurrent scenario) within the current task lifecycle and the total residual resources. 

\textbf{Sufficient residual resources.} 
When the total residual resources across the K8s cluster are abundant for concurrent tasks within the current task lifecycle, 
we only think about $B_{1}$ and $B_{2}$. 
When the CPU and memory resource requests of the current $task^{req}$ are smaller than the maximum remaining resources of 
a node~(meets $B_1$ and $B_2$), the algorithm allocates resources in the light of the current task request $task^{req}$~(lines 6-8). 
If the maximum remaining CPU resource on the cluster node fails to suffice for the current task request $task^{req}$, 
the algorithm allocates $Re_{max}^{cpu} \cdot \alpha$ of the maximum remaining CPU resource on cluster node~(lines 10-12). 
Conversely, so is memory~(lines 14-16). When neither of the two remaining resources on cluster node~(CPU and memory) 
can satisfy $task^{req}$, both types of allocated resources scale down by multiplying  $\alpha$~(lines 17-19).

  \textbf{Insufficient residual CPU resource.} 
 When the total residual CPU resource across the K8s cluster cannot satisfy the total resource demand of concurrent tasks, 
 conditions $C_{1}$ and $B_{2}$ are considered. 
 The algorithm acquires the $cpu_{cut}$ through the resource scaling method~(Eq.~(\ref{eq:rate})). 
 In case of conditions $C_{1}$ and $B_{2}$, we allocate resources according to $cpu_{cut}$ and $task^{req}.mem$~(lines 26-28). 
 When the maximum CPU remaining resources of a node fail to accommodate $cpu_{cut}$, the algorithm 
 adopts $Re_{max}^{cpu} \cdot \alpha$ as the allocated CPU resource. 
Due to sufficient memory capacities, the algorithm suffices for the current task memory request $task^{req}.mem$~(lines 30-32). 
In case of conditions $C_1$ $\wedge$ $\neg$$B_2$, the algorithm allocates $cpu_{cut}$ CPU resource and $Re_{max}^{mem} \cdot \alpha$ memory 
resource, which is due to that the current task memory request is greater than the maximum residual memory resource on cluster node~(lines 34-36). 
Instead, the algorithm allocates resources~(CPU and memory) according to the $\alpha$ scale factor of the largest node's remaining 
resources~(lines 37-39).

\textbf{Insufficient residual memory resource.} 
  When the total residual memory resource across the K8s cluster cannot satisfy the total resource demand of concurrent tasks, 
  conditions $B_{1}$ and $C_{2}$ are considered. 
  The algorithm acquires the $mem_{cut}$ through the resource scaling method~(Eq.~(\ref{eq:rate})). 
  The operations under conditions $B_{1}$ $\wedge$ $C_{2}$, $\neg$$B_{1}$ $\wedge$ $C_{2}$, $B_{1}$ $\wedge$ $\neg$$C_{2}$ 
  and $\neg$$B_{1}$ $\wedge$ $\neg$$C_{2}$ are similar to the above, except that here we are talking about memory resource~(lines 45-63).

\textbf{Insufficient residual CPU and memory resources.} 
In the case of $\neg$$A_1$ $\wedge$ $\neg$$A_2$, meaning that the total remaining CPU and memory resources 
across the K8s cluster fail to suffice for the CPU and memory resource requests of concurrent tasks, 
the algorithm allocates CPU and memory resources according to $cpu_{cut}$ and $mem_{cut}$ obtained by 
resource scaling method~(lines 65-67). 
Finally, \textit{ResourceEvaluationAlgorithm} returns allocated resources $allocated.cpu$ and $allocated.mem$.

\section{Experimental Evaluation}
In the following, we evaluate the proposed ARAS for different evaluation metrics and discuss 
the benefits of the proposed ARAS under three distinct arrival patterns compared with the baseline. 

\subsection{Experimental setup and design}
For the evaluation, we apply a setting employed in our former work~\cite{shan2022kubeadaptor} and make adaptations for the proposed ARAS within 
KubeAdaptor discussed here. In this subsection, we briefly introduce experimental scenarios, 
workflow examples, workflow instantiation, workflow arrival patterns, evaluation metrics, and baseline algorithm.

\subsubsection{Experimental scenarios}
\label{sec:senario}
The K8s cluster used in our experiments consists of one Master node and six nodes. 
Each node equips with an 8-core AMD EPYC 7742 2.2GHz CPU and 16GB of RAM, running Ubuntu 20.4 and K8s v1.19.6 and Docker version 18.09.6. 
The Redis database~v5.0.7 is installed on the Master node. \textit{Workflow Injector Module} and \textit{Containerized Workflow Builder} are 
containerized and deployed into the K8s cluster through \textit{Service}~\footnote[11]{https://kubernetes.io/docs/concepts/} 
and \textit{Deployment}~\footnote[12]{https://kubernetes.io/docs/concepts/workloads/}. 
We explore the performances of the proposed ARAS and baseline by running four scientific 
workflows on the K8s cluster.

\begin{figure*}[htbp]
\centering
\includegraphics[scale=0.4]{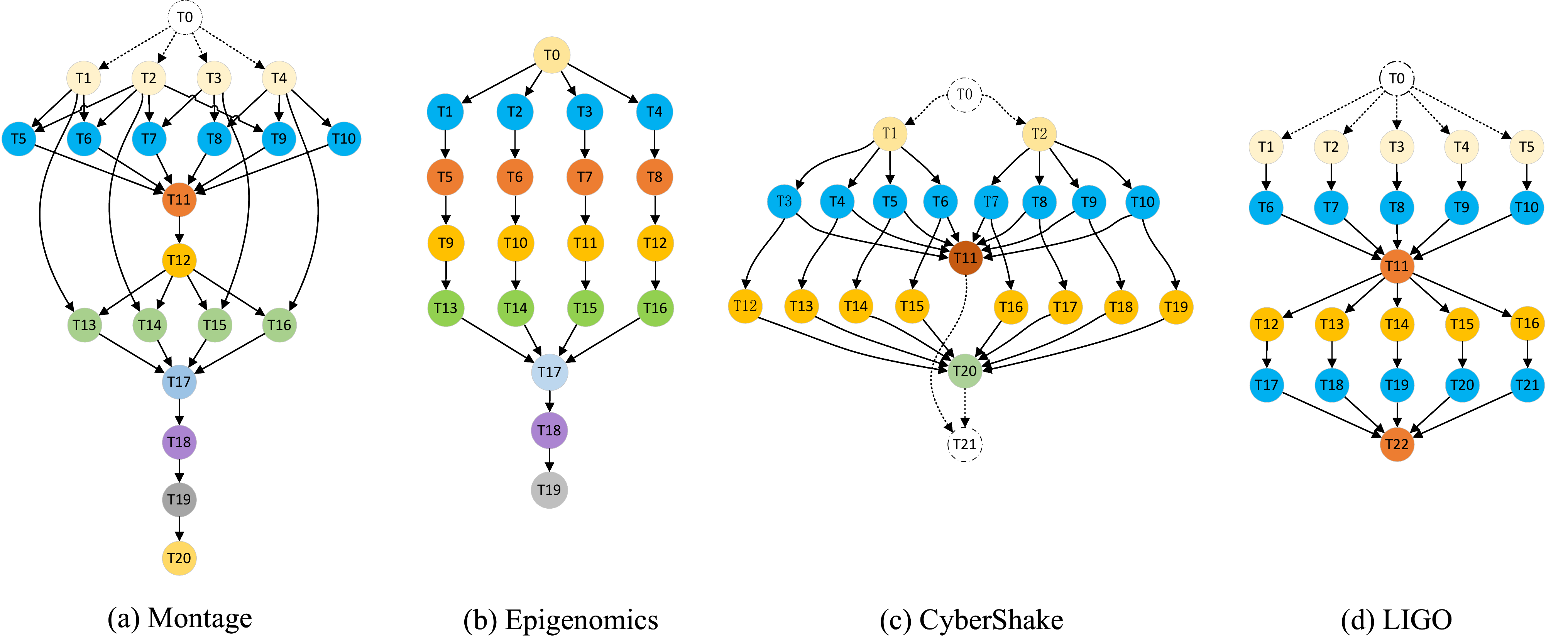}
\caption{The topology diagram of four scientific workflow applications.}
\vspace{-0.5cm}
\label{fig_four}
\end{figure*}

\subsubsection{Workflow examples}
To verify the adaptations of our proposed ARAS within KubeAdaptor, four scientific workflows, such as~Montage~(astronomy), 
Epigenomics~(genome sequence), CyberShake~(earthquake science), and LIGO Inspiral~(gravitational physics), 
are used to run on K8s cluster in a containerized manner~\cite{juve2013characterizing}. 
We make a few tweaks to the workflow structure and add virtual entrance and exit nodes of workflows to form the workflow structure with 
the DAG diagram. 
As for each type of scientific workflow, we uniformly adopt a small-scale workflow~(about 20 tasks) in our experiments, 
as shown in Fig.~\ref{fig_four}, derived from the Pegasus Workflow repository~\cite{workflow:gallery}. 
Structurally, four types of workflows cover all the structural features regarding composition and components~(in-tree, out-tree, fork-join, 
and pipeline) that serve to illustrate the universality and complexity of workflows. 
Herein, we only consider the topologies of four scientific workflows and do not focus on the real-world data processing of the tasks, 
which does not affect verifying the adaptability of our ARAS.
For ease of performance comparisons among resource allocation solutions, we assume that four classes of scientific workflows consist of 
the same tasks. 
Each node of workflow DAGs uses resource load~(CPU and memory utilization) and service runtime to simulate workflow tasks in the experiments. 
Note that the KubeAdaptor schedules workflow tasks topologically in a top-down fashion according to task dependencies.

\subsubsection{Workflow instantiation}
\label{sec:instant}
As for resource load in workflow tasks, we employ several parameters together with \textit{Stress} to work on simulating scientific 
workflow tasks. 
In each task program, we use the \textit{Stress}~\footnote[13]{https://linux.die.net/man/1/stress} tool to set several CPU forks, a memory of $1000Mi$~(is equal to $mem_{min}$ of 
Eq.~(\ref{eq_s_ij})), and random duration~(user's definition ahead of time in Eq.~(\ref{eq_s_ij})). 
CPU forking and memory allocation operations in the task pod last twice as long as $duraion$. 
The total duration of each task pod is random and falls between $10s$ and $20s$.
Then we pack the Python application with \textit{Stress} program into a task Image file through Docker 
Engine~\footnote[14]{https://github.com/IsaacKlop/task-emulator}, 
store the task Image file in local Harbor~\footnote[15]{https://goharbor.io/} or remote Docker Hub 
repository~\footnote[16]{https://docs.docker.com/docker-hub/repos/}. 
We can import container parameters~(refer to Eq.~(\ref{eq_s_ij})) defined in the ConfigMap file of 
in \textit{Workflow Injection Module} into the task container hosted in the task pod. 
For resource setting within task pod, we uniformly set the resource \textit{requests} and \textit{limits} to $2000$ Milli 
cores~(i.e.,~$2000m$) CPU and $4000Mi$ memory. 
Note that the \textit{requests} field has the same parameter as the \textit{limits} field, which ensures that this task pod has the 
highest priority, namely \textit{Guaranteed}~\cite{quality:service}.

\subsubsection{Workflow arrival patterns}
We make use of three distinct workflow request arrival patterns:

\textbf{Constant Arrival Scenario:} In this scenario, workflow requests arrive in a constant manner. 
\textit{Workflow Injector Moduler} together with \textit{CLI} sends $5$ workflow requests simultaneously to 
the \textit{Containerized Workflow Builder} every $300$ seconds, i.e., $y=5$. Send six times for a total of $30$ workflows.
A graphical depiction of this arrival curve is depicted in Fig.~\ref{fig:montage}(a).

 \textbf{Linear Arrival Scenario:} In this scenario, the workflow requests are injected into \textit{Containerized Workflow Builder} 
 in a linear rising function, i.e., $y=k*x+d$, where $y$ is the amount of concurrent workflow request and $d$ is the initial value $2$. 
 Concurrent workflow requests increase by $k=2$ every $300$ seconds. This linear arrival curve is depicted in Fig.~\ref{fig:montage}(b).   
 Send five times for a total of $30$ workflows.

  \textbf{Pyramid Arrival Scenario:} In this scenario, workflow requests are sent to \textit{Containerized Workflow Builder} in line with 
  a pyramid-like function. 
  We start with a small number of concurrent workflow requests~(is equal to $2$), till it grows to a randomly selected large 
  number~(is equal to $6$ for each type of workflow), which can be seen in Fig.~\ref{fig:montage}(c).
  Concurrent workflow requests grow every $300$ seconds by $2$ until the peak is reached. 
  Once the peak reaches, we immediately reduce this number to the small initial value in the same manner and repeat this process until 
  the total number of workflow requests is reached~(herein, $34$). 

 The deployment of these three scenarios aims to maximize coverage of the ever-changing resource needs and sudden peaks of workflow requests in a production environment. 
  Even if there is some predictability in the Constant and Linear Arrival scenarios, the Pyramid function follows an unpredictable arrival pattern.

 \begin{table*}
 	\caption{Evaluation results}
  \label{table:evaluation}
   \centering
 \renewcommand\arraystretch{1} %增加表格行距
\begin{tabular}{|m{1.4cm}<{\centering}|m{5.3cm}<{\centering}|m{1.3cm}<{\centering}|m{1.3cm}<{\centering}|m{1.3cm}<{\centering}|m{1.3cm}<{\centering}|m{1.3cm}<{\centering}|m{1.3cm}<{\centering}|}

%     %保持水平\垂直居中
 		\hline 
 	\multirow{4}{*}{\bfseries{Workflow}}&\multirow{2}{*}{\bfseries{Metrics}}& \multicolumn{2}{c|}{Constant Arrival} & \multicolumn{2}{c|}{Linear Arrival} & \multicolumn{2}{c|}{Pyramid Arrival} \\ 
 	\cline{3-8}
    &  & Adaptive & Baseline & Adaptive & Baseline & Adaptive & Baseline  \\ 
   	\cline{2-8}
    & Number of Workflow Rquests & \multicolumn{2}{c|}{30} & \multicolumn{2}{c|}{30} & \multicolumn{2}{c|}{34} \\  
   	\cline{2-8}
 \bfseries{Types} & Interval between two Requests Bursts~(in Seconds) & \multicolumn{2}{c|}{300} & \multicolumn{2}{c|}{300} & \multicolumn{2}{c|}{300} \\ 
  \hline 
%%%%%%%%%%%%%%%%
 \multirow{10}{*}{Montage} &	\multirow{2}{*}{Total Duration of All Workflows in Minutes} & \multirow{2}{*}{33.18} & \multirow{2}{*}{36.79} & \multirow{2}{*}{26.95} & \multirow{2}{*}{36.45} & \multirow{2}{*}{49.31} &\multirow{2}{*}{54.69} \\ [0.7ex]
 
  & (Standard Deviation) & ($\delta = 0.21$) & ($\delta = 2.26$) & ($\delta = 0.38$) & ($\delta = 6.31$) & ($\delta = 2.46$) & ($\delta = 1.74$) \\ [0.7ex]

  \cline{2-8}
  & \multirow{2}{*}{Average Workflow Duration in Minutes}  & \multirow{2}{*}{5.74} & \multirow{2}{*}{7.80} & \multirow{2}{*}{5.41} & \multirow{2}{*}{11.33} & \multirow{2}{*}{7.22} & \multirow{2}{*}{11.73} \\ [0.7ex]
  
  & (Standard Deviation) & ($\delta = 0.49$) & ($\delta = 0.36$) & ($\delta = 0.26$) & ($\delta = 4.28$) & ($\delta = 1.36$) & ($\delta = 0.88$) \\ [0.7ex]
  
  \cline{2-8}
  &	\multirow{2}{*}{CPU resource Usage}  & \multirow{2}{*}{0.28} & \multirow{2}{*}{0.27} & \multirow{2}{*}{0.35} & \multirow{2}{*}{0.31} & \multirow{2}{*}{0.26} & \multirow{2}{*}{0.20} \\ [0.7ex]
   
  & (Standard Deviation) & ($\delta = 0.00$) & ($\delta = 0.02$) & ($\delta = 0.01$) & ($\delta = 0.07$) & ($\delta = 0.03$) & ($\delta = 0.01$) \\ [0.7ex]
  
  \cline{2-8}
  & \multirow{2}{*}{Memory resource Usage}  & \multirow{2}{*}{0.28} & \multirow{2}{*}{0.27} & \multirow{2}{*}{0.35} & \multirow{2}{*}{0.31} & \multirow{2}{*}{0.26} & \multirow{2}{*}{0.20} \\ [0.7ex]
 	
  & (Standard Deviation) & ($\delta = 0.00$) & ($\delta = 0.13$) & ($\delta = 0.01$) & ($\delta = 0.07$) & ($\delta = 0.03$) & ($\delta = 0.01$) \\ [0.7ex]
\hline
  %%%%%%%%%%%%%%%%%
 \multirow{10}{*}{Epigenomics} &	\multirow{2}{*}{Total Duration of All Workflows in Minutes} & \multirow{2}{*}{30.55} & \multirow{2}{*}{39.06} & \multirow{2}{*}{34.3} & \multirow{2}{*}{43.66} & \multirow{2}{*}{51.42} &\multirow{2}{*}{62.12} \\ [0.7ex]
 
  & (Standard Deviation) & ($\delta = 0.19$) & ($\delta = 1.84$) & ($\delta = 7.29$) & ($\delta = 4.37$) & ($\delta = 4.28$) & ($\delta = 4.32$) \\ [0.7ex]

  \cline{2-8}
  & \multirow{2}{*}{Average Workflow Duration in Minutes}  & \multirow{2}{*}{4.24} & \multirow{2}{*}{9.35} & \multirow{2}{*}{9.81} & \multirow{2}{*}{16.53} & \multirow{2}{*}{9.65} & \multirow{2}{*}{19.41} \\ [0.7ex]
  & (Standard Deviation) & ($\delta = 0.05$) & ($\delta = 1.56$) & ($\delta = 5.11$) & ($\delta = 4.41$) & ($\delta = 3.33$) & ($\delta = 6.04$) \\ [0.7ex]
  
  \cline{2-8}
  &	\multirow{2}{*}{CPU resource Usage}  & \multirow{2}{*}{0.34} & \multirow{2}{*}{0.27} & \multirow{2}{*}{0.32} & \multirow{2}{*}{0.25} & \multirow{2}{*}{0.21} & \multirow{2}{*}{0.20} \\ [0.7ex]
   
  & (Standard Deviation) & ($\delta = 0.02$) & ($\delta = 0.01$) & ($\delta = 0.06$) & ($\delta = 0.00$) & ($\delta = 0.00$) & ($\delta = 0.00$) \\ [0.7ex]
  
  \cline{2-8}
  & \multirow{2}{*}{Memory resource Usage}  & \multirow{2}{*}{0.34} & \multirow{2}{*}{0.27} & \multirow{2}{*}{0.32} & \multirow{2}{*}{0.25} & \multirow{2}{*}{0.21} & \multirow{2}{*}{0.20} \\ [0.7ex]
 	
  & (Standard Deviation) & ($\delta = 0.02$) & ($\delta = 0.01$) & ($\delta = 0.06$) & ($\delta = 0.00$) & ($\delta = 0.01$) & ($\delta = 0.01$) \\ [0.7ex]
\hline
  %%%%%%%%%%%%%
 \multirow{10}{*}{CyberShake} &	\multirow{2}{*}{Total Duration of All Workflows in Minutes} & \multirow{2}{*}{38.30} & \multirow{2}{*}{50.29} & \multirow{2}{*}{34.06} & \multirow{2}{*}{49.46} & \multirow{2}{*}{46.76} &\multirow{2}{*}{66.41} \\ [0.7ex]
 
  & (Standard Deviation) & ($\delta = 5.77$) & ($\delta = 5.29$) & ($\delta = 6.16$) & ($\delta = 1.18$) & ($\delta = 4.02$) & ($\delta = 6.56$) \\  [0.7ex]

  \cline{2-8}
  & \multirow{2}{*}{Average Workflow Duration in Minutes}  & \multirow{2}{*}{9.19} & \multirow{2}{*}{17.29} & \multirow{2}{*}{9.41} & \multirow{2}{*}{20.61} & \multirow{2}{*}{4.94} & \multirow{2}{*}{19.47} \\  [0.7ex]
  & (Standard Deviation) & ($\delta = 3.72$) & ($\delta = 2.89$) & ($\delta = 4.27$) & ($\delta = 0.86$) & ($\delta = 2.07$) & ($\delta = 6.50$) \\ [0.7ex]
  
  \cline{2-8}
  &	\multirow{2}{*}{CPU resource Usage}  & \multirow{2}{*}{0.26} & \multirow{2}{*}{0.24} & \multirow{2}{*}{0.27} & \multirow{2}{*}{0.24} & \multirow{2}{*}{0.22} & \multirow{2}{*}{0.19} \\ [0.7ex]
   
  & (Standard Deviation) & ($\delta = 0.03$) & ($\delta = 0.02$) & ($\delta = 0.04$) & ($\delta = 0.01$) & ($\delta = 0.03$) & ($\delta = 0.01$) \\ [0.7ex]
  
  \cline{2-8}
  & \multirow{2}{*}{Memory resource Usage}  & \multirow{2}{*}{0.26} & \multirow{2}{*}{0.24} & \multirow{2}{*}{0.27} & \multirow{2}{*}{0.23} & \multirow{2}{*}{0.22} & \multirow{2}{*}{0.19} \\ [0.7ex]
 	
  & (Standard Deviation) & ($\delta = 0.03$) & ($\delta = 0.02$) & ($\delta = 0.04$) & ($\delta = 0.01$) & ($\delta = 0.03$) & ($\delta = 0.01$) \\  [0.7ex]
\hline
  %%%%%%%%%%%%%%%%%%%%%%%%
 \multirow{10}{*}{LIGO} &	\multirow{2}{*}{Total Duration of All Workflows in Minutes} & \multirow{2}{*}{30.82} & \multirow{2}{*}{52.17} & \multirow{2}{*}{44.02} & \multirow{2}{*}{53.87} & \multirow{2}{*}{45.26} &\multirow{2}{*}{63.56} \\ [0.7ex]
  & (Standard Deviation) & ($\delta = 0.38$) & ($\delta = 3.99$) & ($\delta = 10.9$) & ($\delta = 4.20$) & ($\delta = 0.52$) & ($\delta = 1.60$) \\ [0.7ex]

  \cline{2-8}
  & \multirow{2}{*}{Average Workflow Duration in Minutes}  & \multirow{2}{*}{4.26} & \multirow{2}{*}{21.15} & \multirow{2}{*}{16.21} & \multirow{2}{*}{28.05} & \multirow{2}{*}{4.20} & \multirow{2}{*}{14.07} \\ [0.7ex]
   & (Standard Deviation) & ($\delta = 0.09$) & ($\delta = 0.44$) & ($\delta = 7.68$) & ($\delta = 7.88$) & ($\delta = 0.15$) & ($\delta = 1.33$) \\ [0.7ex]
  
  \cline{2-8}
  &	\multirow{2}{*}{CPU resource Usage}  & \multirow{2}{*}{0.40} & \multirow{2}{*}{0.24} & \multirow{2}{*}{0.28} & \multirow{2}{*}{0.23} & \multirow{2}{*}{0.31} & \multirow{2}{*}{0.23} \\  [0.7ex]
   
  & (Standard Deviation) & ($\delta = 0.00$) & ($\delta = 0.02$) & ($\delta = 0.08$) & ($\delta = 0.02$) & ($\delta = 0.01$) & ($\delta = 0.00$) \\ [0.7ex]
  
  \cline{2-8}
  & \multirow{2}{*}{Memory resource Usage}  & \multirow{2}{*}{0.40} & \multirow{2}{*}{0.24} & \multirow{2}{*}{0.28} & \multirow{2}{*}{0.23} & \multirow{2}{*}{0.31} & \multirow{2}{*}{0.23} \\[0.7ex]
 	
  & (Standard Deviation) & ($\delta = 0.00$) & ($\delta = 0.02$) & ($\delta = 0.08$) & ($\delta = 0.02$) & ($\delta = 0.01$) & ($\delta = 0.00$) \\ [0.7ex]

  \hline
\end{tabular}
	\vspace{-0.3cm}
 \end{table*}

\subsubsection{Evaluation metrics}
\label{sec:metric}
To evaluate our ARAS, we conduct each arrival pattern three times and analyze the results against the following 
quantitative metrics:

\textbf{Total Duration of All Workflows~(in Minutes):} This metric is the average total duration of all injected workflows, i.e., 
the elapsed time from the arrival of the first workflow request to the moment when the last workflow request is complete.

  \textbf{Average Workflow Duration~(in Minutes):} This metric reflects the average execution time of individual workflow, 
  which is the time that each workflow takes from the start of the first task to the end of the last.

 \textbf{Resource Usage:} Resource usage contains CPU and memory resource utilization, reflecting the average resource 
  utilization throughout the total duration of all injected workflows across the K8s cluster.
 The greater the resource utilization, the closer to our optimization goal. 
  The resource usage comparison covering four types of scientific workflows with the baseline algorithm further verifies the 
  better performance of our ARAS solution. 

\subsubsection{Baseline}
\label{sec:baseline}
In experiments, we used our recent resource allocation strategy~\cite{shan2022kubeadaptor} as a baseline method, which does not take into 
account the potential future task requests throughout the current task's lifecycle. 
It means that the resource allocation strategy in the baseline follows the First Come First Serve~(FCFS) and relies on the adequacy of 
residual resources on cluster nodes. 
If enough, the resource allocation is complete. Otherwise, wait for other task pods to complete and release resources to meet 
the resource reallocation for the current task request. 
\begin{figure*}
	\centering
    \vspace{-0.5cm}
	\subfigure[Constant Arrival Pattern]{
		\begin{minipage}[t]{0.33\linewidth}
			\centering
			\includegraphics[width=2.4in]{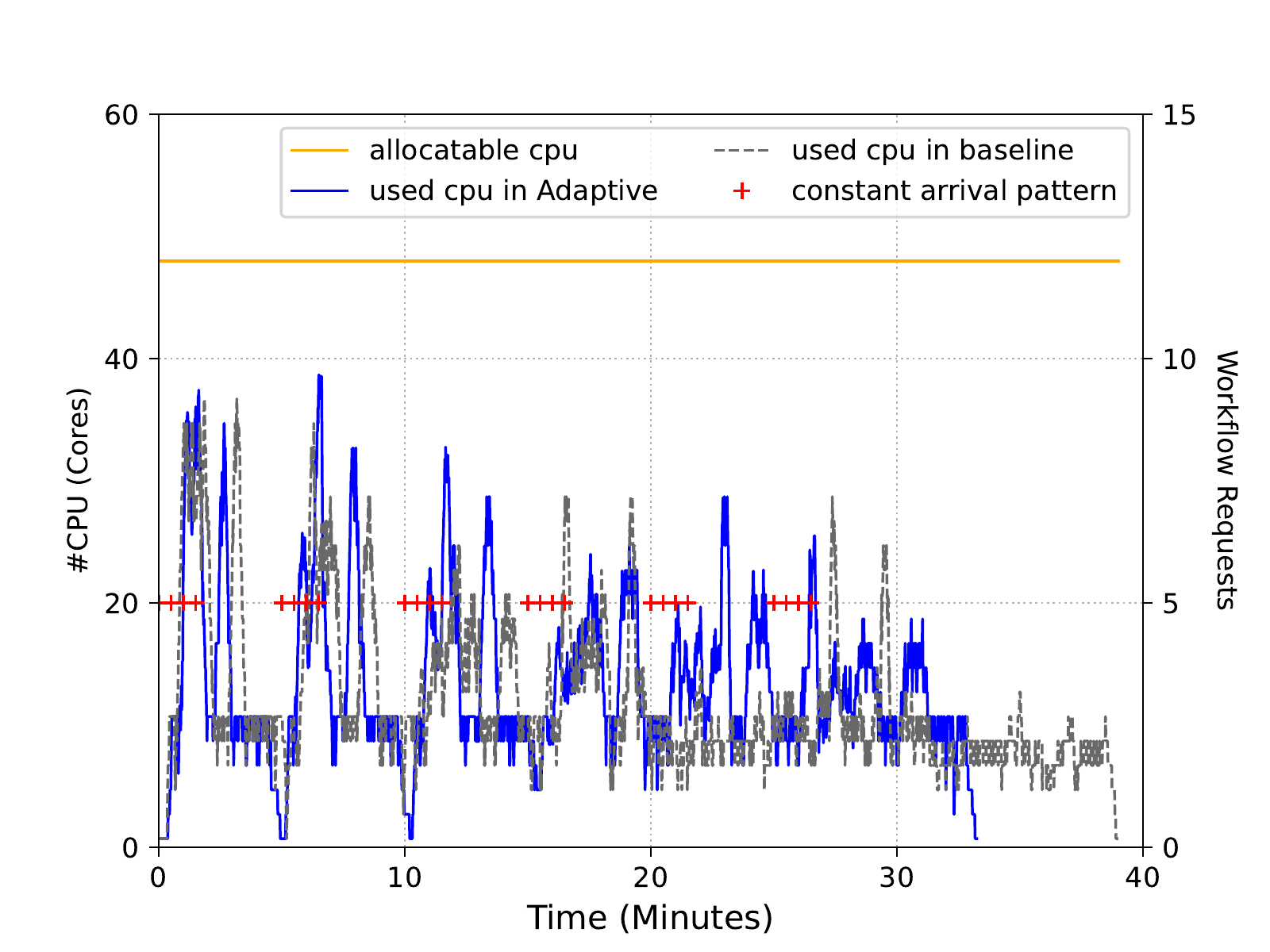}\\
			\vspace{0.02cm}
			\includegraphics[width=2.4in]{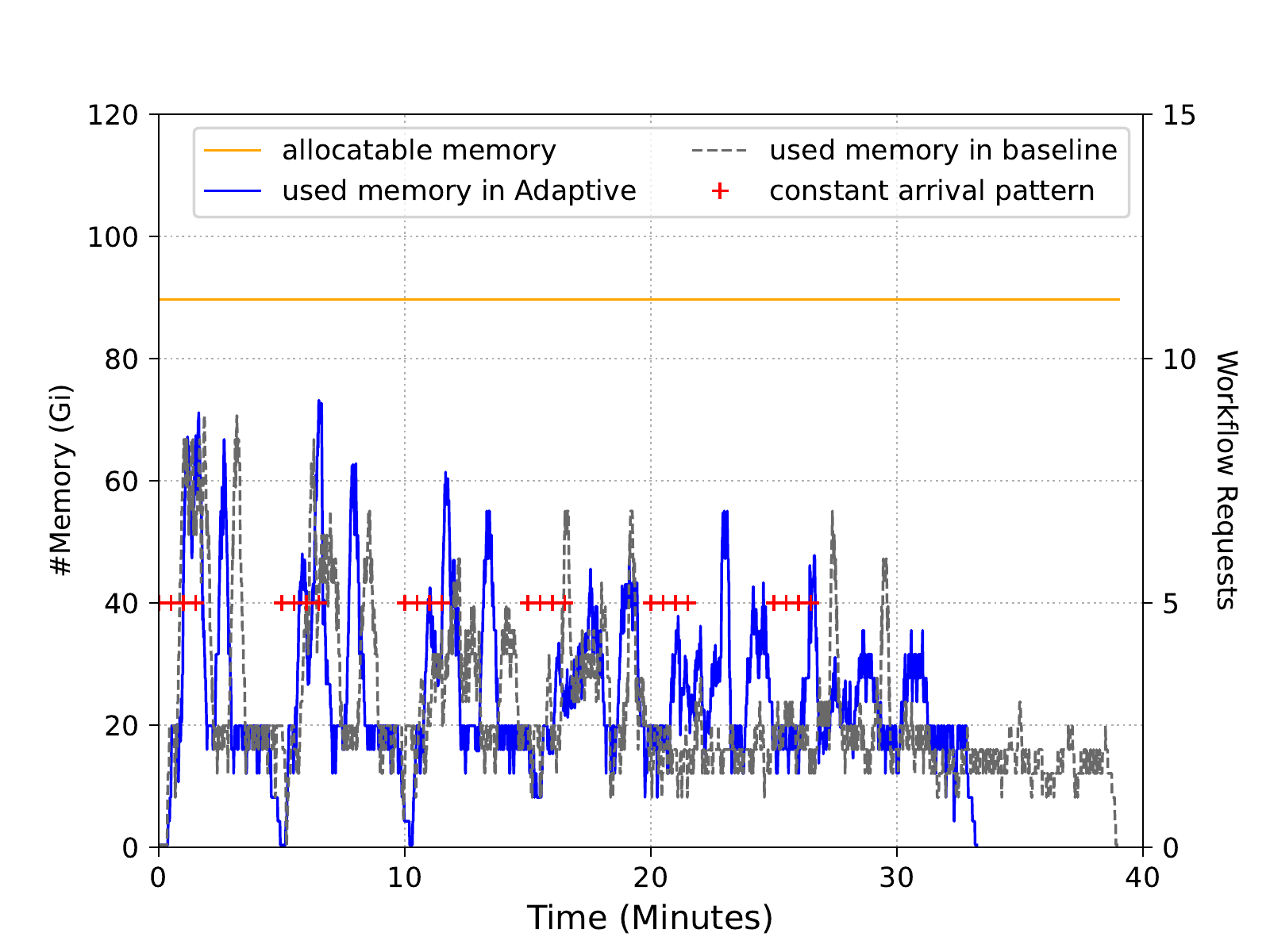}\\
			\vspace{0.02cm}
		\end{minipage}%
	}%
	\subfigure[Linear Arrival Pattern]{
		\begin{minipage}[t]{0.33\linewidth}
			\centering
			\includegraphics[width=2.4in]{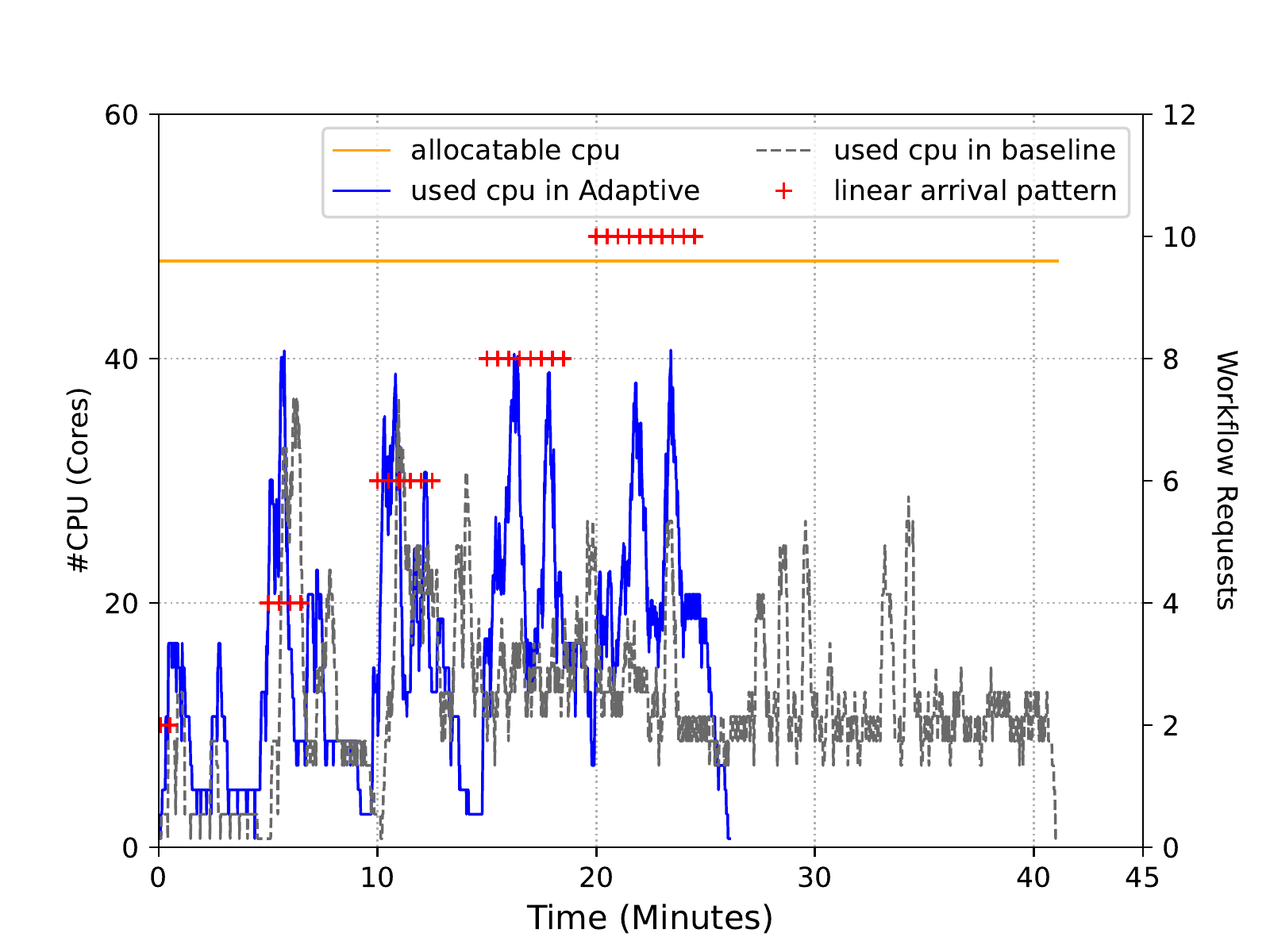}\\
			\vspace{0.02cm}
			\includegraphics[width=2.4in]{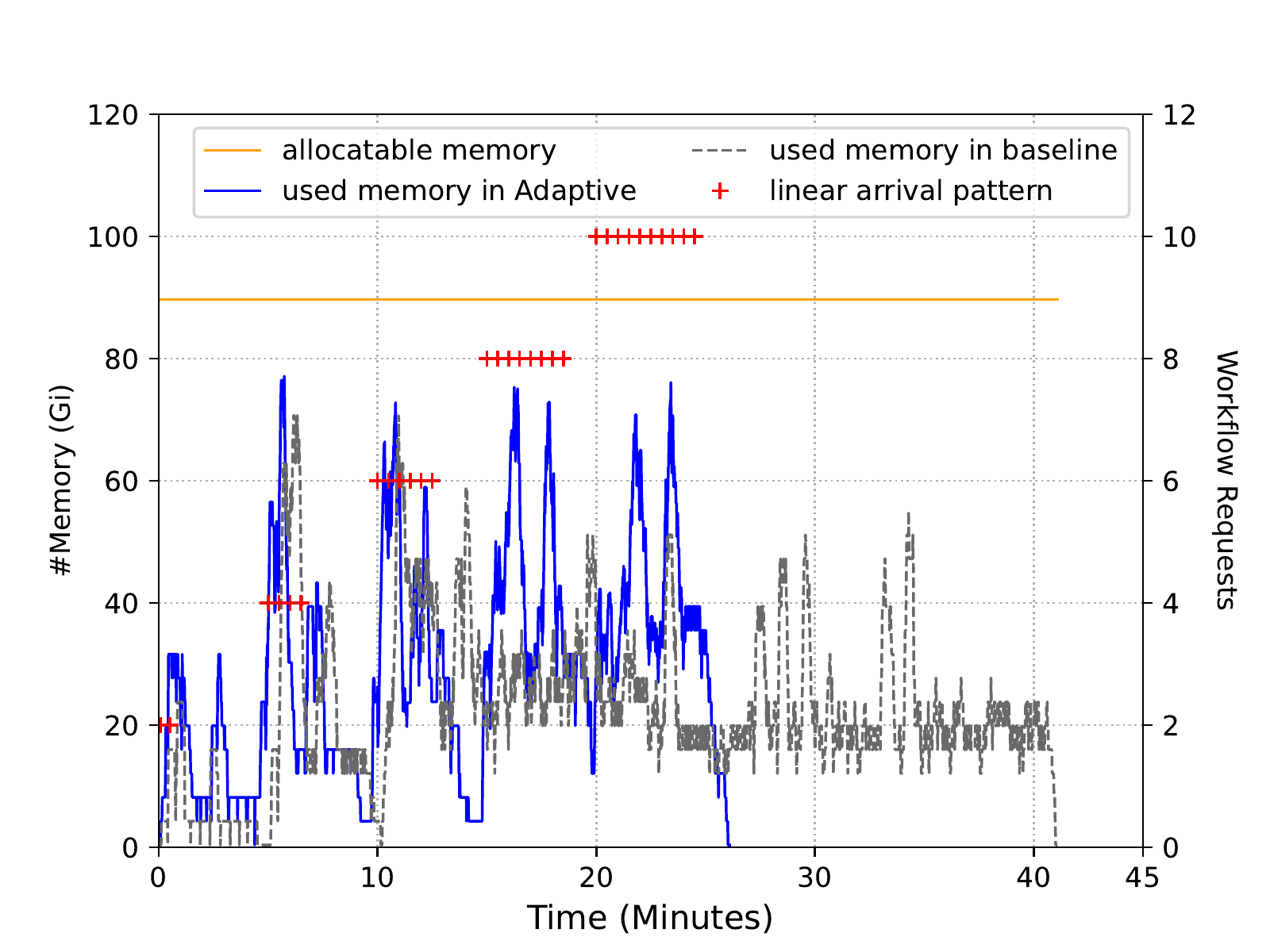}\\
			\vspace{0.02cm}
		\end{minipage}%
	}%
	\subfigure[Pyramid Arrival Pattern]{
		\begin{minipage}[t]{0.33\linewidth}
			\centering
			\includegraphics[width=2.4in]{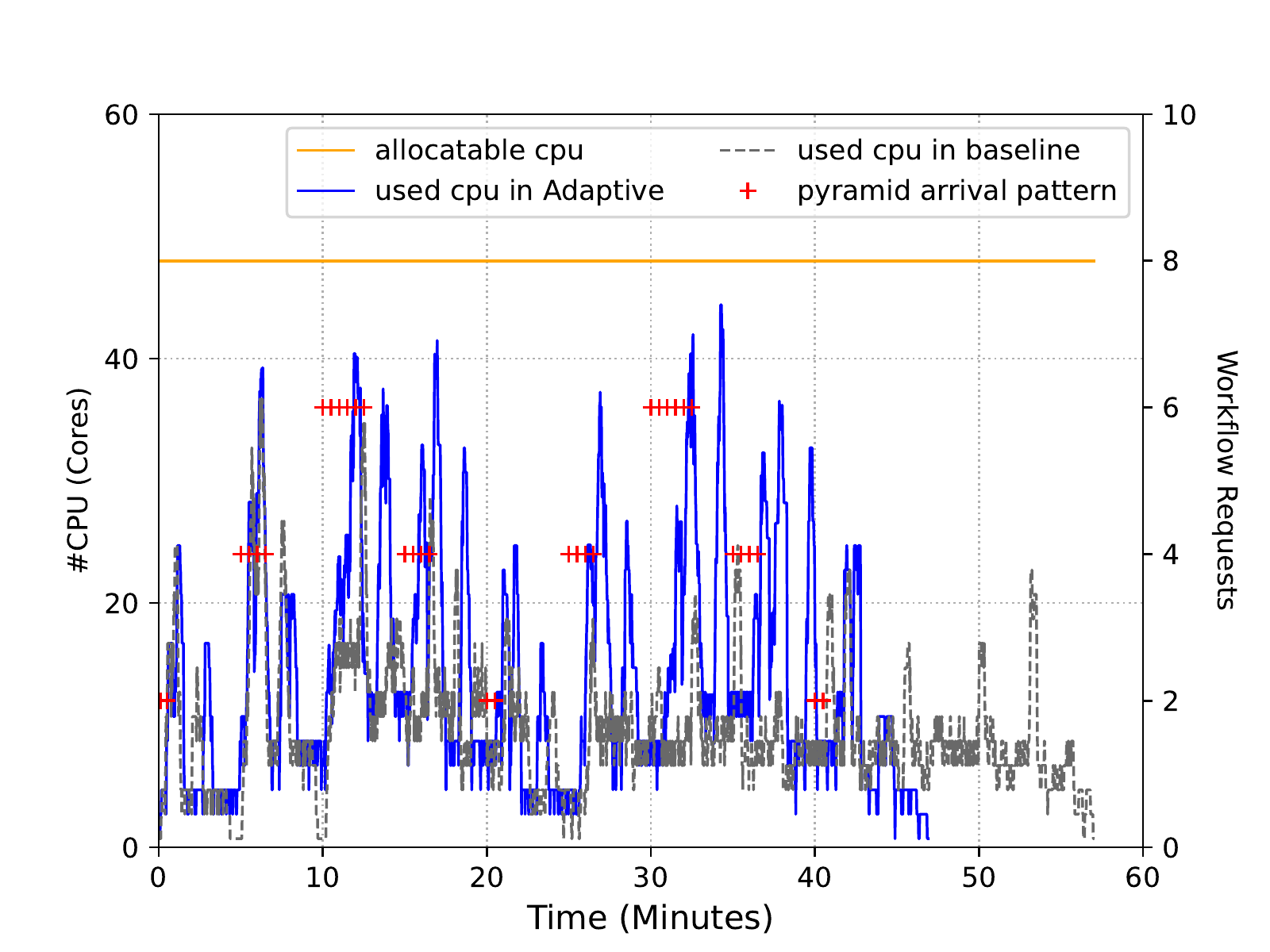}\\
			\vspace{0.02cm}
			\includegraphics[width=2.4in]{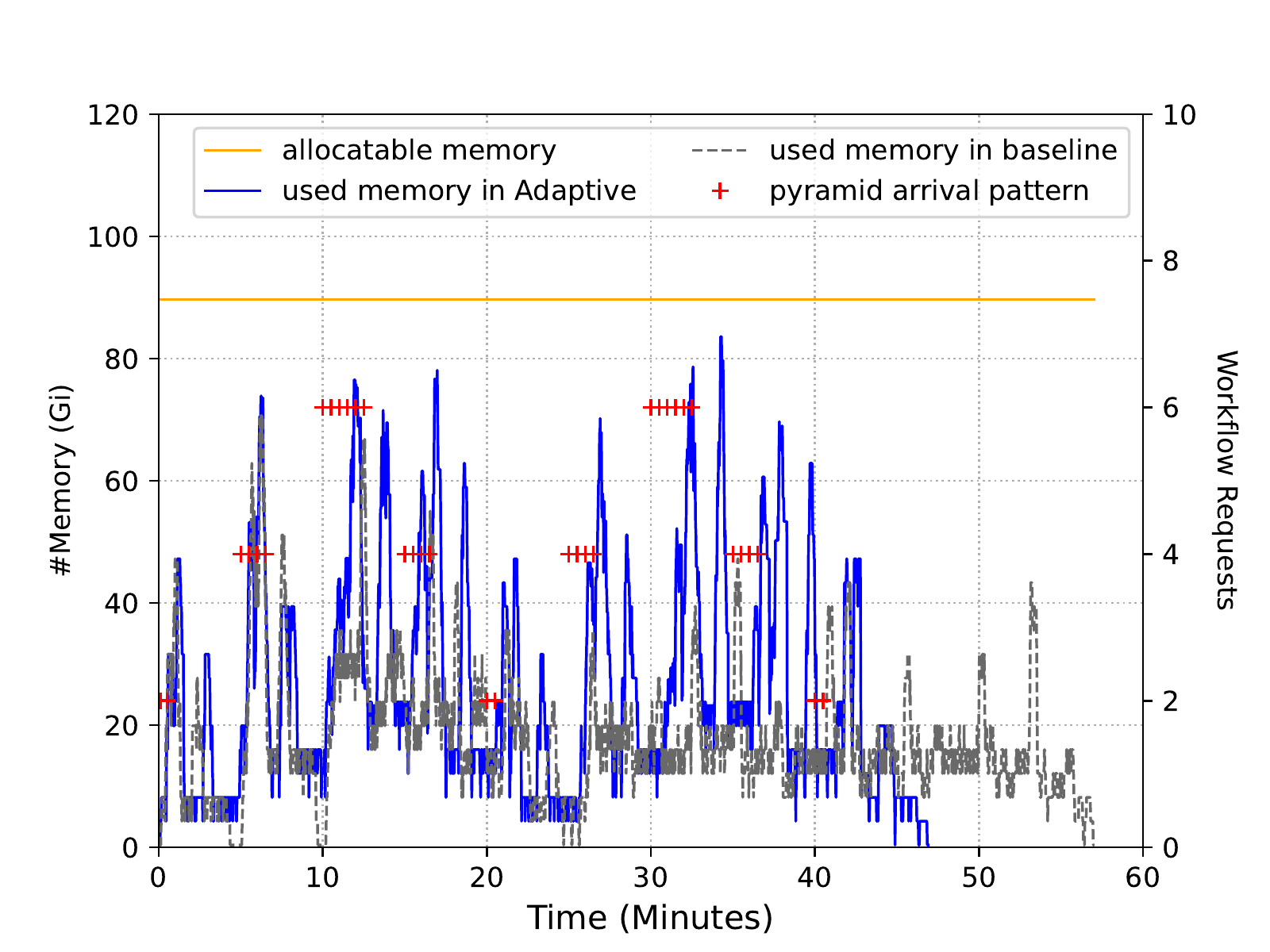}\\
			\vspace{0.02cm}
		\end{minipage}%
	}%
	\centering
	\caption{The CPU and memory resource usage rate under three distinct arrival patterns for Montage workflows.}
  \vspace{-0.2cm}
	\label{fig:montage}
\end{figure*}

\begin{figure*}
	\centering
  \vspace{-0.5cm}
	\subfigure[Constant Arrival Pattern]{
		\begin{minipage}[t]{0.33\linewidth}
			\centering
			\includegraphics[width=2.4in]{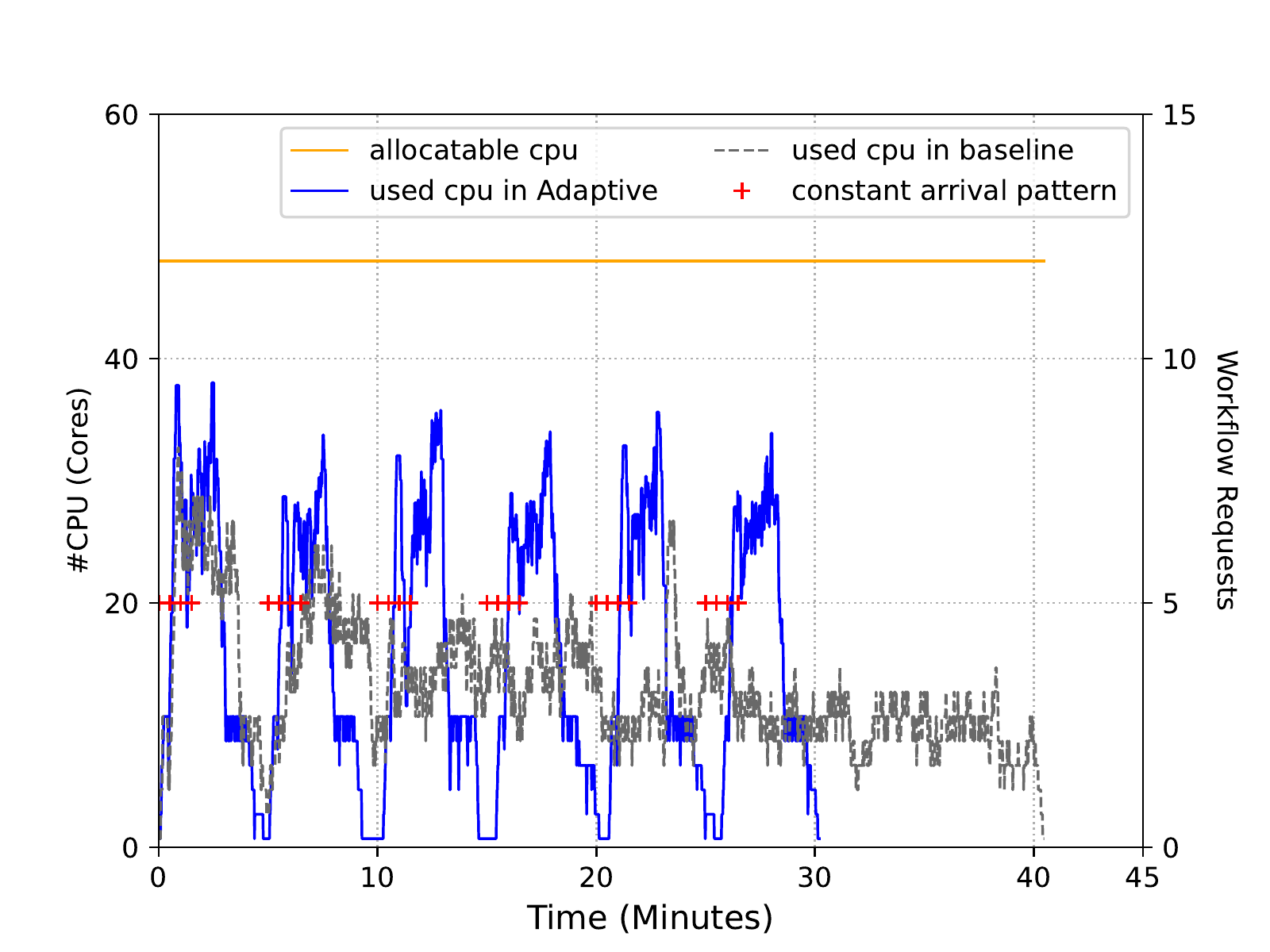}\\
			\vspace{0.02cm}
			\includegraphics[width=2.4in]{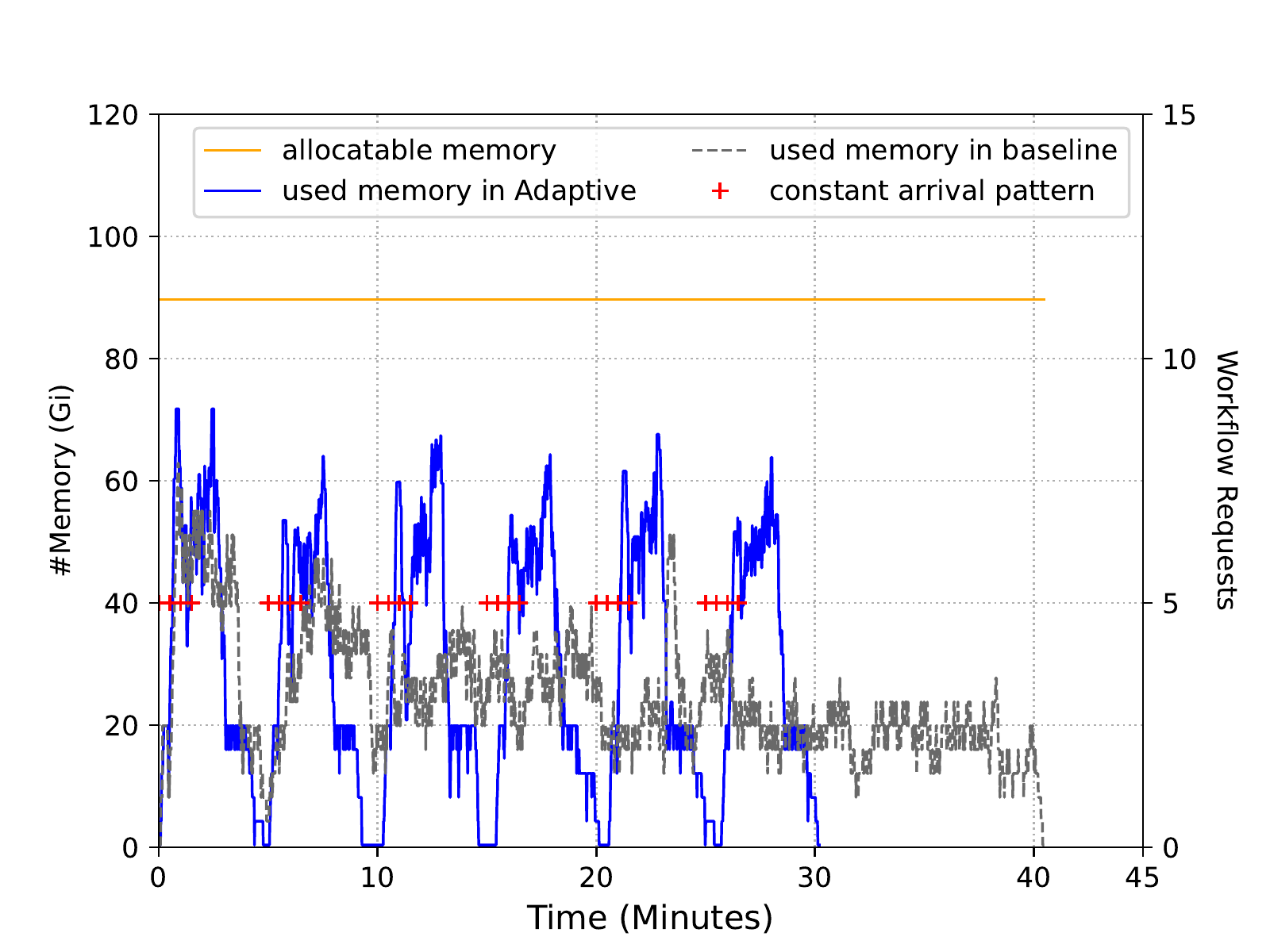}\\
			\vspace{0.02cm}
		\end{minipage}%
	}%
	\subfigure[Linear Arrival Pattern]{
		\begin{minipage}[t]{0.33\linewidth}
			\centering
			\includegraphics[width=2.4in]{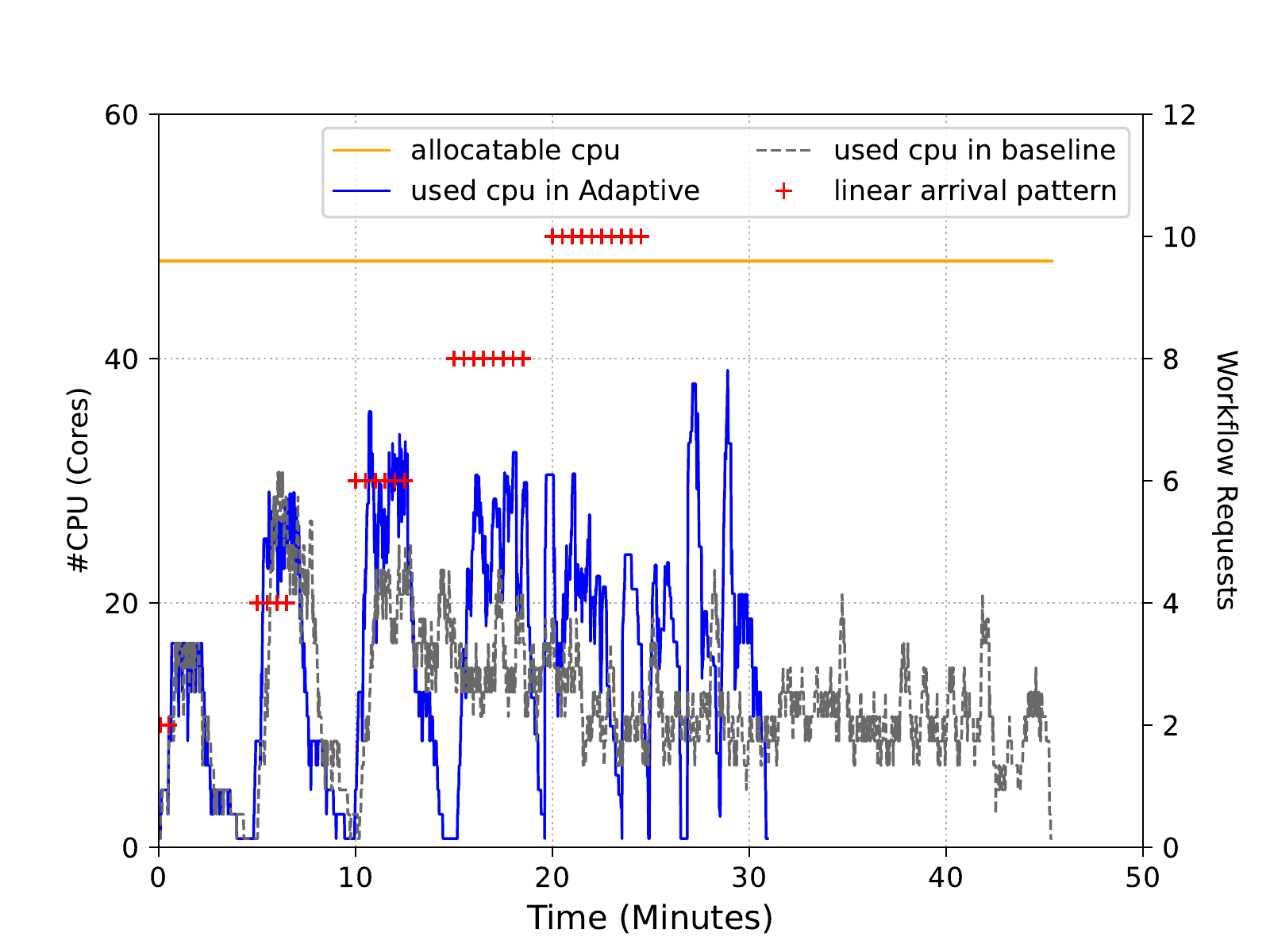}\\
			\vspace{0.02cm}
			\includegraphics[width=2.4in]{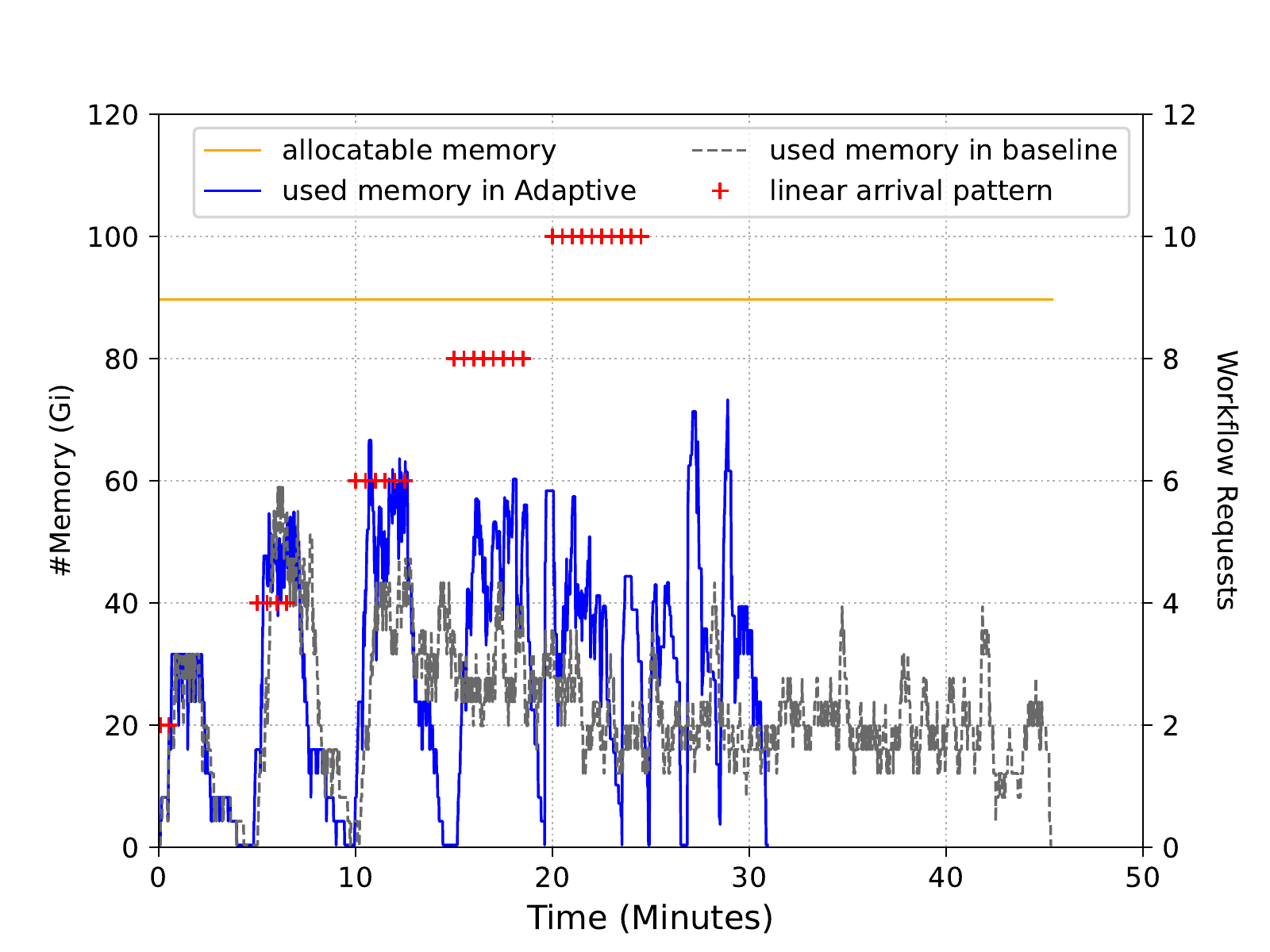}\\
			\vspace{0.02cm}
		\end{minipage}%
	}%
	\subfigure[Pyramid Arrival Pattern]{
		\begin{minipage}[t]{0.33\linewidth}
			\centering
			\includegraphics[width=2.4in]{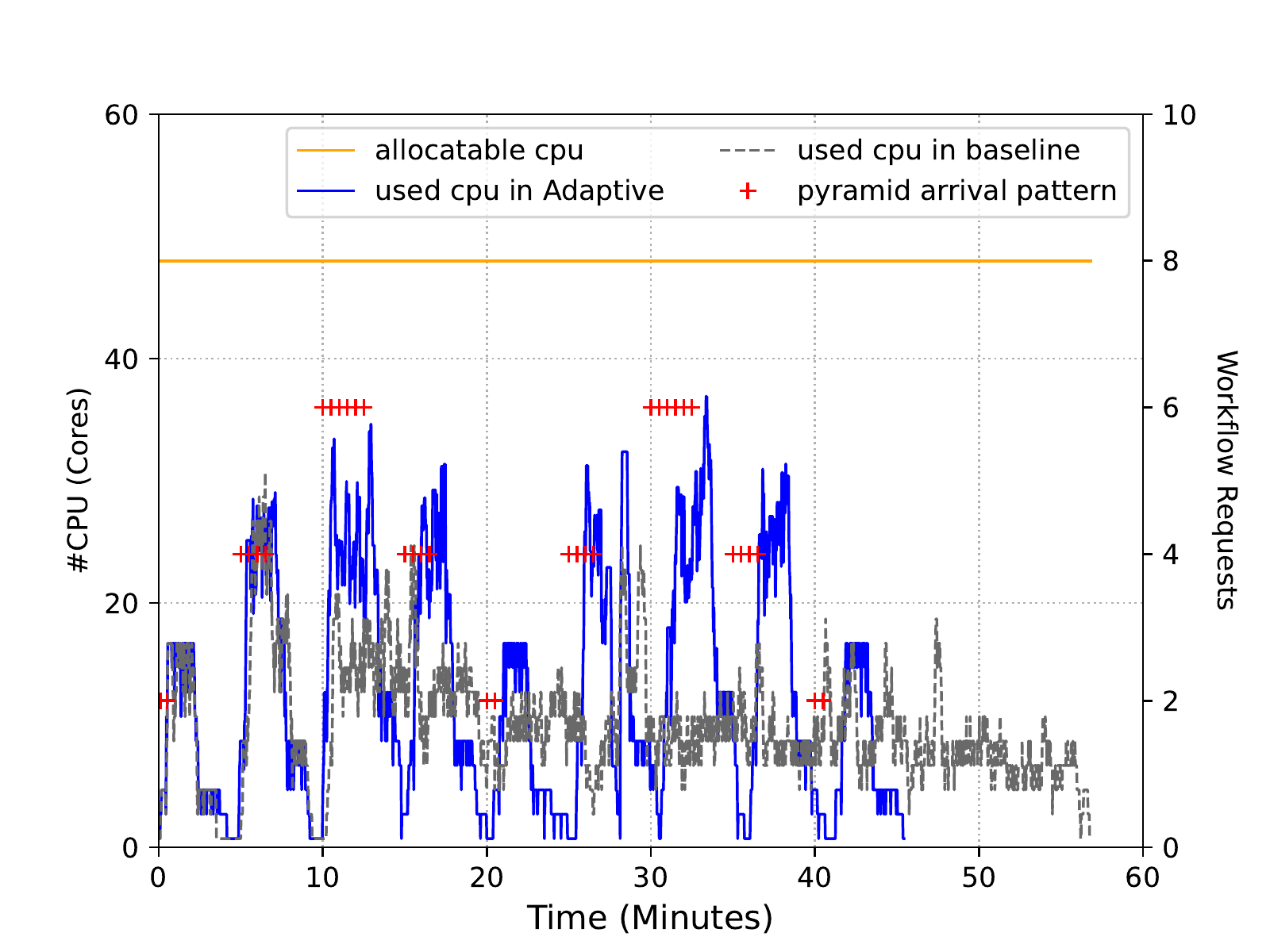}\\
			\vspace{0.02cm}
			\includegraphics[width=2.4in]{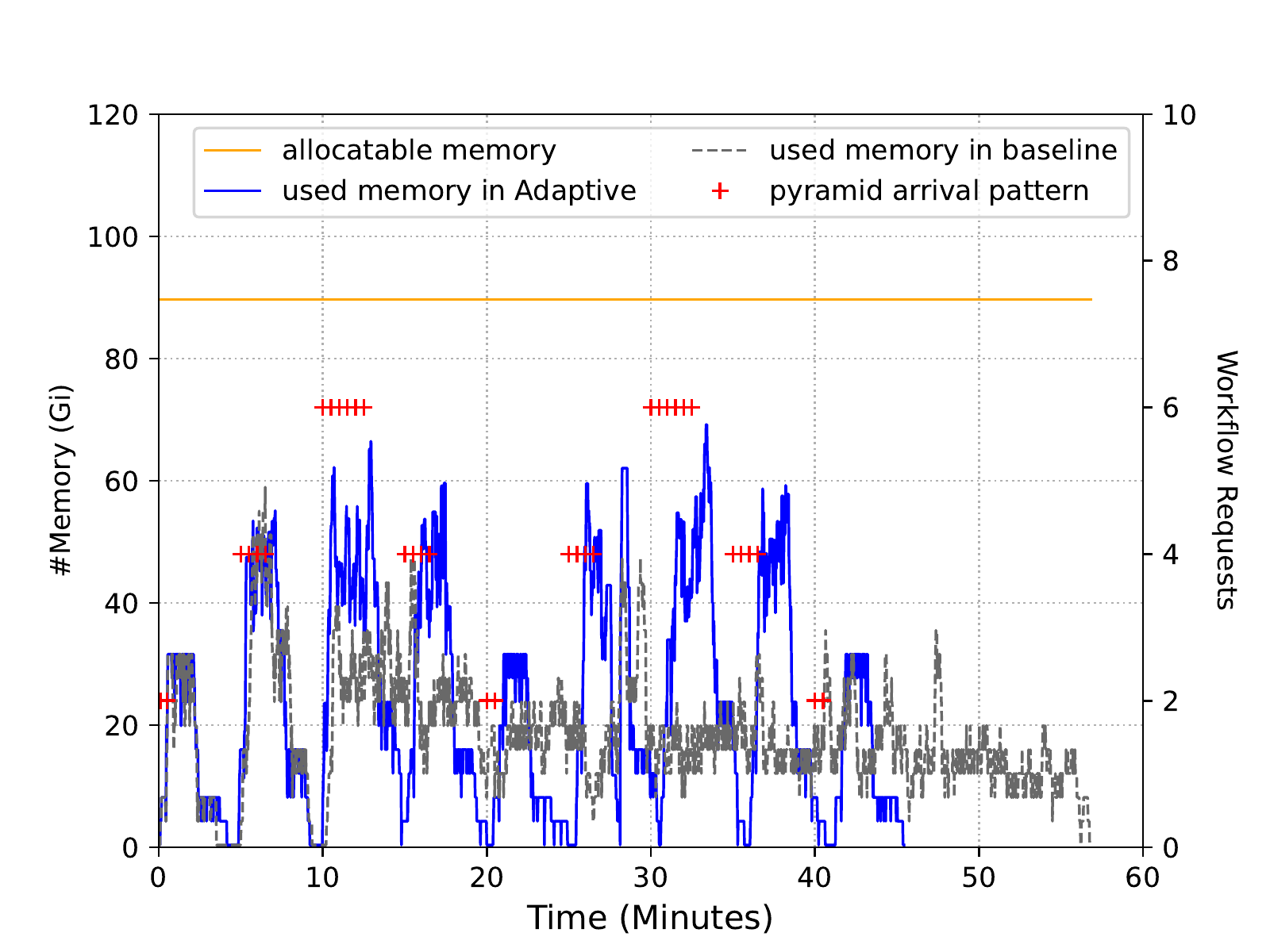}\\
			\vspace{0.02cm}
		\end{minipage}%
	}%
	\centering
	\caption{The CPU and memory resource usage rate under three distinct arrival patterns for Epigenomics workflows.}
	\vspace{-0.5cm}
	\label{fig:epigenomics}
\end{figure*}

\subsection{Results and analysis}
To fully evaluate KubeAdaptor together with our ARAS, we present a general evaluation and the evaluation of 
resource allocation failure, and discuss the evaluation results. 
In order to minimize external influences, our K8s cluster has no other application load, 
and we execute each evaluation three times at different times of one day.

\subsubsection{General evaluation}
In the following, we use the KubeAdaptor with our ARAS and the baseline to run four 
scientific workflows against three distinct workflow arrival patterns three times and compare our ARAS with 
the baseline on experimental results.
We calculate the mean value and the standard deviation $\delta$ for all metrics. 

Table~\ref{table:evaluation} presents the resulting mean values and the standard deviation from the conducted evaluation runs. 
In general, the observed standard deviation is low and therefore indicates a low dispersion in the results of the different evaluations. 
As presented in Table~\ref{table:evaluation}, "Adaptive" denotes our ARAS, while "Baseline" marks the 
application of the baseline algorithm~(mentioned in \ref{sec:baseline}).
The interval between two workflow request bursts is set to $300$ seconds for three arrival patterns, 
and the amount of injected workflows for three arrival patterns is set to $30$, $30$, and $34$, respectively. 

Generally, our ARAS is superior to the baseline algorithm on each observation metric against four different 
workflow types under distinct workflow arrival patterns. 
In addition, The CPU and memory resources set in the task pod are constant, the allocatable cluster resources are fixed, 
and the resource scaling method scales down resources according to Eq.~(\ref{eq:rate}). So no matter in each arrival pattern and each resource 
allocation algorithm, the utilization rate of CPU and memory resources are the same, 
and both resource usage curves in each workflow arrival pattern are similar.  
In the following, we elaborate on the evaluation metrics of each workflow arrival pattern in the light of workflow types.

\begin{figure*}
	\centering
   \vspace{-0.5cm}
	\label{fig:}
	\subfigure[Constant Arrival Pattern]{
		\begin{minipage}[t]{0.33\linewidth}
			\centering
			\includegraphics[width=2.4in]{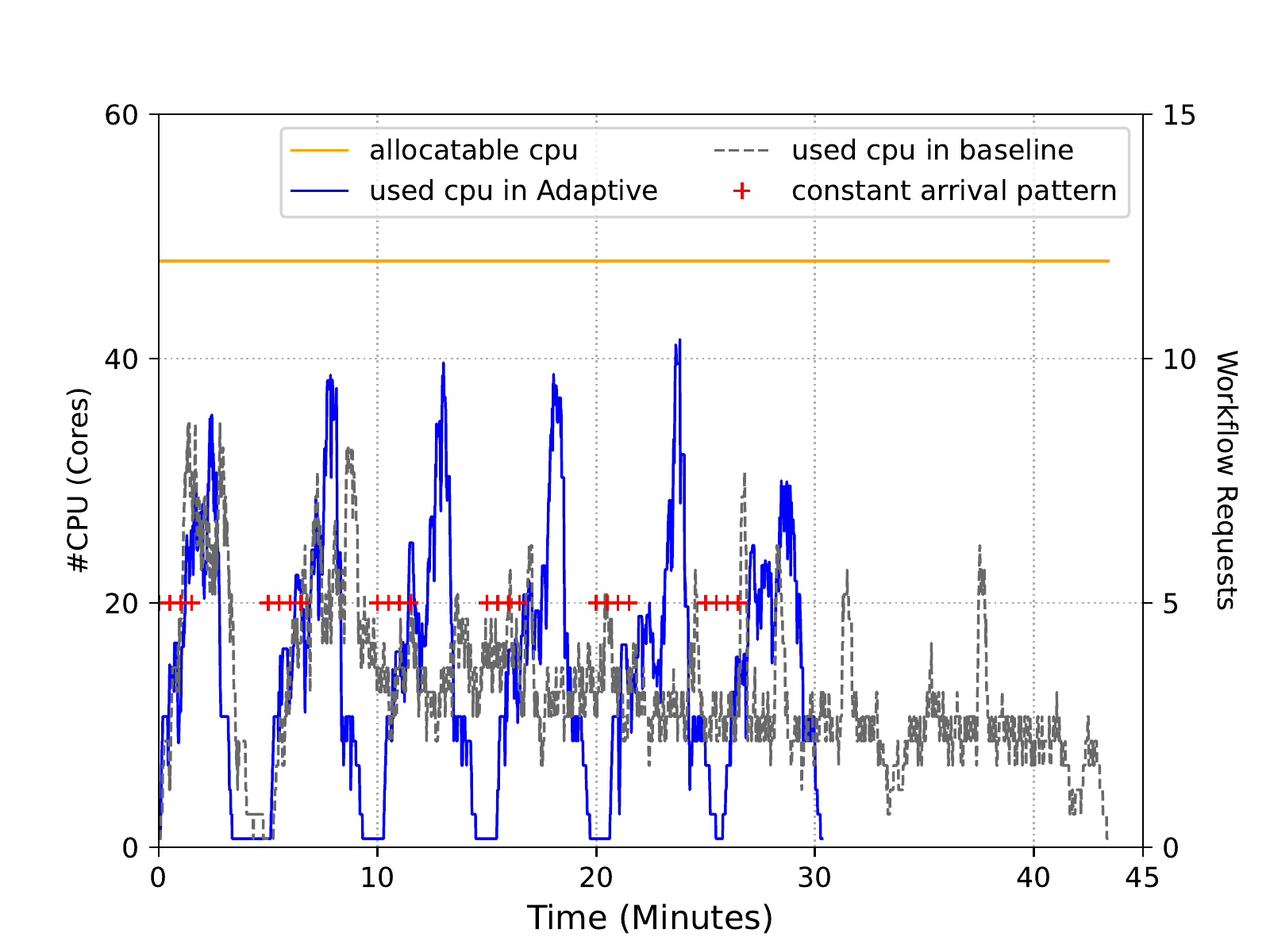}\\
			\vspace{0.02cm}
			\includegraphics[width=2.4in]{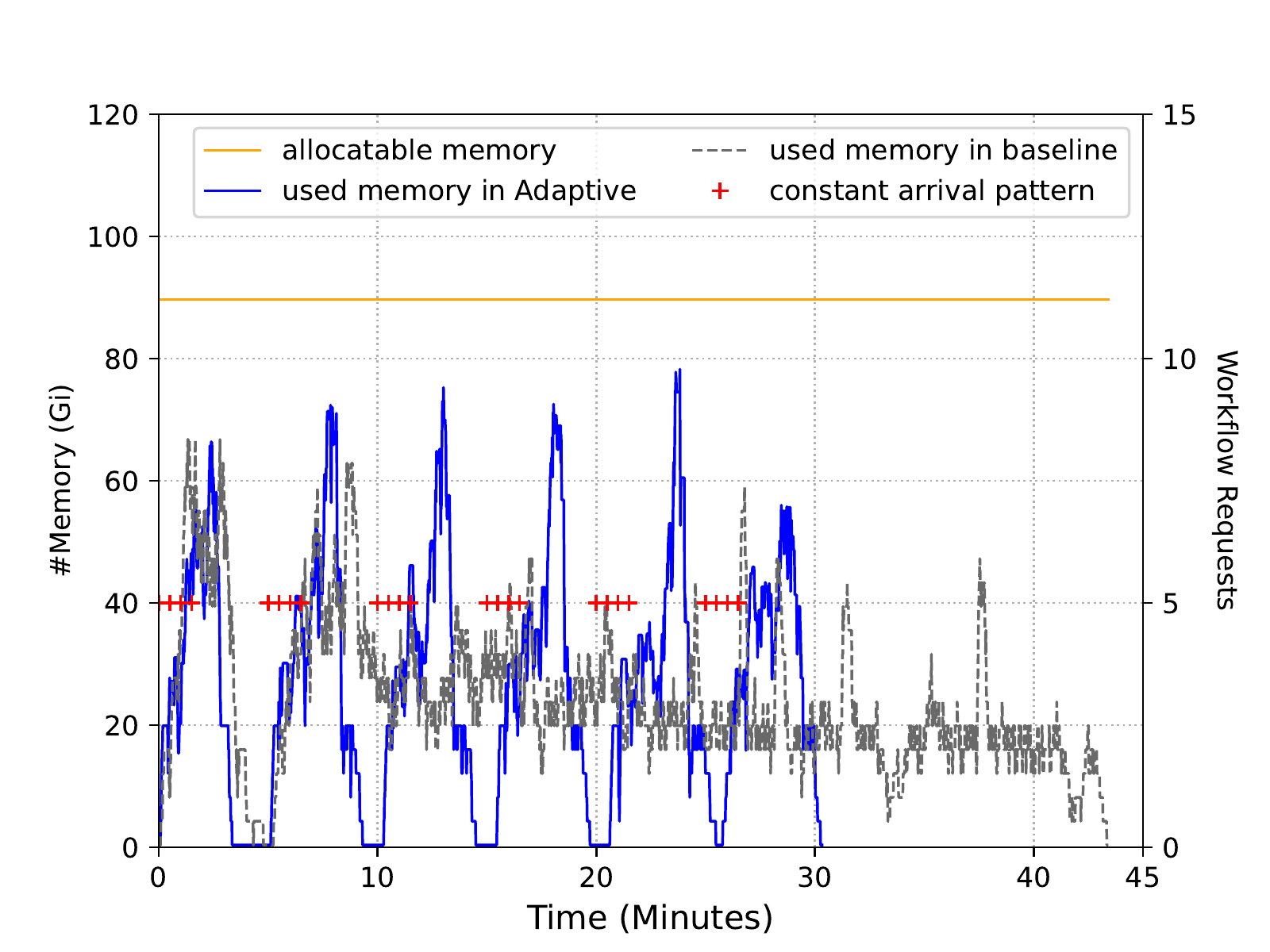}\\
			\vspace{0.02cm}
		\end{minipage}%
	}%
	\subfigure[Linear Arrival Pattern]{
		\begin{minipage}[t]{0.33\linewidth}
			\centering
			\includegraphics[width=2.4in]{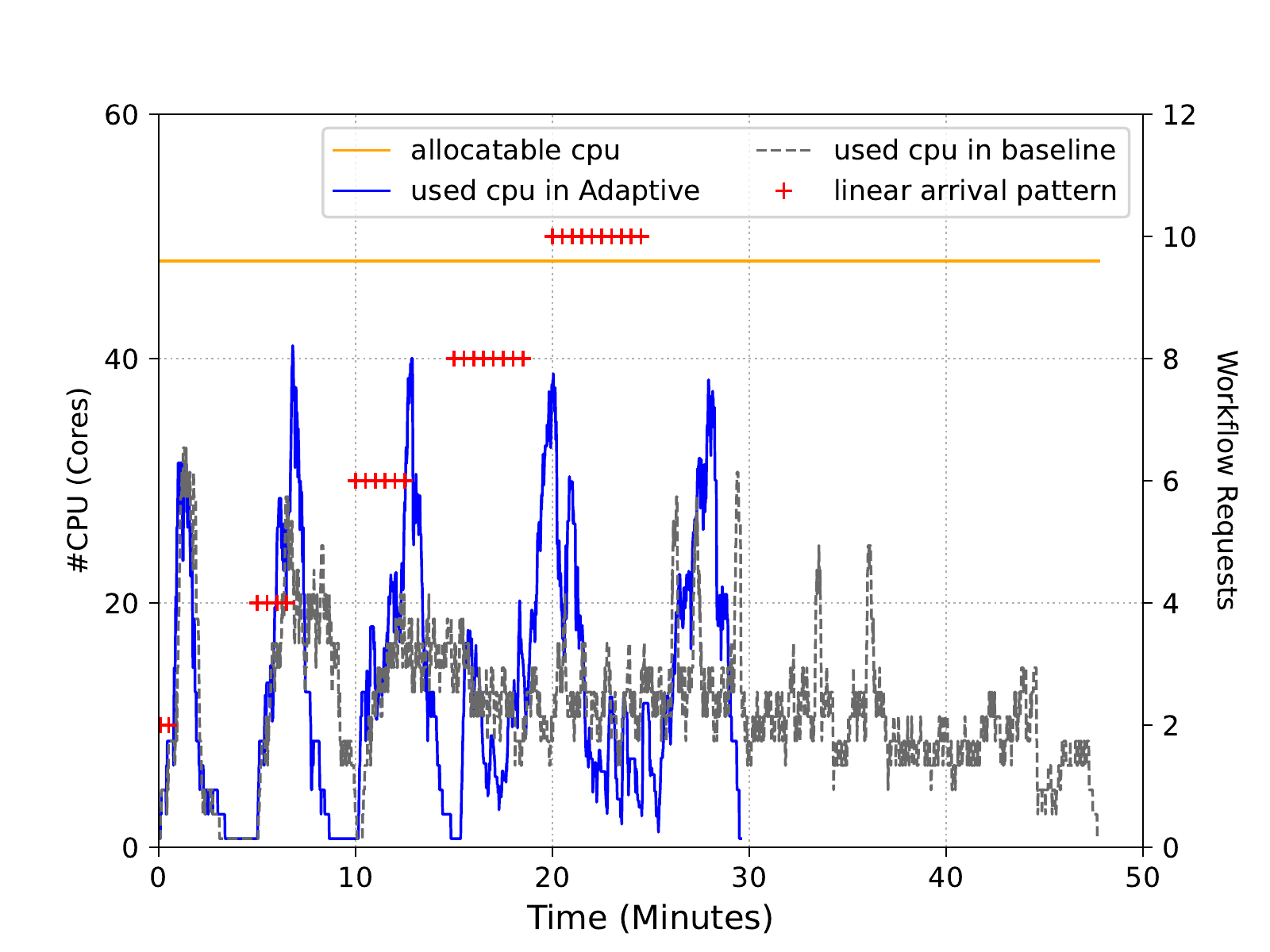}\\
			\vspace{0.02cm}
			\includegraphics[width=2.4in]{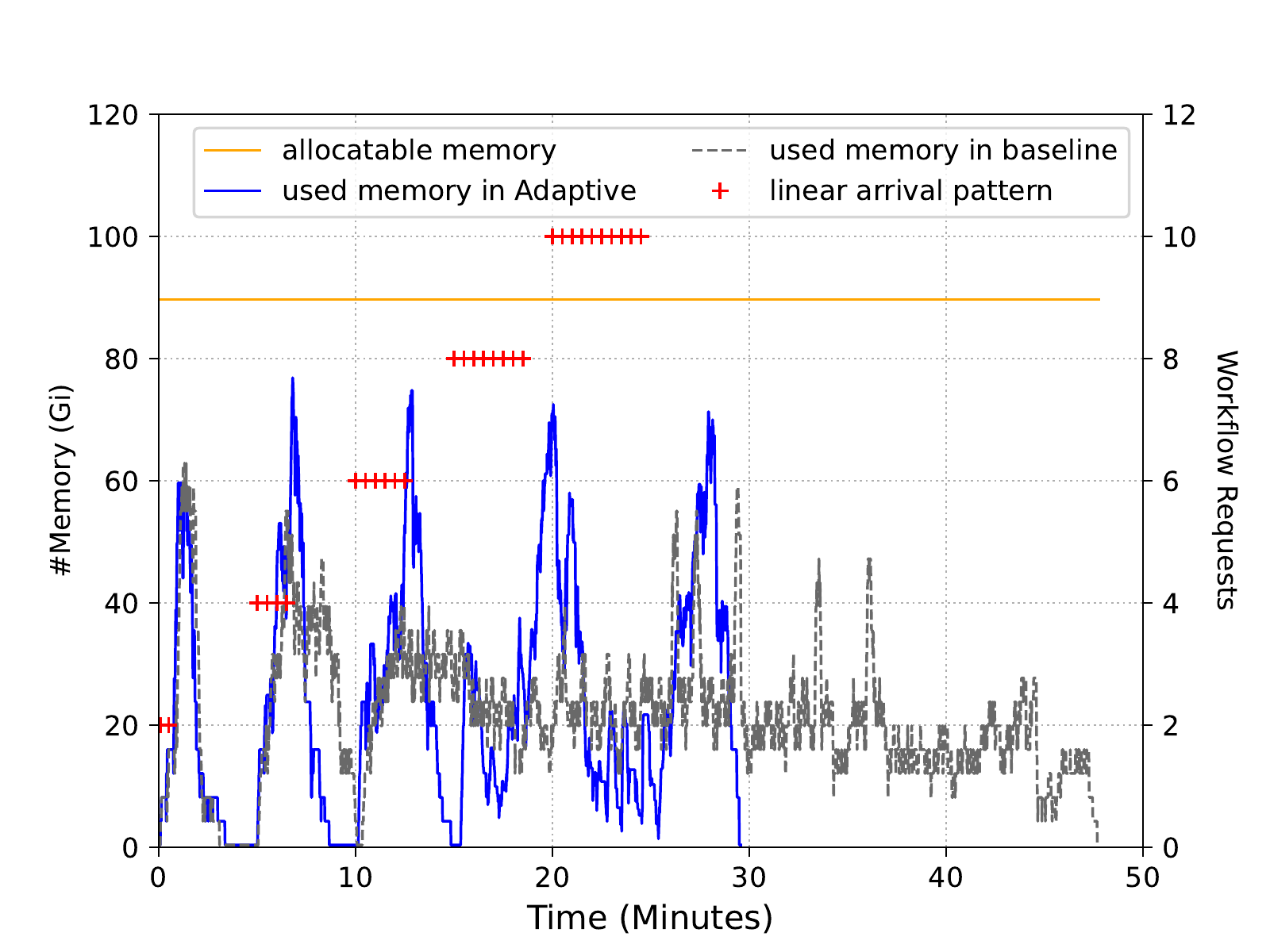}\\
			\vspace{0.02cm}
		\end{minipage}%
	}%
	\subfigure[Pyramid Arrival Pattern]{
		\begin{minipage}[t]{0.33\linewidth}
			\centering
			\includegraphics[width=2.4in]{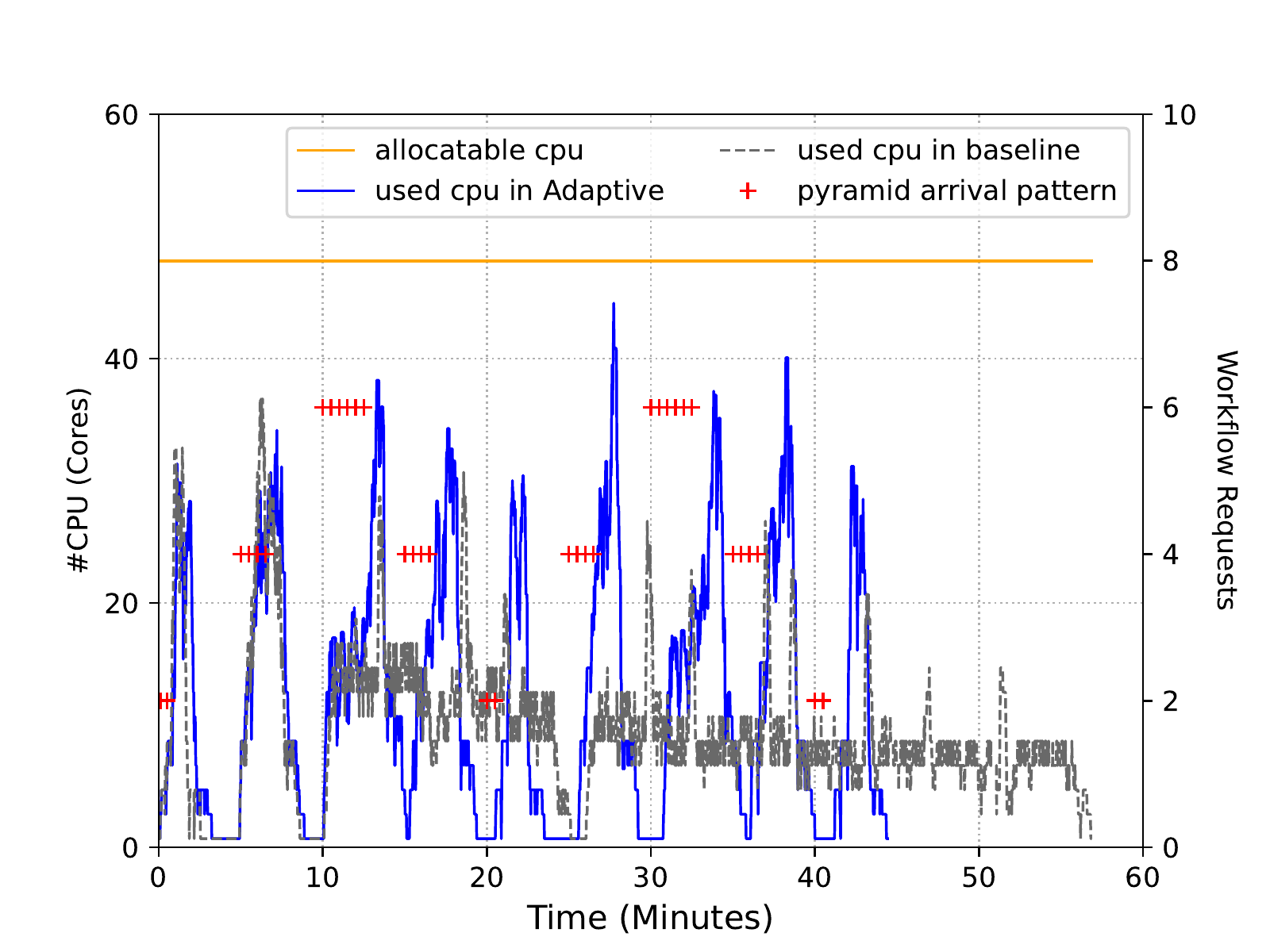}\\
			\vspace{0.02cm}
			\includegraphics[width=2.4in]{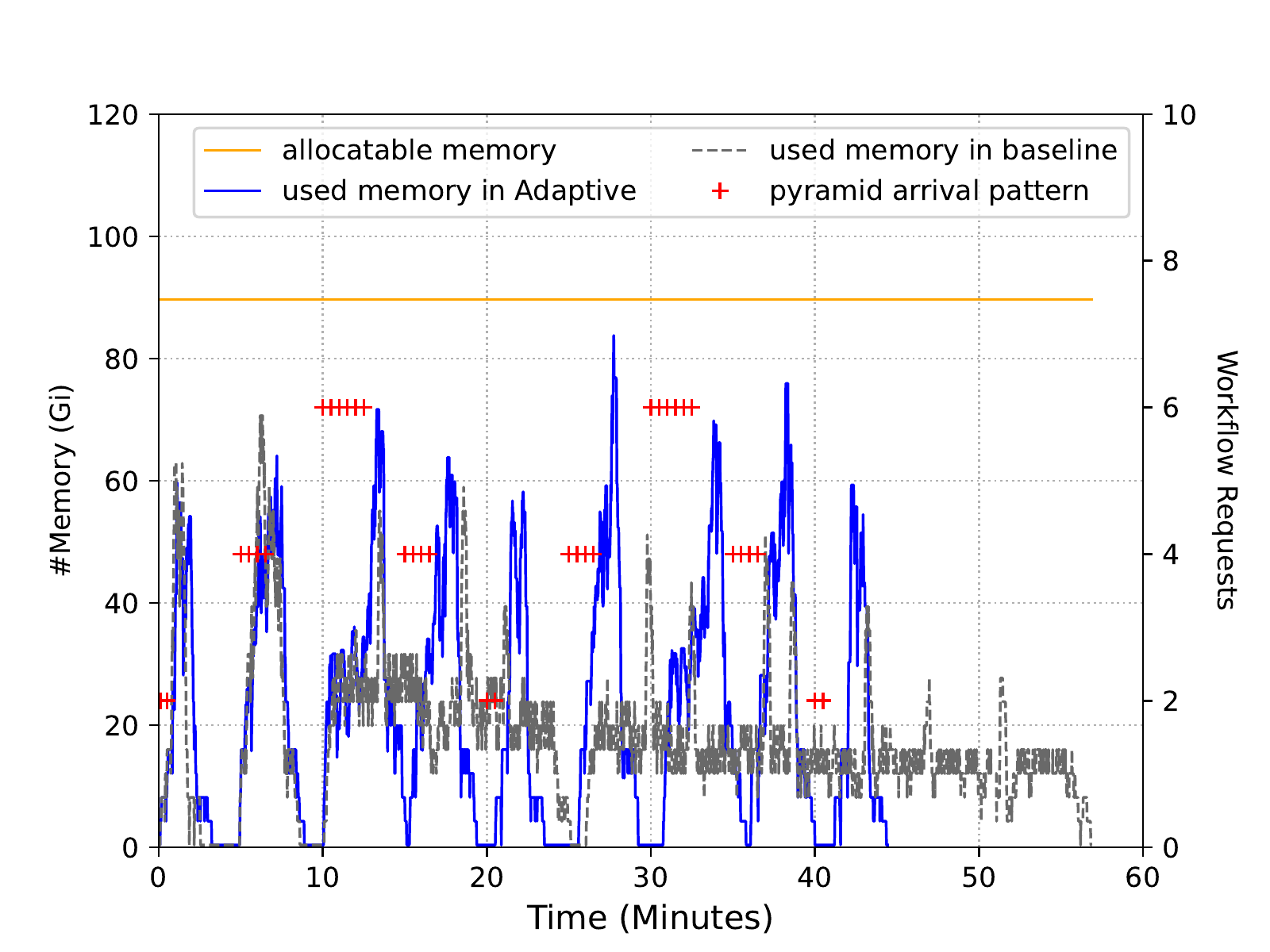}\\
			\vspace{0.02cm}
		\end{minipage}%
	}%
	\centering
	\caption{The CPU and memory resource usage rate under three distinct arrival patterns for CyberShake workflows.}
	\vspace{-0.3cm}
	\label{fig:cybershake}
\end{figure*}

\begin{figure*}
	\centering
     \vspace{-0.5cm}
	\subfigure[Constant Arrival Pattern]{
		\begin{minipage}[t]{0.33\linewidth}
			\centering
			\includegraphics[width=2.4in]{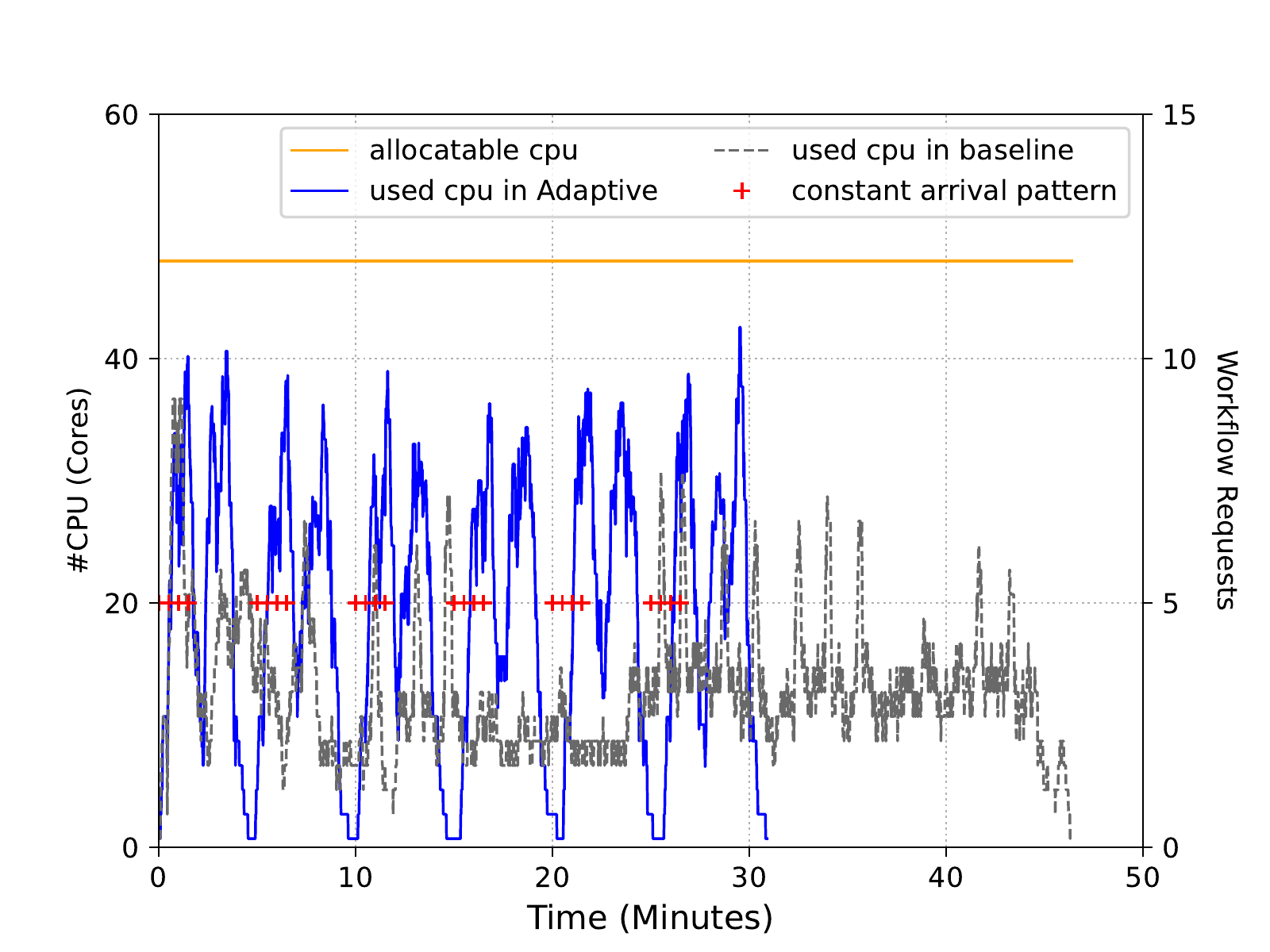}\\
			\vspace{0.02cm}
			\includegraphics[width=2.4in]{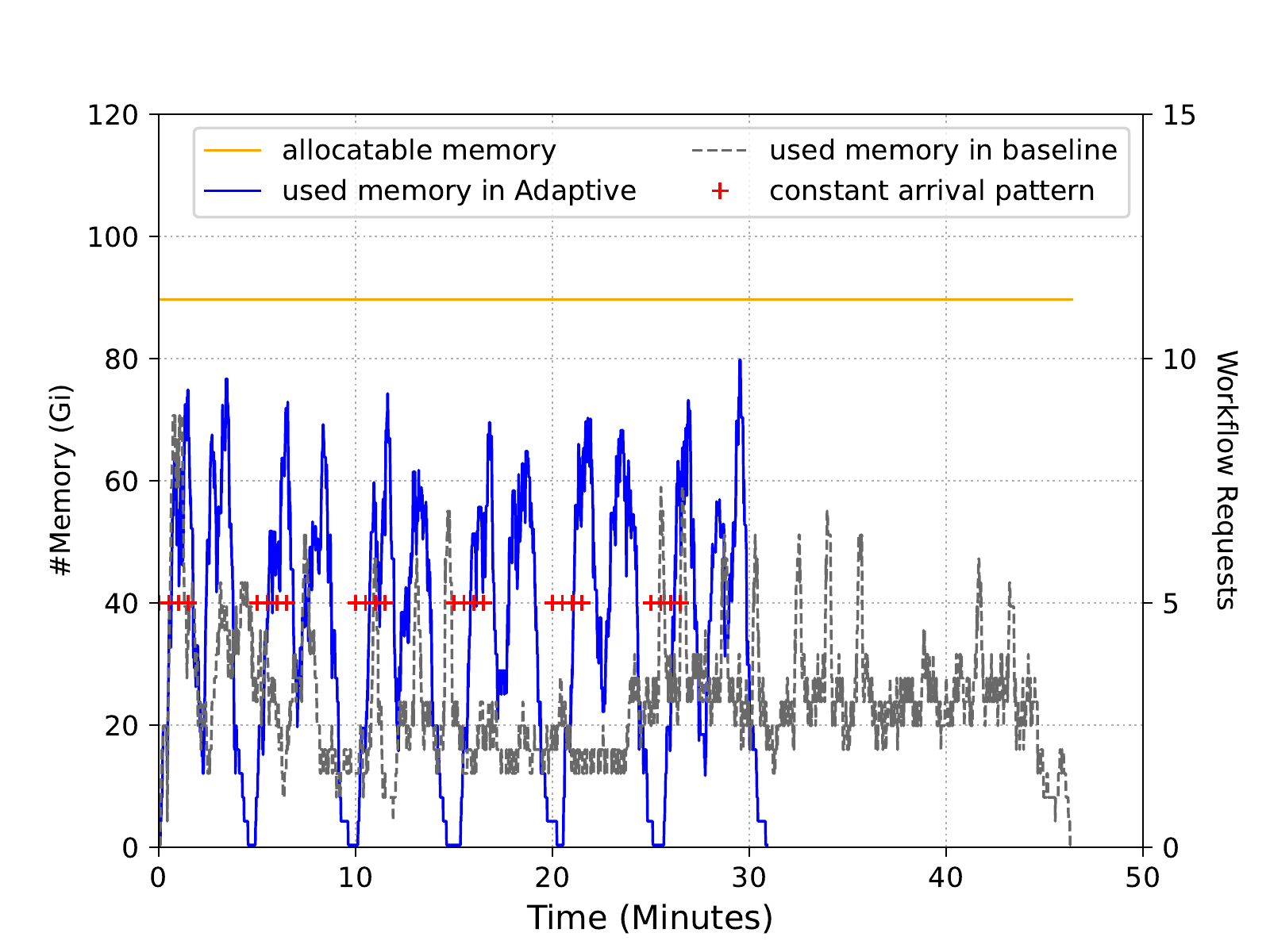}\\
			\vspace{0.02cm}
		\end{minipage}%
	}%
	\subfigure[Linear Arrival Pattern]{
		\begin{minipage}[t]{0.33\linewidth}
			\centering
			\includegraphics[width=2.4in]{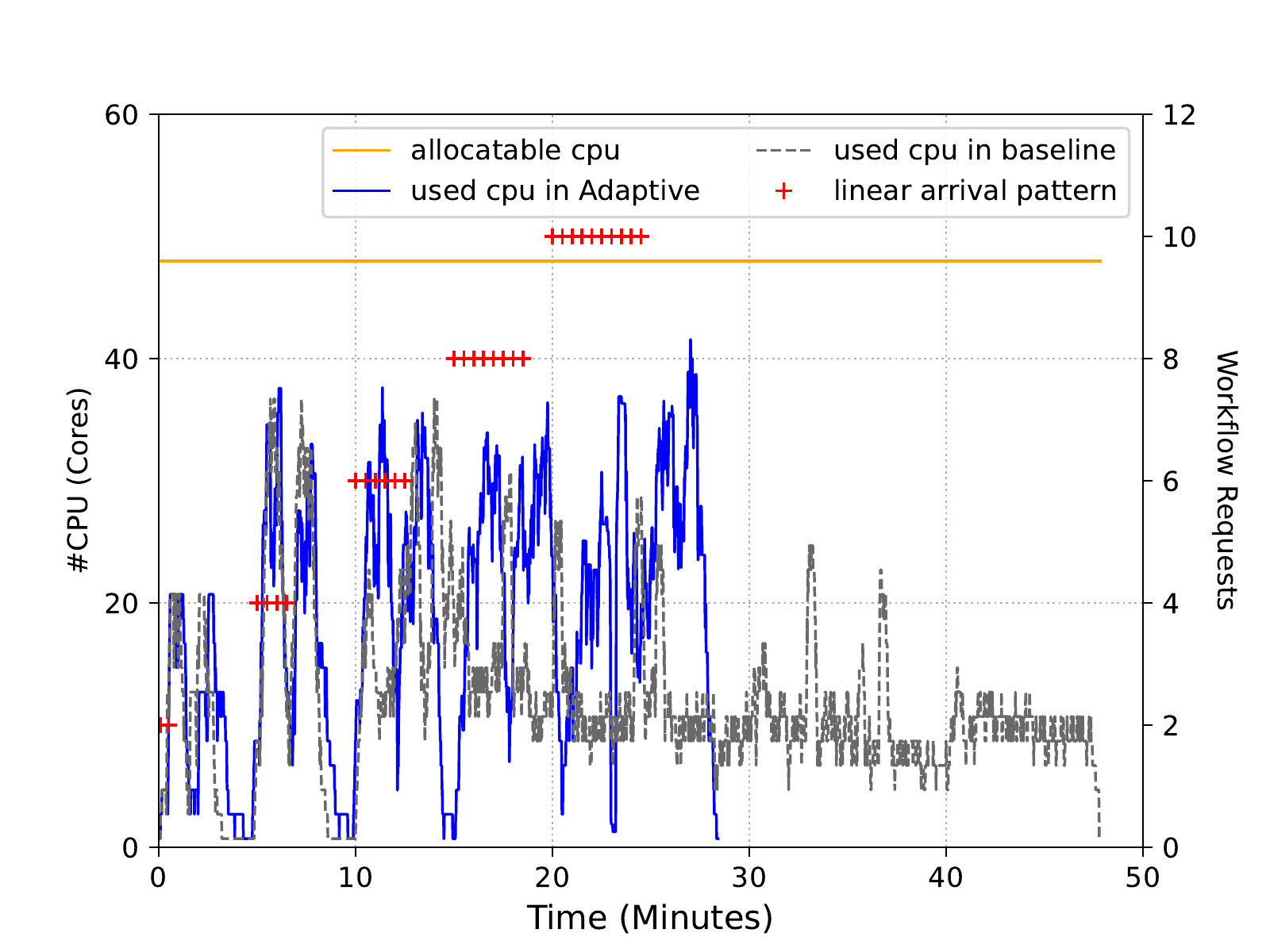}\\
			\vspace{0.02cm}
			\includegraphics[width=2.4in]{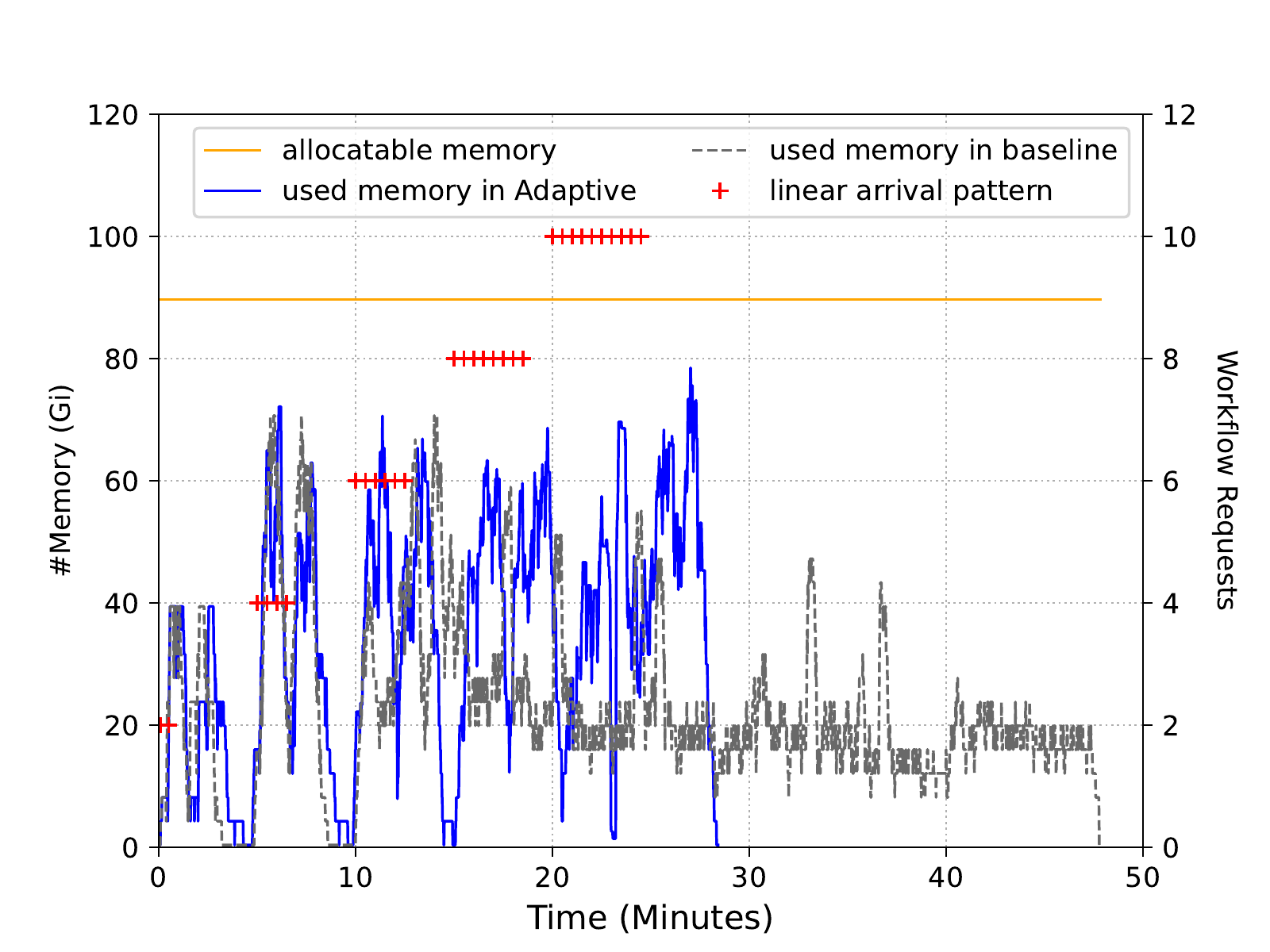}\\
			\vspace{0.02cm}
		\end{minipage}%
	}%
	\subfigure[Pyramid Arrival Pattern]{
		\begin{minipage}[t]{0.33\linewidth}
			\centering
			\includegraphics[width=2.4in]{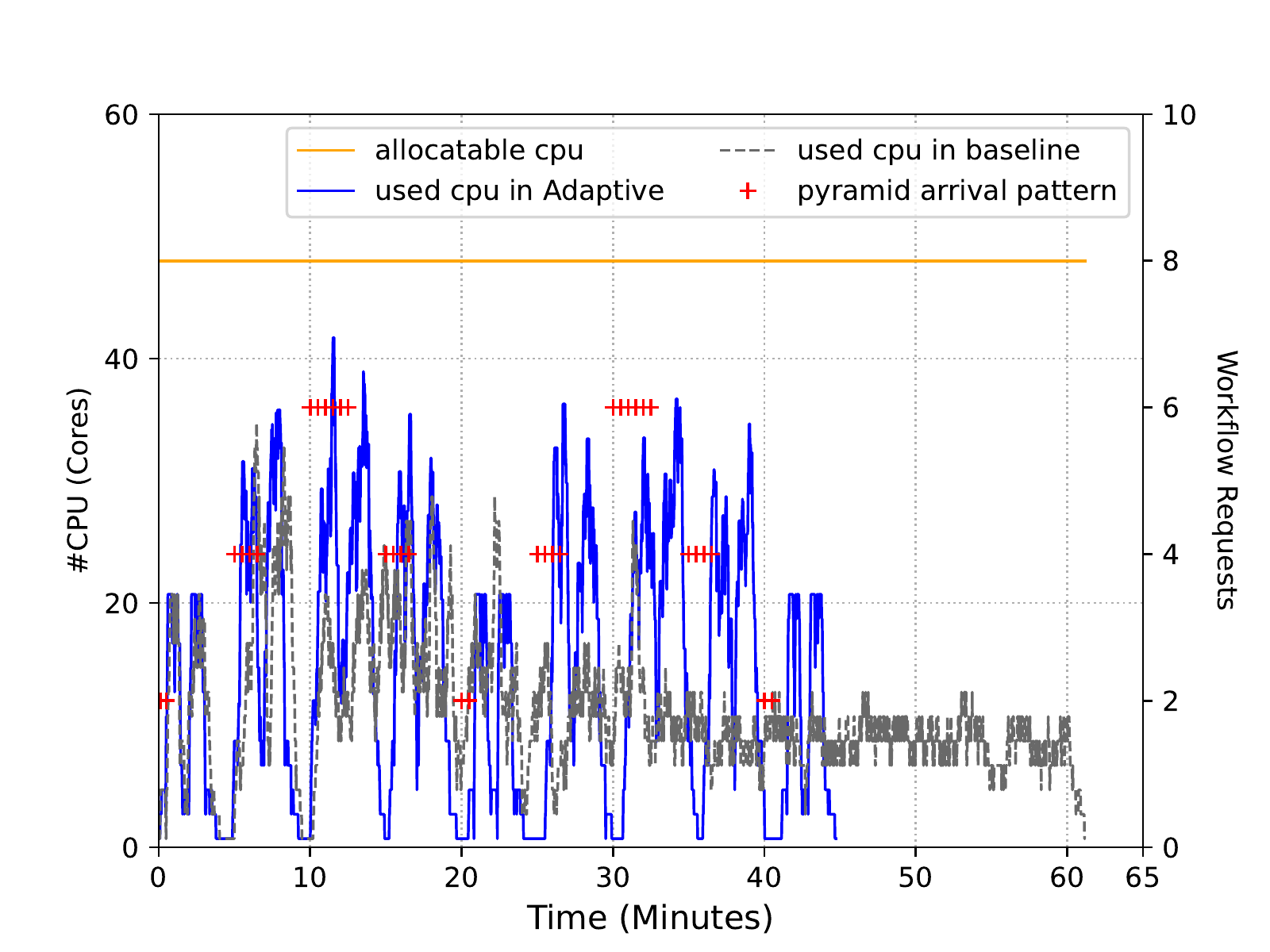}\\
			\vspace{0.02cm}
			\includegraphics[width=2.4in]{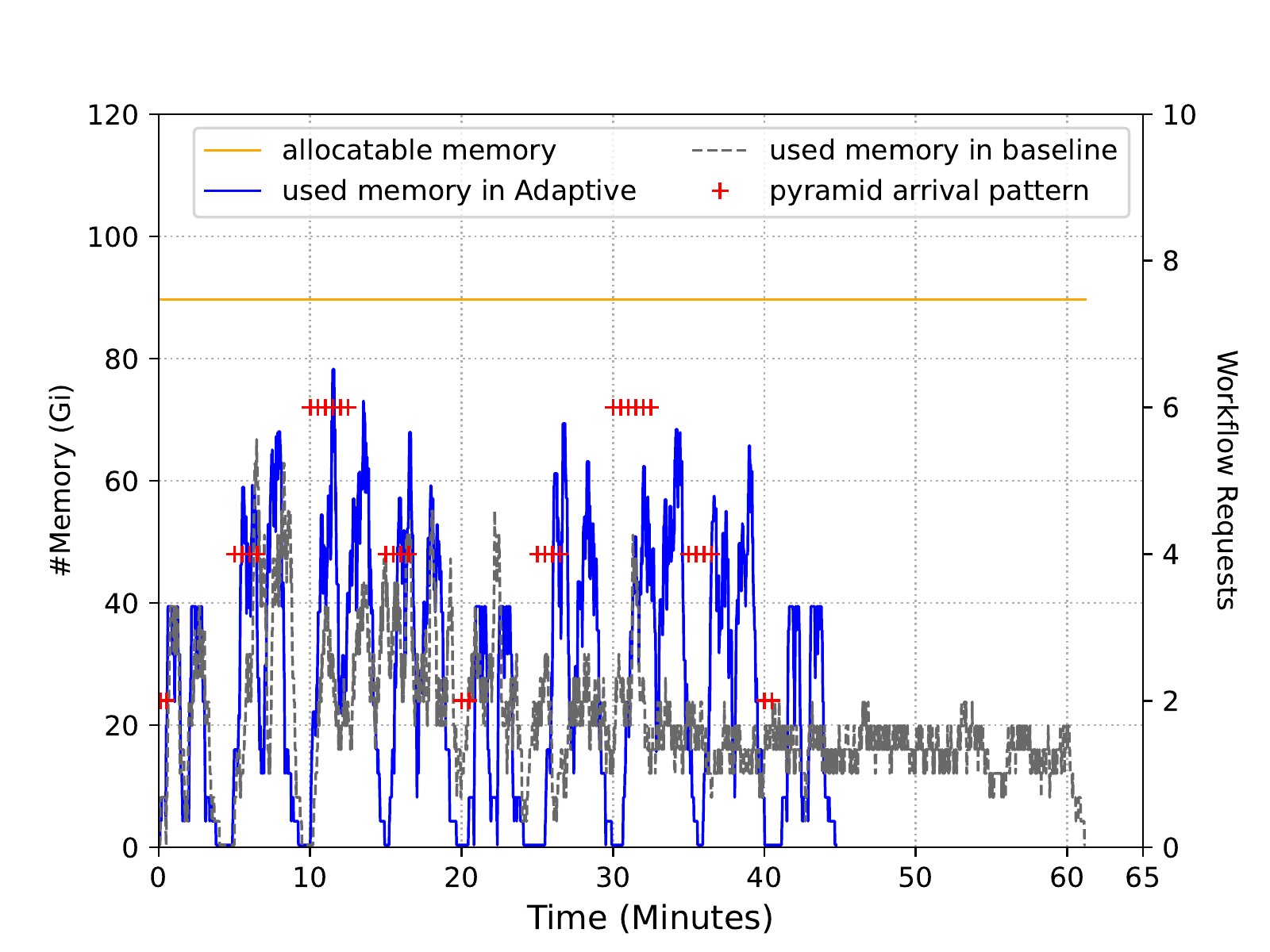}\\
			\vspace{0.02cm}
		\end{minipage}%
	}%
	\centering
	\caption{The CPU and memory resource usage rate under three distinct arrival patterns for LIGO workflows.}
	\vspace{-0.5cm}
	\label{fig:ligo}
\end{figure*}

From Fig.~\ref{fig:montage} to Fig.~\ref{fig:ligo} illustrate the presentation of the average evaluation results 
by depicting the arrival patterns~(Workflow Requests) over time and the number of used computational resources~(CPU and memory) 
until all workflow requests have been served. 
Note that the used resource curve in each workflow arrival pattern usually ends later than the workflow request curve. 
It can be traced to the fact that each workflow has a deadline in the future, and some workflows are still waiting in queue for execution.

\textbf{Montage}:
 In our experimental setup, a small-scale Montage workflow consists of $21$ tasks~(refers to Fig.~\ref{fig_four}(a)). 
Compared with the baseline, as for the total duration of all workflows in Table~\ref{table:evaluation}, our ARAS 
leads to time savings of $9.8\%$ for the constant arrival pattern and time savings of $26.06\%$ for the linear arrival pattern, 
while in the pyramid arrival pattern, time savings amounts to $9.8\%$. 
Similarly, as for average workflow duration, our ARAS, in comparison to the baseline, 
gains a time saving of $26.4\%$, time savings of $52.3\%$, and time savings of $38.5\%$ for three different workflow arrival patterns 
from left to right, respectively. 
Fig.~\ref{fig:montage} broadly reflects the consistency of the above data with the total duration of all injected workflows, 
even with a set of evaluation data.
Our ARAS achieves better performances in total workflow duration and average workflow duration in linear 
arrival patterns. 
It can be deduced that the concurrent degree of received workflow requests is directly related to the total duration of all injected workflows 
and the average duration of a single workflow. The higher the concurrent degree of received workflow requests, the more workflow tasks are 
executed per unit time, so the shorter the total duration of all injected workflows and the average duration of individual workflow.

Regarding the CPU and memory resource usage in Fig.~\ref{fig:montage}, our ARAS outperforms the baseline for 
each arrival pattern. 
Herein, the linear arrival pattern features a maximum value of $35\%$ for our ARAS, $4\%$ higher than the 
baseline algorithm. 
Our ARAS outperforms the baseline algorithm by $1\%$ and $6\%$, respectively, in the other two patterns.
It can be traced back to the fact that over time the linear arrival pattern requests more task pods to be performed in parallel in response 
to more and more workflow requests and gains a maximum resource usage rate. 

Looking at the resource usage curves (CPU and memory) of three workflow arrival patterns in Fig.~\ref{fig:montage}, the resource usage 
peak of our ARAS is higher than that of the baseline algorithm for most of the time.
It can be further observed that the peak of the resource usage curve is consistent with the centralized arrival of workflow requests.
It is because our ARAS can use a resource scaling strategy to adjust the resource limits of potential future 
task requests within the current task’s lifecycle. 
This scheme launches task pods as many as possible on the premise of the smooth operation of task pods, thus speeding up the execution efficiency 
of workflows. 
However, the baseline algorithm depends on the adequacy of residual resources on cluster nodes. 
In high concurrency scenarios, the insufficient remaining resources of nodes will make the baseline algorithm lead to endless waiting and 
much time-wasting and prolong the total duration of workflows and the average duration of a single workflow.

\textbf{Epigenomics}: We adopt a small-scale Epigenomics workflow with $20$ tasks in experimental evaluations. 
As can be seen from Fig.~\ref{fig_four}(b), the topology of Epigenomics workflows is mostly pipeline structure. 
As for the total duration of all workflows, our ARAS obtains time savings of $21.8\%$ for the constant arrival 
pattern, time savings of $21.4\%$ for the linear arrival pattern, and time savings of $17.2\%$ for the pyramid arrival pattern compared 
with the baseline.
As for average workflow duration, our ARAS, in comparison to the baseline, gains time savings of $54.65\%$, 
time savings of $40.65\%$, and time savings of $50.28\%$ for three arrival patterns from left to right, respectively. 
Note that the Epigenomics workflow is substantially more significant in performance improvement than the Montage workflow in terms of average 
total workflow duration and average duration of individual workflow for three arrival patterns. 
Because the pipeline topology in the Epigenomics workflow is better suited for high concurrency scenarios, our ARAS
scheme saves more time than the baseline in response to continuous workflow requests. 

Regarding the resource usage in Fig.~\ref{fig:epigenomics}, the constant arrival pattern features a maximum value of $34\%$ for our ARAS, 
$7\%$ higher than the baseline. 
The linear arrival pattern features a value of $32\%$ for our ARAS, which is also $7\%$ higher than the baseline.  
The higher resource utilization of these two patterns can be attributed to the fact that $30$ workflows were injected in the first $25$ minutes, 
totaling more than $600$ tasks. 
The higher density of workflow requests results in higher CPU and memory resource utilization.
In addition, the resource scaling method enables our ARAS to adjust the resource limits of the task pods 
in time and cope with the continuous workflow requests on the premise of the normal execution of workflow tasks.
It also shows that the peak time of resource usage in our ARAS is longer than that of the baseline. 
The baseline algorithm waits for resource release in response to resource shortage on cluster nodes, so it consumes too much time and results 
in a longer total workflow duration and average duration of a single workflow.

\textbf{CyberShake}: A small-scale CyberShake workflow in our experiments comprises $22$ tasks~(refers to Fig.~\ref{fig_four}(c)). 
Our ARAS, in comparison to the baseline, leads to time savings of $23.8\%$~(for the constant arrival pattern), 
time savings of $31.1\%$~(for the linear arrival pattern), and time savings of $29.6\%$~(for the pyramid arrival pattern) for the total 
duration of all workflows. 
Similarly, for average workflow duration, our ARAS, in comparison to the baseline, 
gains time savings of $46.85\%$, time savings of $54.34\%$, and time savings of $74.63\%$ for three arrival patterns from left to right, 
respectively. 

Due to the topology structure of the CyberShake workflow with smaller depth and greater width, the CyberShke workflow features a higher degree 
of inherent parallelism, which is easier to take advantage of our ARAS in response to continuous workflow 
request arrivals. 
Compared with the baseline algorithm, the ARAS has prominent performance advantages on metrics of total workflow 
duration and duration of a single workflow. 
As for CPU and memory resource usage, our ARAS obtains $26\%$, $27\%$, and $22\%$ for three distinct arrival 
patterns, respectively, and slightly higher than the baseline. 
Combined with the resource utilization curve in Fig.~\ref{fig:cybershake}, it can be observed that our ARAS 
benefiting from the resource scaling method outperforms the baseline in all performance metrics under three different workflow 
arrival patterns.

\textbf{LIGO}: A small-scale LIGO workflow in our experiments consists of $23$ tasks~(refers to Fig.~\ref{fig_four}(d)). 
Compared with the baseline, our ARAS gains time savings of $40.92\%$~(for the constant arrival pattern), 
time savings of $18.28\%$~(for the linear arrival pattern), and time savings of $28.79\%$~(for the pyramid arrival pattern) for the total 
duration of all workflows. 
Similarly, for average workflow duration, our ARAS, in comparison to the baseline, gains time savings 
of $79.86\%$, time savings of $42.21\%$, and time savings of $70.15\%$ for three arrival patterns from left to right, respectively. 
With unique concurrent topology, LIGO workflows, like Epigenomics and CyberShake workflows, enable our ARAS
to perform better in the total workflow duration and individual workflow duration metrics than the baseline algorithm under three different 
arrival patterns.

As for resource usage in Fig.~\ref{fig:ligo}, our ARAS obtained $40\%$ for the constant arrival pattern, 
$28\%$ for linear arrival pattern and $31\%$ for pyramid arrival pattern, respectively, much higher than the baseline algorithm. 
In combination with the resource utilization curve trend, resource scaling strategy and workflow's unique concurrency topology once again 
help our ARAS outperform the baseline under three different workflow arrival patterns.

\subsubsection{Evaluation of resource allocation failure}
\label{sec:failure}
In this evaluation, we analyze the behavior of the KubeAdaptor in a failure situation of resource allocation. 
This situation means that our ARAS allocates resource quotas less than $min_{mem}+\beta$ through the resource 
scaling method against a high-concurrency scenario. 
So the task pods cannot smoothly execute and turn to \verb|OOMKilled| status due to insufficient memory resources. 
Accordingly, the \verb|OOMKilled| task pods make the workflow running get stuck. 
The source code of evaluation of resource allocation failture is available at \footnote[17]{https://github.com/CloudControlSystems/OOM-Test}.

In the following, we investigate how KuberAdaptor responds to \verb|OOMKilled| task pods, reallocates resources to execute task pods, 
and resumes workflow execution under our ARAS.
For this evaluation, we inject $10$ Montage workflows into our K8s cluster~(mentioned in \ref{sec:senario}) at a time under 
the constant arrival pattern. 
We fine-tune $min_{cpu}$~and~$min_{mem}$ to be less than the amount of memory required by the Stress tool in the task pod. 
Subsequently, our ARAS tries to reduce the allocated resource quota by the resource scaling method 
in response to continuous workflow requests. 
When the allocated resource is less than $min_{mem}+\beta$, \verb|OOMKilled| task pods will appear due to running resource shortage.

\begin{figure}[!t]
\centering
\includegraphics[width=3.5in]{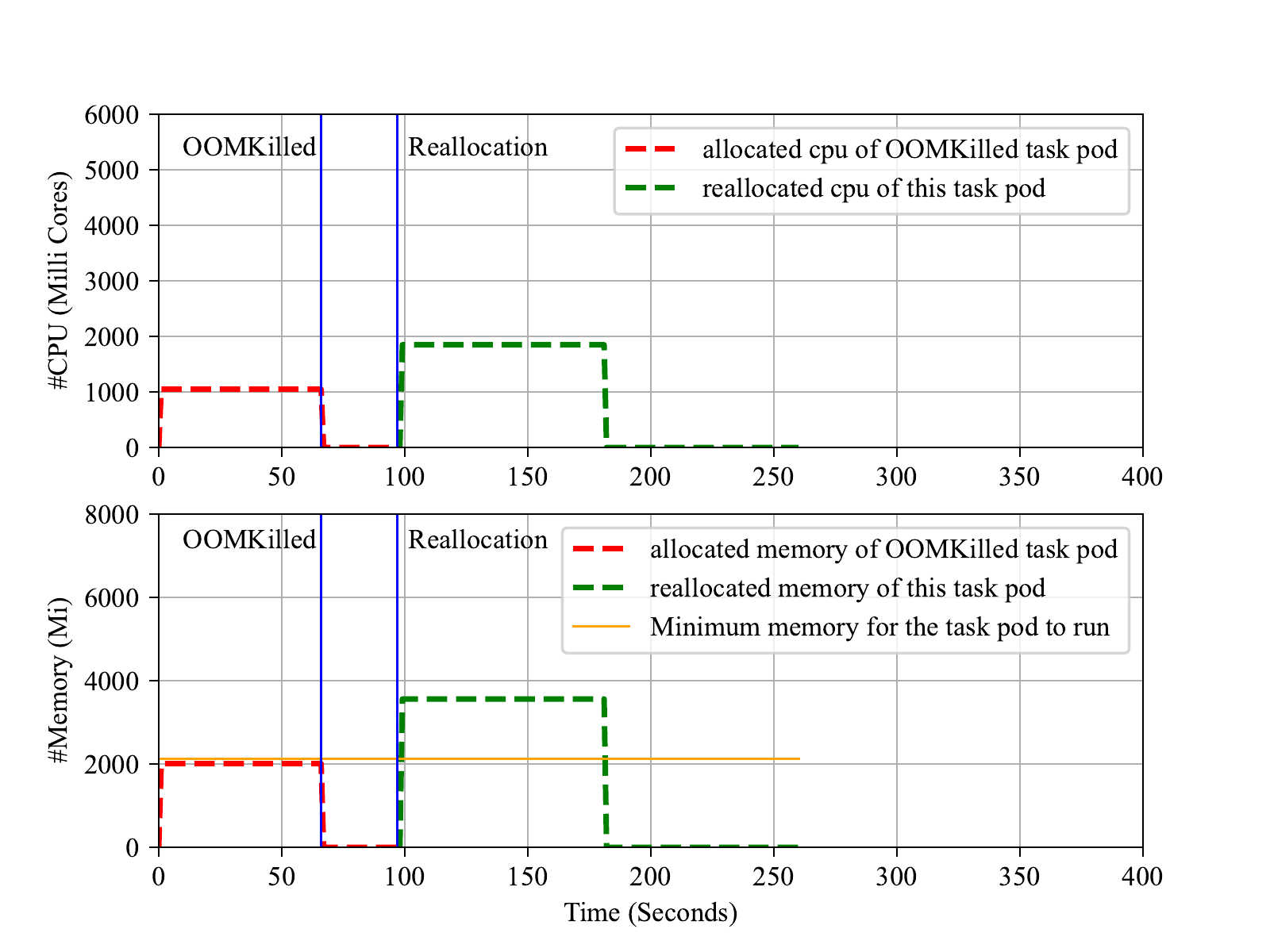}

\caption{Evaluation results of resource allocation failure. 
The first vertical blue line labeled OOMKilled indicates that this task pod encounters OOMKilled due to insufficient allocated memory 
resources. Another vertical blue line labeled Reallocation indicates that the task pod previously OOMKilled is recreated due 
to sufficient allocated memory resources.}
\label{fig_oom}
\vspace{-0.5cm}
\end{figure}

Fig.~\ref{fig_oom} depicts the results of this evaluation. 
In Fig.~\ref{fig_oom} the first annotation marker, labeled \verb|OOMKilled|, signalizes when the current task pod encounters 
the \verb|OOM|~(Out of Memory) event, and the other annotation marker, labeled \verb|Reallocation|, signalizes when the current task pod 
is regenerated by using the reallocated computational resources. 
As can be seen in Fig.~\ref{fig_oom}, at the beginning~(second 0), our ARAS uses the resource scaling method to 
allocate CPU of $1048$ Milli cores and memory of $2009Mi$. 
In this evaluation, the minimum memory for a task pod to run, i.e., the amount of memory operated by the \textit{Stress} tool in the task pod 
is set to $2000Mi$. 
Herein, we only focus on memory resources because memory resources are incompressible resources, and insufficient memory resources will 
trigger the task pod \verb|OOMKilled|, while CPU resources as compressible resources do not. 
Once the allocated memory resource fails to reach $min_{mem}+\beta$~(i.e., $2000Mi+20Mi$), the current task pod turns to \verb|OOMKilled| at $66s$. 
Meanwhile, the workflow with the current \verb|OOMKilled| task pod also terminates execution. 
KubeAdaptor can capture the \verb|OOMKilled| task pod and delete the task pod at $66s$. 
In our experimental evaluation, up to $210$ task pods~($10$ Montage workflows) possess frequent created and deleted operations, 
leading to operation delay of deleting \verb|OOMKilled| task pod.
At $97s$, KubeAdaptor triggers the regeneration of the current task pod, reallocates computational resources, and launches the task pod. 
Since the second allocation of resources is sufficient for the smooth execution of the task pod~($1849m$ CPU and $3560Mi$ memory), 
the task pod is completed at $181s$. At $258s$, KubeAdaptor deletes the completed task pod.

KubeAdaptor equipped with our ARAS in this paper can watch \verb|OOMKilled| events, delete these 
\verb|OOMKilled| task pods, reallocate computational resources, and regenerate these \verb|OOMKilled| task pods, ensuring continuous execution of workflows. 
In production practice, the users inevitably misestimate the resource quota of the main program inside workflow tasks, 
resulting in a large number of \verb|OOMKilled| task pods and termination of workflow execution. 
These countermeasures ensure continuous executions of workflows and keep the KubeAdaptor stable and robust. 
 It also reflects the self-healing and self-configuration abilities of the KubeAdaptor~(mentioned in \ref{sec:mape}). 

\subsubsection{Concluding discussion}
Finally, it can be observed that the KubeAdaptor with our ARAS always achieves better results 
regarding each metric~(Sections \ref{sec:metric}). 
From Montage workflows to LIGO workflows, our ARAS outperforms the baseline against different metrics 
under three distinct workflow arrival patterns. 

Most of the time savings from the total workflow duration and average duration of individual workflow result from the fact that 
the resource scaling method enables our ARAS to maximize resource utilization on cluster nodes 
according to our optimized functions while ensuring the smooth running of workflow pods. 
In addition, the workflow topology with concurrent characteristics also plays a positive role. 

Concerning resource allocation failure and workflow recovery after termination, we have shown in Section~\ref{sec:failure} that 
the KubeAdaptor has abilities to watch the state changes of task pods in real-time, delete the \verb|OOMKilled| task pods, 
and reallocate computational resources for the task pod, followed by the re-creation of this \verb|OOMKilled| task pod and 
recover of workflow executing.

\section{Conclusion}
In this paper, we propose an ARAS for our tailored workflow management engine. 
With the novel architecture of KubeAdaptor and the integration between KubeAdaptor and K8s, our ARAS 
enables the KubeAdaptor to maximize resource utilization through the resource scaling method in response to complex and changing workflow 
requests. 
Experimental evaluations show that our ARAS, ranging from Montage to LIGO workflows, obtain better performances than the baseline algorithm for 
various metrics under different workflow arrival patterns~(Table~\ref{table:evaluation}).
Furthermore, we have shown that the KubeAdaptor detects and handles failure situations of resource allocation in Section~\ref{sec:failure}. 
The self-healing and self-configuration abilities of the KubeAdaptor~(mentioned in \ref{sec:mape}) are also fully verified. 
In our future work, we intend to use KubeAdaptor to analyze different resource allocating algorithms and try to use deep reinforcement learning 
method to investigate cloud resource allocation for cloud workflows. 
In addition, we will study resource allocation strategies suitable for a cloud-edge cooperation environment and provide a practical solution 
for cloud-edge task scheduling.

\ifCLASSOPTIONcompsoc
  \section*{Acknowledgments}
\else
  \section*{Acknowledgment}
\fi

This work was supported in part by the National Natural Science Foundation of China 
(Grant No. 61873030 and No. 62002019), and the 
Beijing Institute of Technology Research Fund Program for Young Scholars.

\ifCLASSOPTIONcaptionsoff
  \newpage
\fi

\bibliographystyle{IEEEtran}
\bibliography{IEEEabrv,references}

\vspace*{-2\baselineskip}
\begin{IEEEbiography}[{\includegraphics[width=1in,height=1.25in,clip,keepaspectratio]{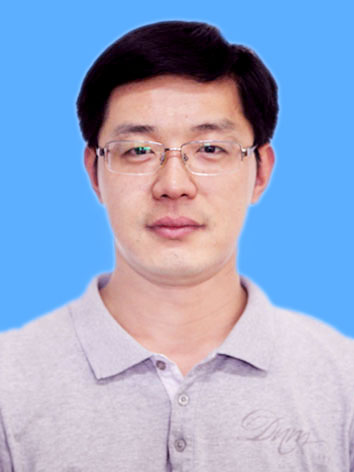}}]{SHAN Chenggang}

was born 1982. He received the M.S. degree in computer applied technology from Qiqihr University, China, in 2007. 
He is working toward the Ph.D. degree with the School of Automation, Beijing Institute of Technology, Beijing, China. 
He was an associate professor with the School of Artificial Intelligence, Zaozhuang University, China, in 2017. 
His research interests include networked control systems, cloud computing, cloud-edge collaboration, wireless networks. 
\\E-mail: uzz\_scg@163.com
\end{IEEEbiography}
\vspace*{-2\baselineskip}

\begin{IEEEbiography}[{\includegraphics[width=1in,height=1.25in,clip,keepaspectratio]{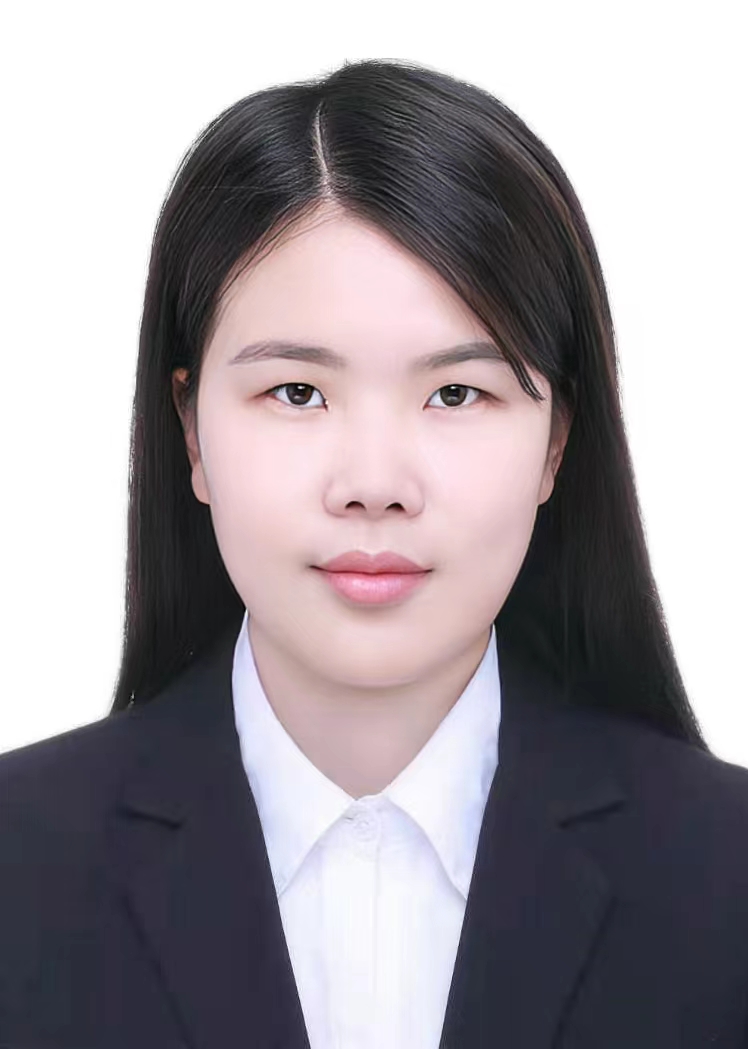}}]{WU Chuge}

was born in 1993. She received the B.E degree in automatic control from Tsinghua University, Beijing, China, in 2015, and the M.S. and Ph.D. degrees in control theory and 
its applications from Tsinghua University, Beijing, China, in 2021. She was a Visiting Scholar with the University of Sydney, NSW, Australia, in 2018. 
She is currently a Assistant Processor for the School of Automation, Beijing Institute of Technology. 
She has published 13 research papers in peer-reviewed journals and conferences. 
Her current research interests include the scheduling and optimization theory and algorithms for cloud computing, fog computing systems, real-time scheduling, 
and evolutionary algorithms.
\\E-mail: wucg@bit.edu.cn
\end{IEEEbiography}

\vspace*{-2\baselineskip}
\begin{IEEEbiography}[{\includegraphics[width=1in,height=1.25in,clip,keepaspectratio]{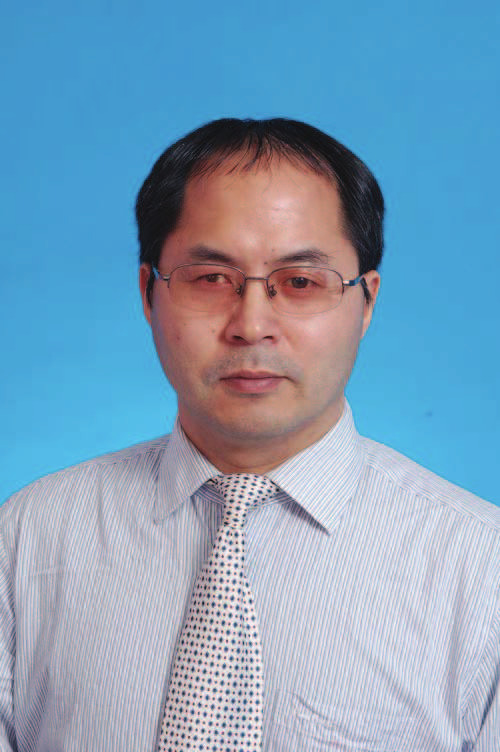}}]{XIA Yuanqing}

was born in 1971. He received his Ph.D. degree in control theory and control engineering from Beijing University of Aeronautics and Astronautics, Beijing, China, in 2001. 
From January 2002 to November 2003, he was a postdoctoral research associate with the Institute of Systems Science, Academy of Mathematics 
and System Sciences, Chinese Academy of Sciences, Beijing, China. From November 2003 to February 2004, he was with National University of Singapore as a research fellow, where he 
worked on variable structure control. From February 2004 to February 2006, he was with University of Glamorgan, Pontypridd, U.K., as a 
research fellow. From February 2007 to June 2008, he was a guest professor with Innsbruck Medical University, Innsbruck, Austria. Since 2004, 
he has been with School of Automation, Beijing Institute of Technology, Beijing, first as an associate professor, then, since 2008, as a professor. 
His research interests include networked control systems, robust control and signal processing, and active disturbance rejection control.
\\E-mail: xia\_yuanqing@bit.edu.cn
\end{IEEEbiography}

\vspace*{-2\baselineskip}

\begin{IEEEbiography}[{\includegraphics[width=1in,height=1.25in,clip,keepaspectratio]{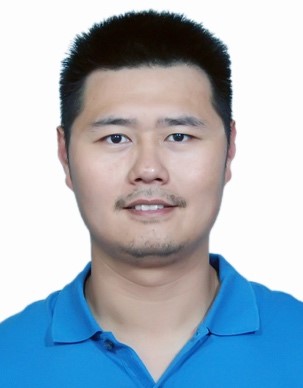}}]{GUO Zehua}

was born in 1985. He received B.S. degree from Northwestern Polytechnical University, Xi’an, China, M.S. degree from Xidian University, Xi’an, China, and Ph.D. degree 
from Northwestern Polytechnical University. He was a Research Fellow at the Department of Electrical and Computer Engineering, 
New York University Tandon School of Engineering, New York, NY, USA, and a Research Associate at the Department of Computer Science and Engineering, 
University of Minnesota Twin Cities, Minneapolis, MN, USA. His research interests include programmable 
networks (e.g., software-defined networking, network function virtualization), machine learning, and network security. 
Dr. Guo is an Associate Editor for IEEE Systems Journal and the EURASIP Journal on Wireless Communications and Networking (Springer), 
and an Editor for the KSII Transactions on Internet and Information Systems. He is serving as the TPC of several journals and conferences 
(e.g., Elsevier Computer Communications, AAAI, IWQoS, ICC, ICCCN, ICA3PP). He is a Senior Member of IEEE, China Institute of Communications, 
Chinese Institute of Electronics, and a Member of ACM and China Computer Federation. 
\\E-mail: guo@bit.edu.cn
\end{IEEEbiography}

\vspace*{-2\baselineskip}

\begin{IEEEbiography}[{\includegraphics[width=1in,height=1.25in,clip,keepaspectratio]{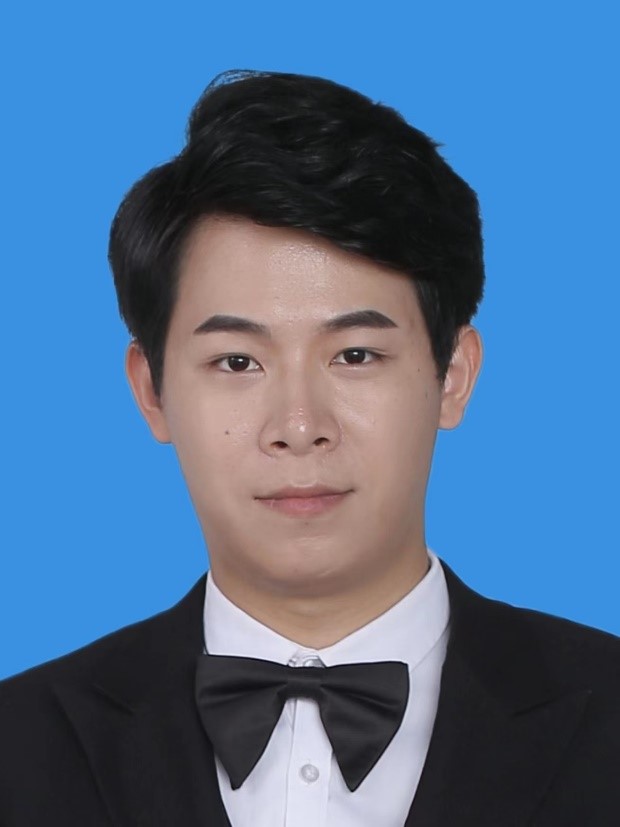}}]{LIU Danyang}

  was born in 1993. He received the B.S. degree in mathematics and information science from the Shijiazhuang University, Shijiazhuang, China, in 2016, and the M.S. degree 
from Hebei University of Science and Technology University, Shijiazhuang, China, in 2020. He is currently pursuing the Ph.D. degree in control science 
and engineering from the School of Automation, Beijing Institute of Technology, Beijing, China. His research interests include cloud computing and data center 
networks. 
\\E-mail: liudanyang093@163.com
\end{IEEEbiography}
\vspace*{-2\baselineskip}
\begin{IEEEbiography}[{\includegraphics[width=1in,height=1.25in,clip,keepaspectratio]{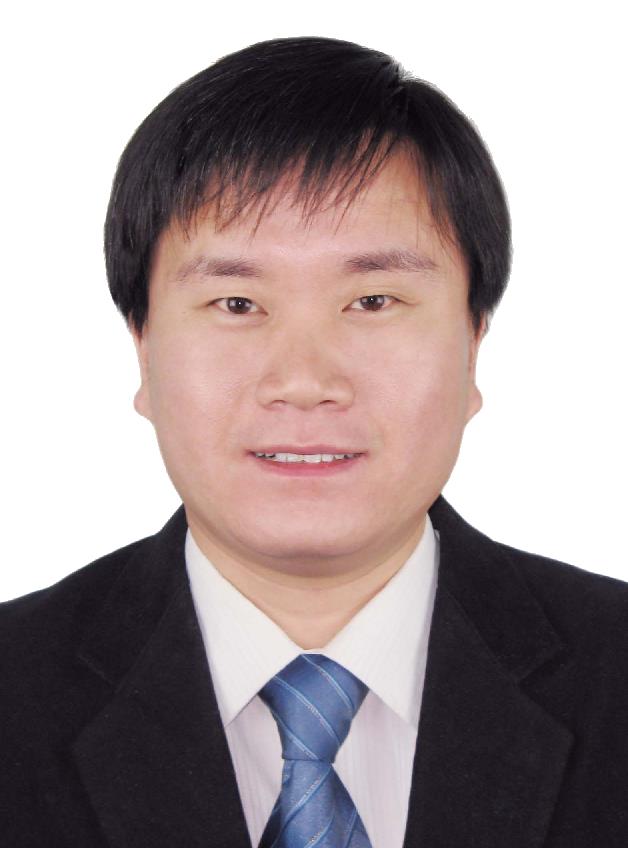}}]{ZHANG Jinhui}

was born in 1982. He received the Ph.D. degree in Control Science and Engineering from Beijing Institute of Technology, Beijing, China, in 2011. 
He was a Research Associate in the Department of Mechanical Engineering, University of Hong Kong, Hong Kong, from February 2010 to May 2010, 
a Senior Research Associate in the Department of Manufacturing Engineering and Engineering Management, City University of Hong Kong, Hong Kong, 
from December 2010 to March 2011, and a Visiting Fellow with the School of Computing, Engineering \& Mathematics, University of Western Sydney, Sydney, 
Australia, from February 2013 to May 2013. He was an Associate Professor in the Beijing University of Chemical Technology, Beijing, from March 2011 to March 2016, 
a Professor in the School of electrical and automation engineering, Tianjin University, Tianjin, from April 2016 to September 2016. He joined Beijing Institute 
of Technology in October 2016, where he is currently an Tenured Professor. His research interests include networked control systems and 
composite disturbance rejection control.
\\E-mail: zhangjinh@bit.edu.cn
\end{IEEEbiography}

\end{document}